\documentclass[amsmath,amssymb,aps,prd,11pt,tightenlines,superscriptaddress,nofootinbib,preprintnumbers,notitlepage]{revtex4-1}

\input{header}

\usepackage{lineno}

\begin{document}

\preprint{{\footnotesize BNL-222142-2021-FORE, CERN-PBC-Notes-2021-025, DESY-21-142, FERMILAB-CONF-21-452-AE-E-ND-PPD-T}}
\preprint{{\footnotesize KYUSHU-RCAPP-2021-01, LU TP 21-36, PITT-PACC-2118, SMU-HEP-21-10, UCI-TR-2021-22}}

\vspace*{0.2in}
\title{{\LARGE The Forward Physics Facility: \\ 
\vspace*{0.05in} Sites, Experiments, and Physics Potential }
\vspace*{0.2in}}

\author{Luis A.~Anchordoqui} 
\affiliation{Department of Physics and Astronomy, Lehman College, City University of New York, Bronx, NY 10468, USA}

\author{Akitaka Ariga}
\affiliation{Albert Einstein Center for Fundamental Physics, Laboratory for High Energy Physics, University of Bern, Sidlerstrasse 5, CH-3012 Bern, Switzerland}
\affiliation{Department of Physics, Chiba University, 1-33 Yayoi-cho Inage-ku, Chiba, 263-8522, Japan}

\author{Tomoko Ariga}
\affiliation{Kyushu University, Nishi-ku, 819-0395 Fukuoka, Japan}

\author{Weidong Bai} 
\affiliation{Sun Yat-sen University, School of Physics, No.~135, Xingang Xi Road, Guangzhou, 510275, P.~R.~China}

\author{Kincso~Balazs}
\affiliation{CERN, CH-1211 Geneva 23, Switzerland}

\author{Brian Batell} 
\affiliation{PITT PACC, Department of Physics and Astronomy, University of Pittsburgh, Pittsburgh, PA 15260, USA}

\author{Jamie Boyd}
\affiliation{CERN, CH-1211 Geneva 23, Switzerland}

\author{Joseph Bramante} 
\affiliation{Department of Physics, Queen's University, Kingston, ON K7L 2S8, Canada}

\author{Mario~Campanelli} 
\affiliation{Department of Physics \& Astronomy,  University College London, Gower Street, London, WC1E 6BT, United Kingdom}

\author{Adrian~Carmona} 
\affiliation{CAFPE and Departamento de F\'isica Teórica y del Cosmos, Universidad de Granada, E18071 Granada, Spain}

\author{Francesco~G.~Celiberto}
\affiliation{European Centre for Theoretical Studies in Nuclear Physics and Related Areas (ECT*), I-38123 Villazzano, Trento, Italy}
\affiliation{Fondazione Bruno Kessler (FBK), I-38123 Povo, Trento, Italy}
\affiliation{INFN-TIFPA Trento Institute of Fundamental Physics and Applications, \sl I-38123 Povo, Trento, Italy}

\author{Grigorios Chachamis}
\affiliation{Laboratorio de Instrumenta\c{c}\~{a}o e Fisica Experimental de Particulas (LIP),  Av.~Prof.~Gama Pinto~2,  P-1649-003 Lisboa, Portugal}

\author{Matthew~Citron}
\affiliation{University of California, Santa Barbara, CA 93106, USA}

\author{Giovanni~De~Lellis}
\affiliation{Dipartimento di Fisica ``E.~Pancini'', Universit\`a Federico II di Napoli, Napoli, Italy}
\affiliation{INFN Sezione di Napoli, via Cinthia, Napoli 80126, Italy}

\author{Albert~De~Roeck}
\affiliation{CERN, CH-1211 Geneva 23, Switzerland}

\author{Hans~Dembinski} 
\affiliation{Exp. Physics 5a, TU Dortmund, Otto-Hahn Str. 4a, 44227 Dortmund, Germany}

\author{Peter B.~Denton} 
\affiliation{High Energy Theory Group, Physics Department, Brookhaven National Laboratory, Upton, NY 11973, USA}

\author{Antonia~Di~Crecsenzo}
\affiliation{Dipartimento di Fisica ``E.~Pancini'', Universit\`a Federico II di Napoli, Napoli, Italy}
\affiliation{INFN Sezione di Napoli, via Cinthia, Napoli 80126, Italy}
\affiliation{CERN, CH-1211 Geneva 23, Switzerland}

\author{Milind~V.~Diwan}
\affiliation{Physics Department, Brookhaven National Laboratory, Upton, NY 11973, USA}

\author{Liam~Dougherty} 
\affiliation{PM Group, Killakee House, Belgard Square, Tallaght, Dublin 24, Ireland}

\author{Herbi~K.~Dreiner} 
\affiliation{BCTP \& Physikalisches Institut, University of Bonn, 53115 Bonn, Germany}

\author{Yong~Du} 
\affiliation{CAS Key Laboratory of Theoretical Physics, Institute of Theoretical Physics, Chinese Academy of Sciences, Beijing 100190, P.~R.~China}

\author{Rikard~Enberg} 
\affiliation{Department of Physics and Astronomy, Uppsala University, Sweden}

\author{Yasaman~Farzan} 
\affiliation{School of physics, Institute for Research in Fundamental Sciences (IPM), P.O. Box 19395-5531, Tehran, Iran}

\author{Jonathan~L.~Feng}
\email[Corresponding author: ]{jlf@uci.edu}
\affiliation{Department of Physics and Astronomy, University of California, Irvine, CA 92697-4575, USA}

\author{Max~Fieg}
\affiliation{Department of Physics and Astronomy, University of California, Irvine, CA 92697-4575, USA}

\author{Patrick~Foldenauer} 
\affiliation{Institute for Particle Physics Phenomenology, Durham University, Durham DH1 3LE, United Kingdom}

\author{Saeid~Foroughi-Abari}
\affiliation{Department of Physics and Astronomy, University of Victoria, Victoria, BC V8W 2Y2, Canada}

\author{Alexander Friedland}
\affiliation{Theory Group, SLAC National Accelerator Laboratory, Menlo Park, CA 94025, USA}

\author{Michael~Fucilla} 
\affiliation{Dipartimento di Fisica, Universit\`a della Calabria, I-87036 Arcavacata di Rende, Cosenza, Italy}
\affiliation{Istituto Nazionale di Fisica Nucleare, Gruppo collegato di Cosenza, I-87036 Arcavacata di Rende, Cosenza, Italy}

\author{Jonathan~Gall}
\affiliation{Turner \& Townsend UK Limited, Low Hall, Calverley Lane, Leeds, United Kingdom}

\author{Maria~Vittoria~Garzelli}
\email[Corresponding author: ]{maria.vittoria.garzelli@desy.de}
\affiliation{II. Institut F\"ur Theoretische Physik, Universit\"at Hamburg, 22761 Hamburg, Germany}

\author{Francesco Giuli}
\affiliation{University of Rome Tor Vergata and INFN, Sezione di Roma 2, Via della Ricerca Scientifica 1, 00133 Roma, Italy}

\author{Victor~P.~Goncalves}
\affiliation{Instituto de F\'{\i}sica e Matem\'atica,  Universidade Federal de Pelotas (UFPel), Caixa Postal 354, CEP 96010-900, Pelotas, RS, Brazil}

\author{Marco~Guzzi} 
\affiliation{Department of Physics, Kennesaw State University, 370 Paulding Ave., Kennesaw, GA 30144, USA}

\author{Francis~Halzen} 
\affiliation{Department of Physics and Wisconsin IceCube Particle Astrophysics Center, University of Wisconsin–Madison, Madison, WI 53706, USA}

\author{Juan~Carlos~Helo} 
\affiliation{Departamento de F\' isica, Facultad de Ciencias, Universidad de La Serena, Avenida Cisternas 1200, La Serena, Chile}
\affiliation{Millennium Institute for Subatomic Physics at High Energy Frontier (SAPHIR), Fernández Concha 700, Santiago, Chile}

\author{Christopher S.~Hill}
\affiliation{The Ohio State University, Columbus, OH 43218, USA}

\author{Ahmed~Ismail}
\affiliation{Department of Physics, Oklahoma State University, Stillwater, OK 74078, USA}

\author{Ameen~Ismail} 
\affiliation{Department of Physics, LEPP, Cornell University, Ithaca, NY 14853, USA}

\author{Richard Jacobsson}
\affiliation{CERN, CH-1211 Geneva 23, Switzerland}

\author{Sudip~Jana} 
\affiliation{Max-Planck-Institut für Kernphysik, Saupfercheckweg 1, 69117 Heidelberg, Germany}

\author{Yu~Seon~Jeong} 
\affiliation{High Energy Physics Center, Chung-Ang University, Seoul 06974, Korea}

\author{Krzysztof Jodłowski} 
\affiliation{National Centre for Nuclear Research, Pasteura 7, 02-093 Warsaw, Poland}

\author{Kevin~J.~Kelly}
\affiliation{Fermilab, Fermi National Accelerator Laboratory, Batavia, IL 60510, USA}

\author{Felix~Kling}
\email[Corresponding author: ]{felix.kling@desy.de}
\affiliation{Theory Group, SLAC National Accelerator Laboratory, Menlo Park, CA 94025, USA}
\affiliation{Deutsches Elektronen-Synchrotron DESY, Notkestrasse 85, 22607 Hamburg, Germany}

\author{Fnu Karan Kumar} 
\affiliation{Physics Department, Brookhaven National Laboratory, Upton, NY 11973, USA}

\author{Zhen Liu}
\affiliation{School of Physics and Astronomy, University of Minnesota, Minneapolis, MN 55455, USA}

\author{Rafa{\l} Maciu{\l}a}
\affiliation{Institute of Nuclear
Physics, Polish Academy of Sciences, Radzikowskiego 152, PL-31-342 Krak{\'o}w, Poland}

\author{Roshan~Mammen~Abraham} 
\affiliation{Department of Physics, Oklahoma State University, Stillwater, OK 74078, USA}

\author{Julien~Manshanden}
\affiliation{II. Institut F\"ur Theoretische Physik, Universit\"at Hamburg, 22761 Hamburg, Germany}

\author{Josh~McFayden}
\affiliation{Department of Physics \& Astronomy, University of Sussex, Sussex House, Falmer, Brighton, BN1 9RH, United Kingdom}

\author{Mohammed~M.~A.~Mohammed} 
\affiliation{Dipartimento di Fisica, Universit\`a della Calabria, I-87036 Arcavacata di Rende, Cosenza, Italy}
\affiliation{Istituto Nazionale di Fisica Nucleare, Gruppo collegato di Cosenza, I-87036 Arcavacata di Rende, Cosenza, Italy}

\author{Pavel~M.~Nadolsky}
\affiliation{Department of Physics, Southern Methodist University, Dallas, TX 75275-0181, USA}

\author{Nobuchika~Okada}
\affiliation{Department of Physics and Astronomy, University of Alabama, Tuscaloosa, AL 35487, USA}

\author{John Osborne}
\affiliation{CERN, CH-1211 Geneva 23, Switzerland}

\author{Hidetoshi Otono} 
\affiliation{Kyushu University, Nishi-ku, 819-0395 Fukuoka, Japan}

\author{Vishvas~Pandey} 
\affiliation{Department of Physics, University of Florida, Gainesville, FL 32611, USA}
\affiliation{Fermilab, Fermi National Accelerator Laboratory, Batavia, IL 60510, USA}

\author{Alessandro Papa}
\affiliation{Dipartimento di Fisica, Universit\`a della Calabria, I-87036 Arcavacata di Rende, Cosenza, Italy}
\affiliation{Istituto Nazionale di Fisica Nucleare, Gruppo collegato di Cosenza, I-87036 Arcavacata di Rende, Cosenza, Italy}

\author{Digesh Raut} 
\affiliation{Bartol Research Institute, Department of Physics and Astronomy, University of Delaware, Newark, DE 19716, USA}

\author{Mary~Hall~Reno} 
\affiliation{Department of Physics and Astronomy, University of Iowa, Iowa City, IA 52242, USA}

\author{Filippo~Resnati}
\affiliation{CERN, CH-1211 Geneva 23, Switzerland}

\author{Adam~Ritz}
\affiliation{Department of Physics and Astronomy, University of Victoria, Victoria, BC V8W 2Y2, Canada}

\author{Juan~Rojo}
\affiliation{Department of Physics and Astronomy, Vrije Universiteit Amsterdam, NL-1081 HV Amsterdam, The Netherlands}

\author{Ina Sarcevic} 
\affiliation{Department of Physics, University of Arizona, Tucson, AZ 85721, USA}

\author{Christiane~Scherb} 
\affiliation{PRISMA+ Cluster of Excellence \& Mainz Institute for Theoretical Physics, Johannes Gutenberg University, 55099 Mainz, Germany}

\author{Holger~Schulz} 
\affiliation{Department of Computer Science, Durham University, South Road, Durham DH1 3LE, UK}

\author{Pedro~Schwaller} 
\affiliation{Institute of Physics and MITP, Johannes Gutenberg University, 55128 Mainz, Germany}

\author{Dipan Sengupta} 
\affiliation{Department of Physics, University of California at San Diego, 9500 Gilman Drive, La Jolla, CA 92093-0319, USA}

\author{Torbj\"orn Sj\"ostrand}
\affiliation{
  Theoretical Particle Physics, Department of Astronomy and Theoretical Physics, Lund University, S\"olvegatan 14A, 223 62 Lund, Sweden}
  
\author{Tyler~B.~Smith} 
\affiliation{Department of Physics and Astronomy, University of California, Irvine, CA 92697-4575, USA}

\author{Dennis~Soldin} 
\affiliation{Bartol Research Institute, Department of Physics and Astronomy, University of Delaware, Newark, DE 19716, USA}

\author{Anna Stasto}
\affiliation{Department of Physics, Penn State University, University Park, PA 16802, USA}

\author{Antoni Szczurek}
\affiliation{Institute of Nuclear Physics, Polish Academy of Sciences, Radzikowskiego 152, PL-31-342 Krak{\'o}w, Poland}

\author{Zahra Tabrizi}
\affiliation{Center for Neutrino Physics, Department of Physics, Virginia Tech, Blacksburg, VA 24061, USA}

\author{Sebastian Trojanowski} 
\affiliation{Astrocent, Nicolaus Copernicus Astronomical Center Polish Academy of Sciences, ul.~Rektorska 4, 00-614 Warsaw, Poland}
\affiliation{National Centre for Nuclear Research, ul.~Pasteura 7, 02-093 Warsaw, Poland}

\author{Yu-Dai~Tsai}
\affiliation{Department of Physics and Astronomy, University of California, Irvine, CA 92697-4575, USA}
\affiliation{Fermilab, Fermi National Accelerator Laboratory, Batavia, IL 60510, USA}

\author{Douglas Tuckler} 
\affiliation{Department of Physics, Carleton University, Ottawa, ON K1S 5B6, Canada}

\author{Martin W.~Winkler} 
\affiliation{The Oskar Klein Centre, Department of Physics, Stockholm University, AlbaNova, SE-10691 Stockholm, Sweden}

\author{Keping Xie}
\affiliation{PITT PACC, Department of Physics and Astronomy, University of Pittsburgh, Pittsburgh, PA 15260, USA}

\author{Yue Zhang \vspace*{0.3in}}
\affiliation{Department of Physics, Carleton University, Ottawa, ON K1S 5B6, Canada}

\begin{abstract}
\vspace*{.3in}
The Forward Physics Facility (FPF) is a proposal to create a cavern with the space and infrastructure to support a suite of far-forward experiments at the Large Hadron Collider during the High Luminosity era. Located along the beam collision axis and shielded from the interaction point by at least 100 m of concrete and rock, the FPF will house experiments that will detect particles outside the acceptance of the existing large LHC experiments and will observe rare and exotic processes in an extremely low-background environment. In this work, we summarize the current status of plans for the FPF, including recent progress in civil engineering in identifying promising sites for the FPF and the experiments currently envisioned to realize the FPF's physics potential.  We then review the many Standard Model and new physics topics that will be advanced by the FPF, including searches for long-lived particles, probes of dark matter and dark sectors, high-statistics studies of TeV neutrinos of all three flavors, aspects of perturbative and non-perturbative QCD, and high-energy astroparticle physics. 
\end{abstract}

\maketitle
\clearpage
\tableofcontents
\clearpage

\section{Introduction}
\label{sec:intro}

Particle physics is at a critical juncture.  The program of discovering Standard Model (SM) particles was completed by the discovery of the Higgs boson nine years ago, but there are still many outstanding questions, from those internal to particle physics to those originating in cosmological observations.  At the same time, the spectacular run of the Large Hadron Collider (LHC), now in its second decade, has not produced evidence of new physics, and there is as yet no consensus plan for the next generation of particle colliders at the energy frontier.  In this context, it is clear that new ideas are welcome and needed, especially if they extend the physics potential of existing facilities, have some guaranteed physics return, and may lead to groundbreaking discoveries that will clarify the path forward in the years to come.  

At present, existing experiments at the LHC are primarily focused on high-$p_T$ physics. For example, most new particle searches target heavy, TeV-scale states, which are produced with pb to fb cross sections and decay to particles traveling at large angles relative to the beamline.  At the same time, these processes are only a small subset of the total inelastic $pp$ collisions, which, at center-of-mass energies of $\sqrt{s} = 13$ to 14 TeV, have a much larger cross section of $\sim$100 mb.  Most of these inelastic collisions produce particles that travel approximately parallel to the beamline and escape through the holes in existing detectors.

In recent years, it has been realized that these ``wasted,'' inelastic collisions may, in fact, contain a treasure trove of useful information. Within the SM, pions, kaons, and other mesons may decay to electron, muon, and tau neutrinos, producing intense beams of highly energetic neutrinos in the far-forward direction.  Modest-sized experiments placed in the far-forward region may detect millions of these neutrinos in the coming decades, enabling precision studies of neutrinos and antineutrinos of all three flavors at the highest human-made energies ever observed.  These neutrino events will also shed light on forward hadron production and both perturbative and non-perturbative QCD, with strong implications for models of collider physics and astroparticle experiments.  For new physics searches, the far-forward region is also promising, because the SM particles produced in the far-forward direction may also decay to new particles.  These exotic decays are typically rare, but in many models, the large fluxes of the parent SM particles predict an intense and highly-collimated beam of light and extremely weakly-interacting particles in the far-forward region, with discovery prospects for new gauge bosons, new scalars, sterile neutrinos, dark matter (DM), millicharged particles (mCPs), and axion-like particles (ALPs).  

The Forward Physics Facility (FPF) is a proposal to realize these physics opportunities by creating space in the far-forward region for a suite of experiments during the High Luminosity LHC (HL-LHC) era.  As noted above, for many years, experiments at the LHC have been dominated by large detectors focused on high-$p_T$ physics.  Very recently, however, the FASER~\cite{Feng:2017uoz, Ariga:2018zuc, Ariga:2018pin, Ariga:2018uku}, FASER$\nu$~\cite{Abreu:2020ddv, Abreu:2019yak, Abreu:2021hol}, and SND@LHC~\cite{Ahdida:2020evc, Ahdida:2750060} detectors have been proposed and approved to operate during LHC Run~3 in the far-forward region, along or very close to the beam collision axis line of sight (LOS).  These detectors are located approximately 480 m from the ATLAS interaction point (IP) and shielded from the ATLAS IP by $\sim 100$ m of concrete and rock. In the far-forward region covered by these experiments, spanning pseudorapidities from $\eta \sim 7$ to $\infty$, a host of interesting SM and beyond the SM (BSM) particle fluxes are maximal, and the shielding of the rock and LHC infrastructure suppresses backgrounds, providing an extremely promising environment for the diverse array of SM and BSM studies noted above.

The FASER, FASER$\nu$, and SND@LHC detectors are currently being constructed to operate in TI12 and TI18, existing injector tunnels that merge with the main LHC tunnel.  These tunnels were excavated for the Large Electron-Positron Collider in the 1980's.  They were never intended to house experiments or provide the necessary services, which are currently being assembled piecemeal for each experiment, and there is no room to expand these detectors or to add additional ones.  The FPF will put the far-forward physics program now underway on a solid footing, either by extending an existing LHC cavern or by building a new, purpose-built facility along the LOS.  The resulting facility will potentially extend coverage to pseudorapidities below 7 and will accommodate larger experiments to more fully realize the physics opportunities provided by the far-forward region. 

The FPF's special location makes its experiments uniquely sensitive to many SM and BSM phenomena, and its physics capabilities are complementary to those of other existing and proposed experiments at the LHC. Besides the large LHC experiments probing high-$p_T$ physics, these include a number of smaller detectors performing SM measurements in the forward region, including ALFA~\cite{AbdelKhalek:2016tiv}, AFP~\cite{Grinstein:2016sen},  CASTOR~\cite{CMS:2020ldm}, LHCf~\cite{Adriani:2008zz}, TOTEM~\cite{Anelli:2008zza}, and CT-PPS~\cite{CMS:2014sdw}.  These are located in or around the LHC beam pipe close to either the ATLAS or CMS IP, but, in contrast to the FPF, are not shielded from these IPs by hundreds of meters of concrete and rock. The FPF is also complementary to MoEDAL~\cite{Acharya:2014nyr} and MilliQan~\cite{Haas:2014dda,Ball:2016zrp}, as well as proposed experiments, such as MATHUSLA~\cite{Chou:2016lxi, MATHUSLA:2018bqv, MATHUSLA:2020uve}, CODEX-b~\cite{Gligorov:2017nwh, Aielli:2019ivi}, and ANUBIS~\cite{Bauer:2019vqk}, which also aim to search for new physics at the LHC, but, in contrast to FPF experiments, are located at large angles relative to the beamline. Last, there are also important synergies of FPF physics with experiments running or proposed at other facilities, including, for example, BSM searches at beam dump experiments, such as SHiP~\cite{Alekhin:2015byh} at the SPS, and SM studies at accelerators, such as the Electron-Ion Collider (EIC)~\cite{AbdulKhalek:2021gbh} at Brookhaven, as will be described below. 

To realize the FPF's promise, physics studies must be carried out to guide which experiments should be placed in the far-forward region, the experiments must be designed and built, and a facility to house the necessary experiments is required.  Of course, all of these aspects must be considered together to maximize the physics output within the constraints of time, space, and funding. Although work on the FPF is in its early stages, in the last year, there has been growing interest in the physics potential of the FPF, leading to a Snowmass Letter of Interest~\cite{SnowmassFPF}, two dedicated workshops~\cite{FPFKickoffMeeting,FPF2ndMeeting}, and numerous studies, many of which will be reviewed in this paper.  The goal of this work is to bring together many of the rapid developments in this area and to summarize the current status of studies for the proposed FPF.  

This paper is organized as follows.  In \secref{facility}, we outline the civil engineering studies that have been done so far, which have identified two leading candidate sites for the FPF.  In \secref{experiments}, we summarize the experiments currently envisioned for the FPF, which span a number of detector technologies to realize the FPF's diverse physics goals.  The physics potential of these experiments for BSM searches, neutrinos, QCD, and astroparticle physics is then described in Secs.~\ref{sec:bsm}, \ref{sec:neutrinos}, \ref{sec:qcd}, and \ref{sec:astro}, respectively.  The division into these four subject areas is somewhat artificial, as there are large overlaps and synergies between the different sections, but the organization highlights the relevance of the FPF to a diverse group of well-established communities.  We conclude in \secref{conclusions} with a brief summary and outlook.

\section{The Facility and Civil Engineering}
\label{sec:facility}




\subsection{Overview}

Civil Engineering (CE) generally represents a significant portion of the effort for physics projects like the FPF.  For this reason, CE studies are of critical importance to ensure a viable and cost-efficient conceptual design. This section provides an overview of the current status of FPF CE studies, including key considerations and the current designs being studied. 

\begin{figure}[tb]
  \centering
 \includegraphics[width=0.98\textwidth]{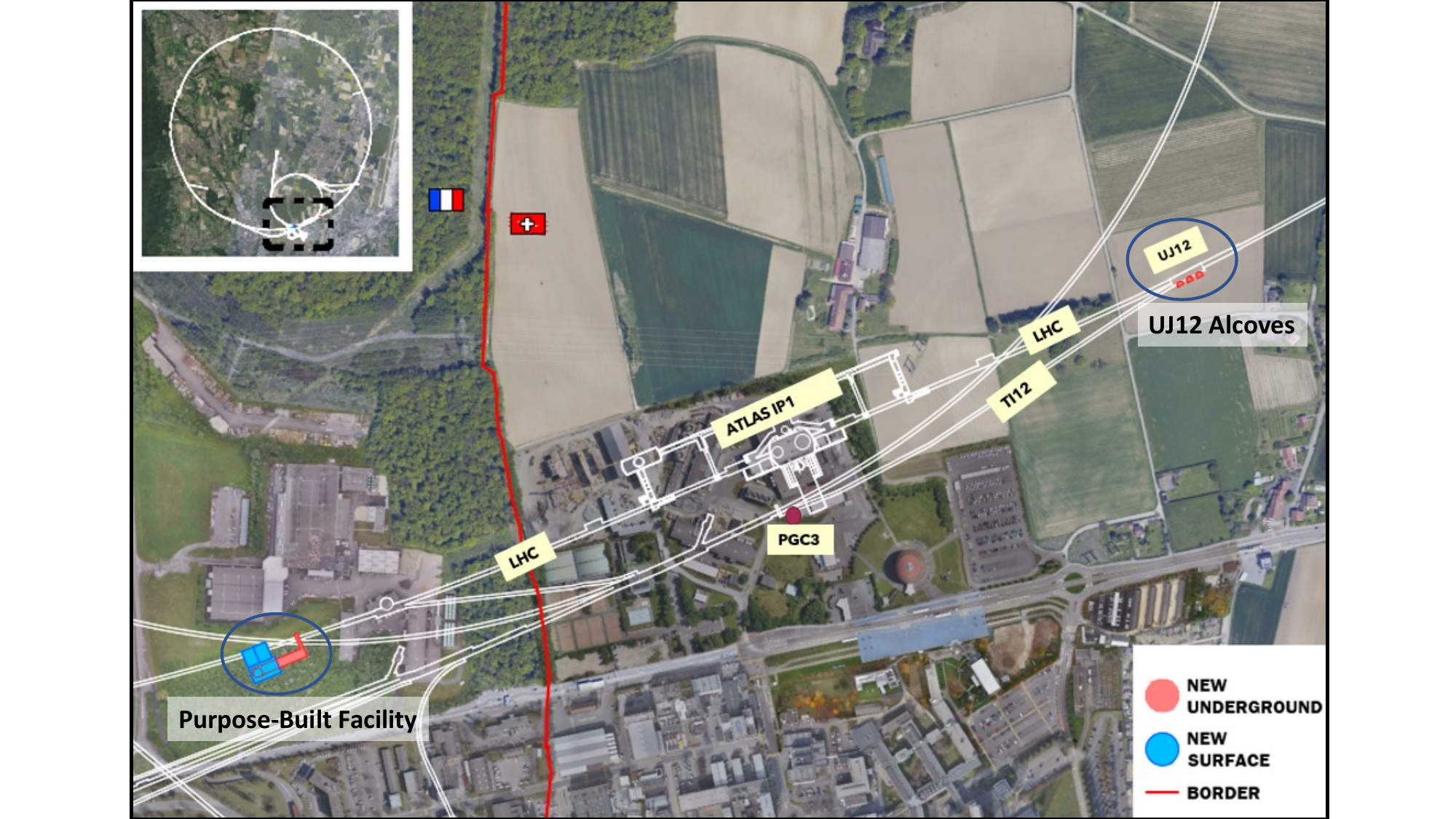}
  \caption{The locations of the two preferred FPF sites currently under consideration. For the UJ12 alcoves option, the existing UJ12 cavern is enlarged with alcoves, providing a site located roughly 480--521 m west of ATLAS IP1 in Switzerland.  For the purpose-built facility option, a new cavern and shaft are excavated, creating a site on the LOS roughly 617--682 m east of ATLAS IP1 on CERN land in France.}
  \label{fig:FPFoptions}
\end{figure}

The CE studies have been based on the requirement that the FPF be approximately 500--600~m away from an LHC IP on the beam collision axis or LOS. Following an initial study of the existing LHC infrastructure and geological conditions, several options were considered to accommodate the facility around the ATLAS IP (IP1). These have been narrowed down to two preferred solutions: alcoves in the UJ12 cavern and a purpose-built facility. The locations of these two options are shown in \cref{fig:FPFoptions}.  In the next two subsections, we present more details of each of these two FPF sites.

\subsection{Alcoves in the UJ12 Cavern}

One of the preferred options is to house the FPF in the location of the existing UJ12 cavern by expanding one side of UJ12 with separate alcoves to accommodate the experiments and to provide the space needed around them.  UJ12 is part of the LHC tunnel system and is situated approximately 480-521 m west of ATLAS IP1 at CERN's site in Switzerland, as shown in \cref{fig:FPFoptions}. 

A drawback of the UJ12 alcoves option is the difficulty of accessing the worksite.  As an access point, it is envisaged to use the existing PGC3 shaft located on the top of the abandoned tunnel TI12, and then passing through the 536 m long TI12, which currently houses the FASER experiment. The PGC3 shaft has an internal diameter of 3 m, which imposes significant space constraints, and the works need to be designed around what can be achieved with only small equipment. 

\begin{figure}[tb]
  \centering
  \includegraphics[width=0.98\textwidth]{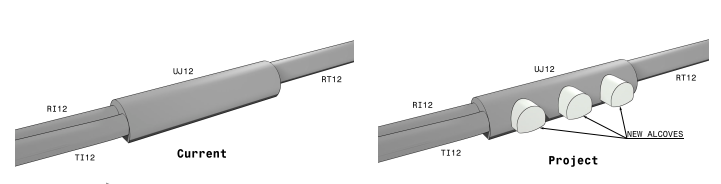}
  \caption{Proposed CE works for the UJ12 alcoves option.  The existing UJ12 cavern would be enlarged by adding three alcoves, each with 6.4 m width, but with different lengths, to accommodate experiments and accompanying services.}
  \label{fig:UJ12CEworks}
\end{figure}

Following the conceptual design studies, the baseline layout includes three alcoves, each with 6.4 m width, but with different lengths, as shown in \figsref{UJ12CEworks}{UJ12alcoves}. It must be noted that the impact of the foreseen works on the existing wall of the cavern and the cavern itself has yet to be fully assessed. All the works must be carried out in a way that minimizes the impact on the existing facility. It is assumed that all the existing services and equipment will be removed from the cavern prior to the works. This would include temporarily removing 4 LHC dipole magnets and a 60 m-long section of the QRL cryogenic line, as well as electrical and ventilation equipment. Initial studies suggest that this would be possible during a multi-year Long Shutdown between LHC runs, but it would be significant work for many CERN teams. 

\begin{figure}[tb]
  \centering
  \includegraphics[width=0.98\textwidth]{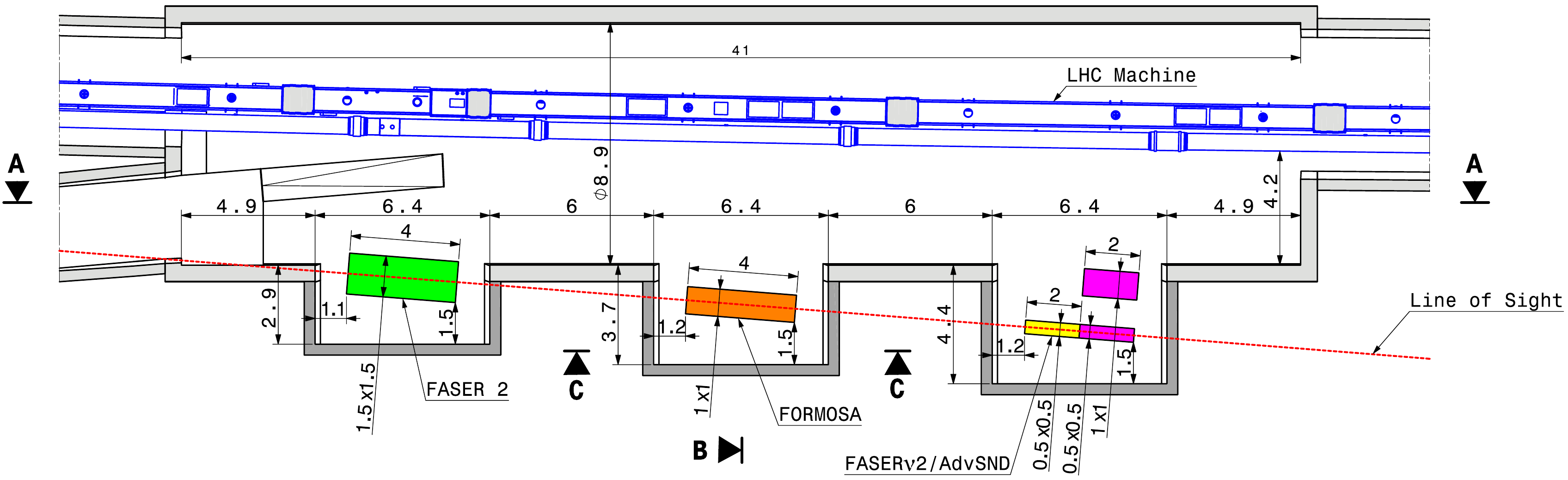}
  \caption{Plan view of the UJ12 alcoves option.  The colored boxes indicate the possible experiments (and their dimensions) that could be installed in the alcoves, including FASER2 to search for long-lived particles, FORMOSA to search for mCPs, and FASER$\nu$2 and AdvSND to detect neutrinos and search for DM.}
  \label{fig:UJ12alcoves}
\end{figure}

\subsection{Purpose-Built Facility}

\begin{figure}[tb]
  \centering
  \includegraphics[width=0.8\textwidth]{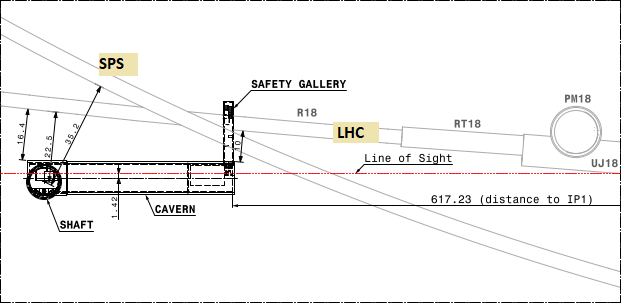}
  \caption{Situation plan of the purpose-built facility option, located approximately 617 m to the west of IP1 on the French side of CERN land, 10 m away from the LHC tunnel.}
  \label{fig:NewFacility-plan}
\end{figure}

\begin{figure}[tb]
  \centering
  \includegraphics[width=0.98\textwidth]{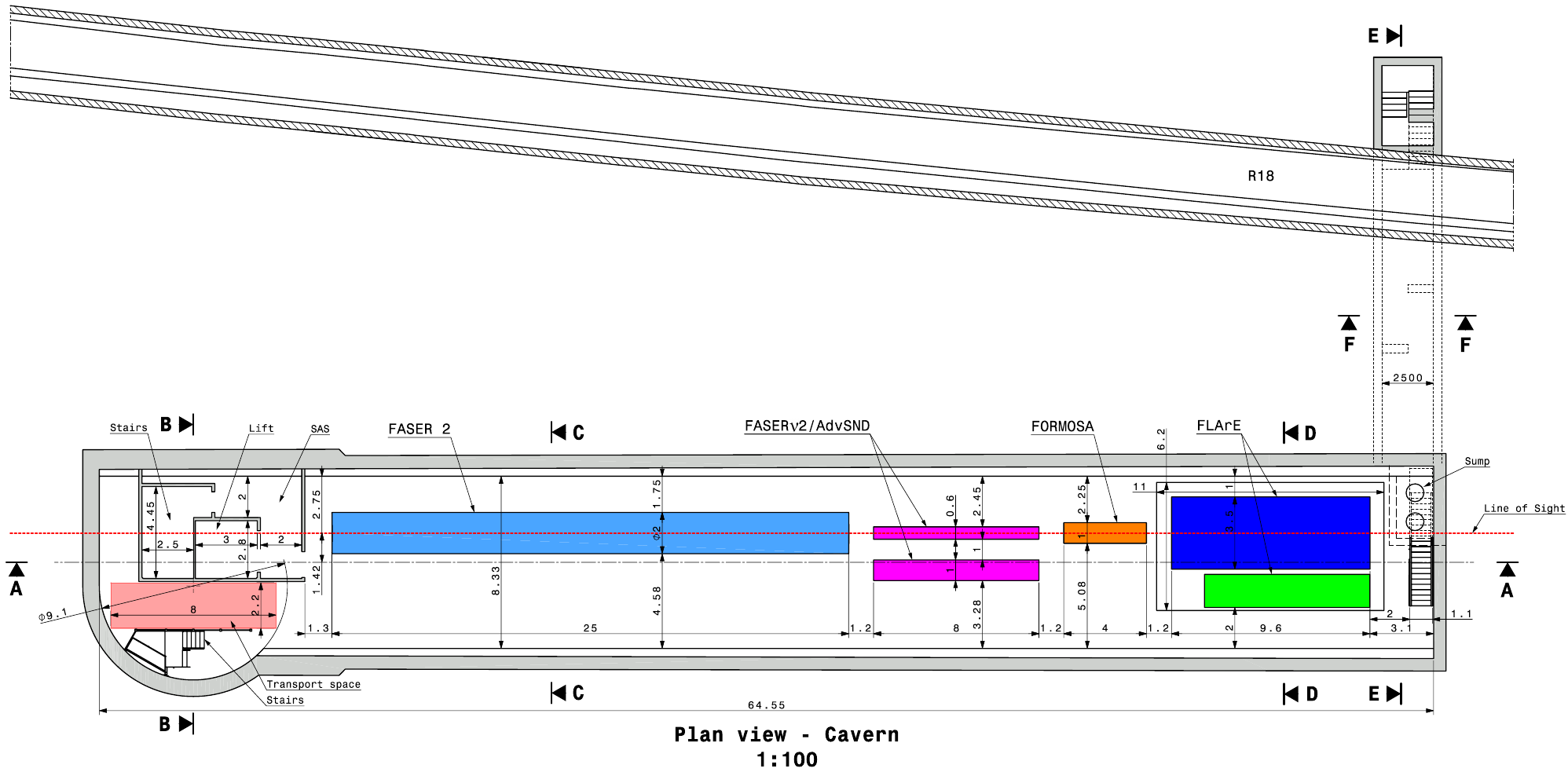}
  \caption{General layout plan of the purpose-built facility option.  The colored boxes indicate the possible experiments (and their dimensions) that could be installed in this option, including FASER2 to search for long-lived particles, FASER$\nu$2 and AdvSND to study neutrinos and search for new particles, FORMOSA to search for mCPs, and FLArE to detect neutrinos and search for DM.  The green box is a possible cooling unit for FLArE. }
  \label{fig:NewFacility-plan2}
\end{figure}

\begin{figure}[tb]
  \centering
  \includegraphics[width=0.52\textwidth]{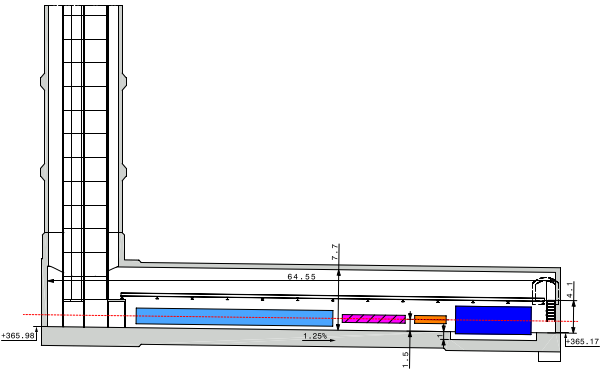}
  \includegraphics[width=0.46\textwidth]{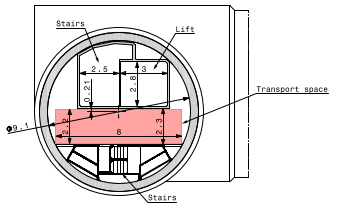}
  \caption{Section through the cavern and access shaft for the purpose-built facility.  The colored boxes are the example experiments shown in \cref{fig:NewFacility-plan2}.}
  \label{fig:NewFacility-section}
\end{figure}

The construction of a new facility is proposed as a second option to implement the FPF at CERN.  The proposed location begins approximately 617 m from IP1 on the French side of CERN land, 10 m away from the LHC tunnel, as shown in \cref{fig:FPFoptions}.  More detailed views are given in Figs.~\ref{fig:NewFacility-plan}, \ref{fig:NewFacility-plan2}, and \ref{fig:NewFacility-section}. 

\begin{figure}[tb]
  \centering
  \includegraphics[width=0.65\textwidth]{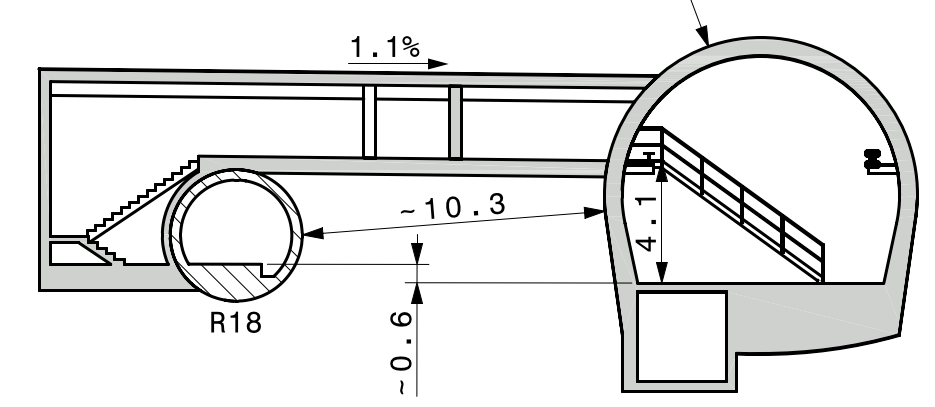}
  \caption{Proposed safety gallery for the purpose-built facility. The safety gallery connects the FPF cavern on the right to the LHC tunnel (R18) on the left.}
  \label{fig:NewFacility-safety}
\end{figure}

\begin{figure}[tb]
  \centering
  \includegraphics[width=0.47\textwidth]{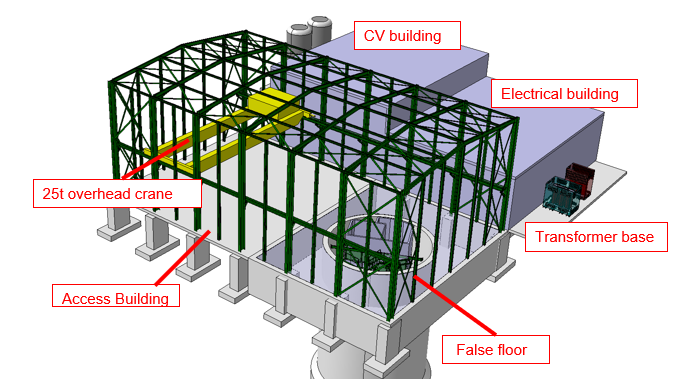}
  \includegraphics[width=0.47\textwidth]{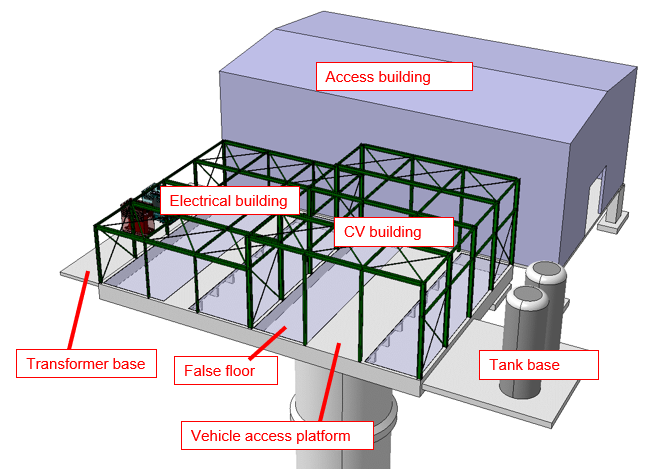}  
  \caption{Surface buildings for the purpose-built facility.}
  \label{fig:NewFacility-SurfaceBuildings}
\end{figure}

The main features of the current layout for the purpose-built facility are:
\vspace*{-0.8em}
\begin{itemize}
\setlength\itemsep{-0.4em}
\item {\em 65 m-long cavern}.  The experimental cavern located on the LOS will be approximately 65 m long and 8.5 m wide and equipped with a crane serving the experiments along the cavern. The floor level is set at 1.5 m under the LOS, with a 1.25\% fall towards IP1, following the inclination of the LOS.  For safety reasons, given the potential of cold gas leakage, a 1~m-deep trench is foreseen under the LAr detector (FLArE).  See \figsref{NewFacility-plan2}{NewFacility-section}. 
\item {\em 9.1~m internal diameter shaft.} The 88 m-deep and 9.1 m-diameter shaft will be located on the top of the experimental cavern. It will be equipped with a lift and staircase for access, having enough space reserved for transport, as shown in \cref{fig:NewFacility-section}. 
\item {\em Safety gallery.}   To comply with CERN's safety requirements and avoid any possible dead ends, a safety gallery will connect the experimental cavern to the LHC, as shown in \cref{fig:NewFacility-safety}. 
\item {\em Surface buildings.} The surface buildings are designed as steel portal frame structures; see \cref{fig:NewFacility-SurfaceBuildings}. The access building located over the shaft will be equipped with a 25-tonne overhead crane to lower the experiments into the cavern. The service buildings for electrical, cooling, and ventilation infrastructure will be adjacent to the access building with a 1.2 m-deep false floor to allow the services to be distributed into the shaft. 
\end{itemize}

\subsection{Civil Engineering Costs}

The cost of construction is difficult to estimate at such an early stage of the study. The variability of ground conditions, inflation, change of scope, and lack of detailed design means that developing a high level of confidence is not possible. For FPF costing purposes, a comparative costing was adopted, based on the presented layouts.  A very preliminary cost estimate suggests that the UJ12 option would cost about 15 MCHF, whereas the purpose-built facility, including the needed services (as discussed in the next subsection), would cost about 40 MCHF.  The accuracy of the estimates is considered Class 4 -- Study or Feasibility, which could be 15--30\% lower or 20--50\% higher~\cite{CEcosting}. Until the project requirements are further developed, it is suggested that a suitable band to adopt would be 20\% lower to 40\% higher for CE costs.

\subsection{Services}

Given the early stage of the project and the lack of designs and requirements for the proposed experiments in the FPF, further work is required for a detailed understanding of the needed services.  For the UJ12 alcoves option, most of the needed services would be available from close by within the LHC infrastructure. On the other hand, based on similar underground facilities at CERN, it is clear that the purpose-built facility would need dedicated services, including electrical distribution, ventilation system, transport/handling infrastructure, communication infrastructure, access and alarm systems, and safety systems. 

\subsection{Sweeper Magnet}
\label{sec:sweepermagnet}

FLUKA simulations and {\em in situ} measurements made by the FASER Collaboration~\cite{Ariga:2018pin} show an expected flux of muons arising from the collisions in IP1 at the level of $2.5~\text{Hz}/\cm^2$ (for the expected HL-LHC luminosity of $5 \times 10^{34}~\cm^{-2}~\text{s}^{-1}$) along the LOS. The rate is expected to rise by up to an order of magnitude in some directions when greater than 1 m from the LOS.  More details on the expected muon flux can be seen in Figs.~4 and 5 of Ref.~\cite{Ariga:2018pin}.  Although this background particle rate is very low compared to the LHC collision rate and to the background rates at the current LHC experiments, it could be problematic for some of the experiments proposed for the FPF.

To reduce the background rate, it may be possible to install a sweeper magnet in the LHC tunnel where the LOS leaves the LHC magnet cryostats, but before it leaves the LHC tunnel.  This location is about 100 m from the UJ12 cavern and about 200 m from the proposed location of the purpose-built facility. Preliminary studies suggest that, with some modifications to the cryogenic infrastructure, there could be space for a 20 cm-diameter magnet with bending power of 7 Tm to be installed in this location. Detailed simulations are needed to see how much such a magnet would reduce the muon background in the FPF, but naively this would bend a 100 GeV muon 4 m (2 m) away from the LOS for the purpose-built facility (UJ12 alcoves) option, significantly reducing the background rate.

\subsection{Conclusions}

From a pure civil engineering point of view, both the UJ12 alcoves and purpose-built facility options are feasible.  The UJ12 alcoves option would be lower in cost, but would allow only 2 to 3 alcoves, with the experiments being designed around what is possible, while having a major impact on the existing infrastructure. In contrast, the purpose-built facility would be higher in cost, but can be designed around the needs of the experiments and has the advantage of not being limited in size and length in comparison to the UJ12 alcoves option. 

Other important advantages of the purpose-built facility over the UJ12 alcoves option include:
\vspace*{-0.5em}
\begin{itemize}
\setlength\itemsep{-0.4em}
\item Much of the excavation work could likely be carried out during LHC operations, allowing more flexible scheduling of when the FPF could be implemented.
\item Initial radio protection studies suggest that people would be able to work in the FPF cavern during LHC operations, which would be very beneficial for installation, commissioning, and maintenance of the experiments (including the exchange of emulsion films).
\item The size of experiment components would be very limited (by the size of the LHC transport corridor) in the UJ12 alcoves option.  
\item The purpose-built facility would allow dedicated safety systems to be included for a large liquid argon-based detector, such as FLArE.  Such a detector would not be possible in the UJ12 alcoves option.
\item If desired by considerations of the physics requirements, the purpose-built facility would allow a detector to be placed up to a few meters off-axis.  Such a positioning would be much more difficult for the UJ12 alcoves option.
\item During the preparatory work for the FASER experiment, the radiation level in UJ12 was measured to be low; however, it could still be at a level that is problematic for experiments installed in the UJ12 alcoves. Due to the shielding provided by at least 10 m of rock, the radiation level will not be a problem for the purpose-built facility.
\item FLUKA simulations and {\em in situ} measurements show that the beam-related physics background for the FASER experiment close to UJ12 is very low. However, for experiments searching for very rare low-energy processes, these backgrounds could still be problematic for the physics goals of experiments in the UJ12 alcoves option. Due to the shielding from the beamline provided by greater than 10 m of rock, this will not be a concern for experiments in the purpose-built facility.
\end{itemize}

\section{Proposed Experiments}
\label{sec:experiments}




\subsection{FASER2 \label{sec:FASER2}}

The existing FASER experiment is already set to probe new parameter space in the search for BSM physics. However, the overall size of FASER, and therefore its possible decay volume, has been heavily constrained by the available space underground ever since the initial stages of planning. This directly affects the sensitivity and reach obtainable by FASER, since, for many representative BSM models, the sensitivity is directly related to the length and radius of the decay volume. This strongly motivates the case for an enlarged detector, FASER2, which was already explored in the FASER Letter of Intent~\cite{Ariga:2018zuc}, Technical Proposal~\cite{Ariga:2018pin}, and physics reach~\cite{Ariga:2018uku} documents. 

In previous studies, the nominal FASER2 design is comprised of a cylindrical decay volume 5~m in length and 2~m in diameter. This results in an angular acceptance of neutral pions that increases from 0.6\% in FASER to 10\% in FASER2, as shown in  Fig.~5 (left) of Ref.~\cite{Ariga:2018uku}.  In addition, there is a significant improvement in sensitivities to long-lived particles (LLPs) produced in decays of heavy mesons, due to the additional acceptance of $B$-meson production, as shown in Fig.~5 (right) of Ref.~\cite{Ariga:2018uku}. The larger decay volume also improves sensitivity to larger LLP masses and longer LLP lifetimes. The combined effect of all these factors, as well as the increased luminosity expected for the HL-LHC over LHC Run~3, is an improvement in reach of 4 orders of magnitude for some models~\cite{Ariga:2018uku}.

There are several key design considerations for FASER2. The larger radius reduces the importance of being directly on-axis. The significant improvement in sensitivity to higher mass LLPs has the consequence of exposing FASER2 to a more complicated mixture of decay channels, which strongly motivates particle identification capabilities to differentiate between, for example, electrons, pions, and kaons. The factor of 10 increase in decay volume radius corresponds to a factor of 100 increase in area, which needs to be instrumented. It therefore becomes much more challenging to accommodate an extended version of the ATLAS SCT tracker module configuration, currently used in FASER, given cost considerations and the services required. However, the marked increase in detector length of FASER2 creates the potential to achieve larger decay product separations with different and possibly cheaper technologies. The overall increase in detector size will also lead to a larger background rate, which is likely to require more complicated trigger and data analysis techniques.

Given these considerations, there is much to be studied in terms of possible detector configurations and technologies. So far, studies have focused on general size/layout optimizations. Several possibilities for decay volume sizes and locations have been considered, based on the constraints imposed by the FPF facility scenarios discussed in \secref{facility}; these are shown in \tableref{FASER2_configs}.

\begin{table}[tb]
    \centering
    \begin{tabular}{c||c|c|c|c|c}
\hline
\hline
 & Distance 	& Available & \ Decay Volume \ & Available & \ Decay Volume \ \\
FPF Scenario &\  to IP [m] \ & \ Length [m] \ & Length [m]	&  \ Diameter [m] \ & Diameter [m] \\
\hline
F2: Original FASER2 &	480 & 15 & 5 & 2 & 2 (/ 1 / 0.5) \\
S1: UJ12 Alcoves &	500 &	5	& 1.5 (/ 2) & 1.52 & 2 / 1 (/ 0.5) \\
\ S2: Purpose-Built Facility \ & 620 &	25	& 10 (/ 15 / 20) &	2 &	2 / 1 (/ 0.5) \\
\hline
\hline
\end{tabular}
\caption{Possible FASER2 design parameters, given the FPF options described in \secref{facility}. Dimensions in parentheses show configurations that have been considered, but are not shown in the figures or discussed in the text.}
\label{table:FASER2_configs}
\end{table}

\begin{figure}[tb]
  \centering
  \includegraphics[width=0.49\textwidth]{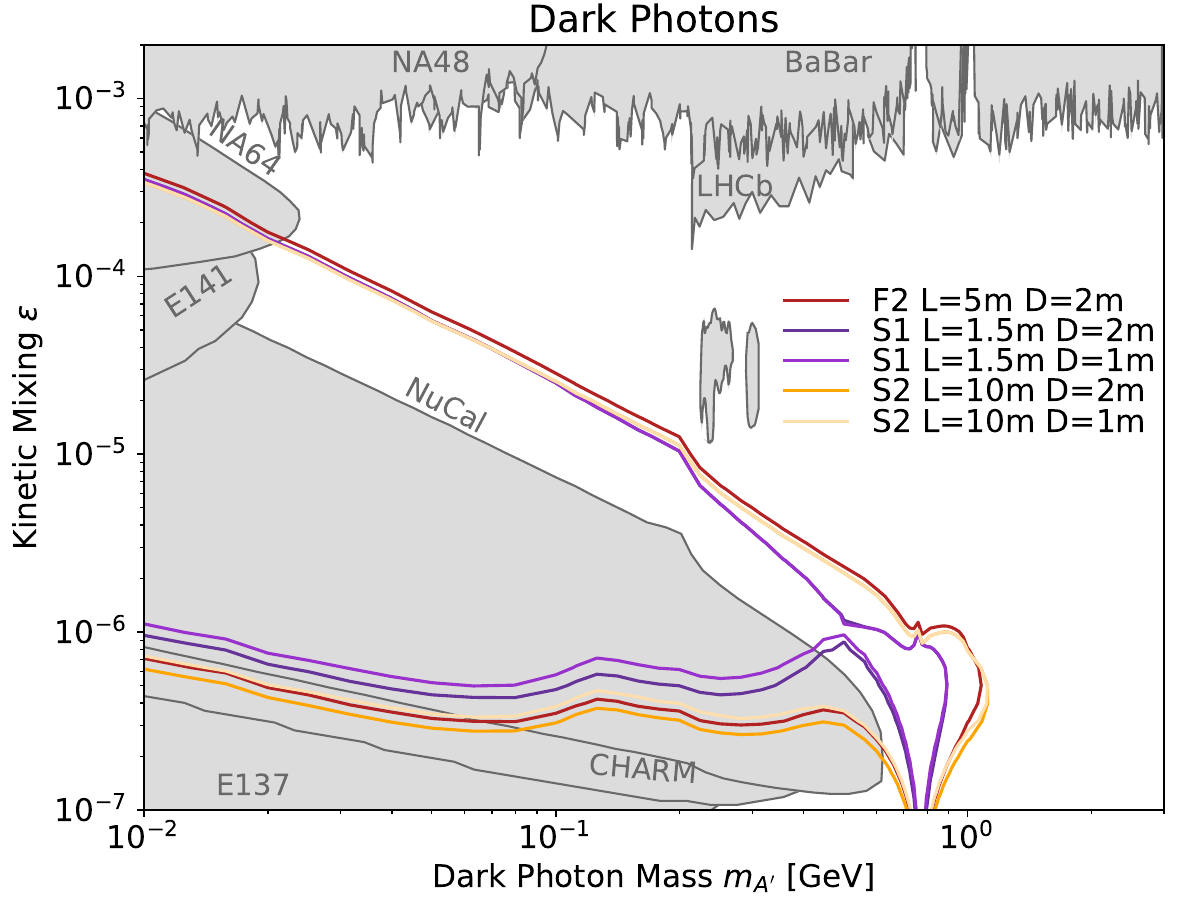}
  \includegraphics[width=0.49\textwidth]{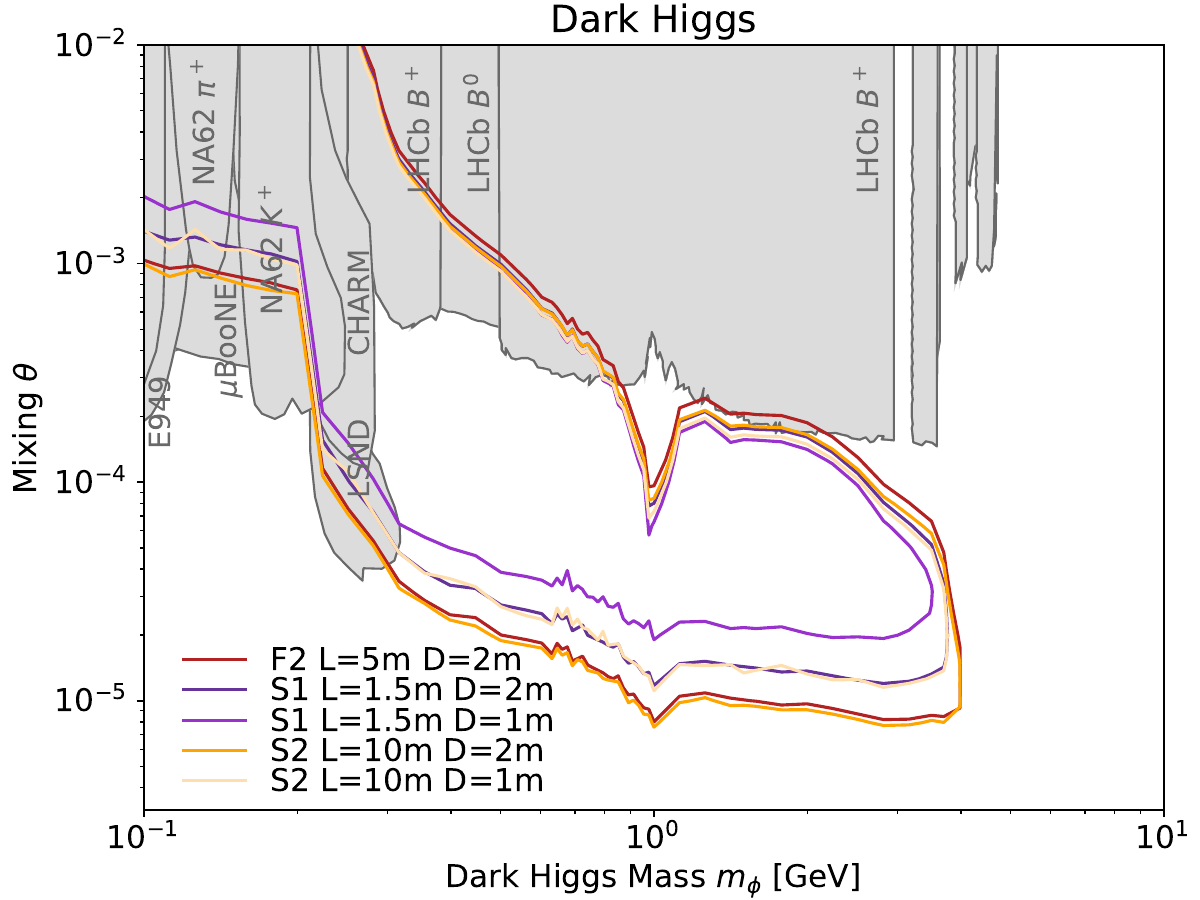}  
  \caption{Projected reach sensitivities for dark photon (left) and dark Higgs (right) models for various FASER2 scenarios described in \tableref{FASER2_configs}.}
  \label{fig:FASER2-Scenarios}
\end{figure}

\cref{fig:FASER2-Scenarios} shows the sensitivity to dark photon (left) and dark Higgs (right) models for a selection of the possible FASER2 scenarios shown in \tableref{FASER2_configs}. The dark photon and dark Higgs models contain new particles with spin-1 and spin-0, respectively, that couple to SM particles through renormalizable couplings; for more details, see \cref{sec:llps}. The sensitivities have been determined using the FORESEE tool~\cite{Kling:2021fwx}. For dark photons, the sensitivity in the nominal FASER2 configuration is significantly greater than for FASER. However, the UJ12 alcoves option does not allow for such a large detector, and \cref{fig:FASER2-Scenarios} shows a significant loss of sensitivity with respect to the nominal. However, FASER2 scenarios located in the purpose-built facility option are able to recover and even improve upon the default FASER2 sensitivity, making it the strongly-preferred scenario. The only downside to the purpose-built facility scenario is the slight shift in sensitivity along the diagonal due to the increased distance from the ATLAS IP, but this is a rather small effect.  Similar conclusions can be drawn for the dark Higgs sensitivity, where the effect of the increased radius is even stronger due to the enhancement in acceptance to $B$-meson decays already discussed.

\begin{figure}[tb]
  \centering
  \includegraphics[width=0.85\textwidth]{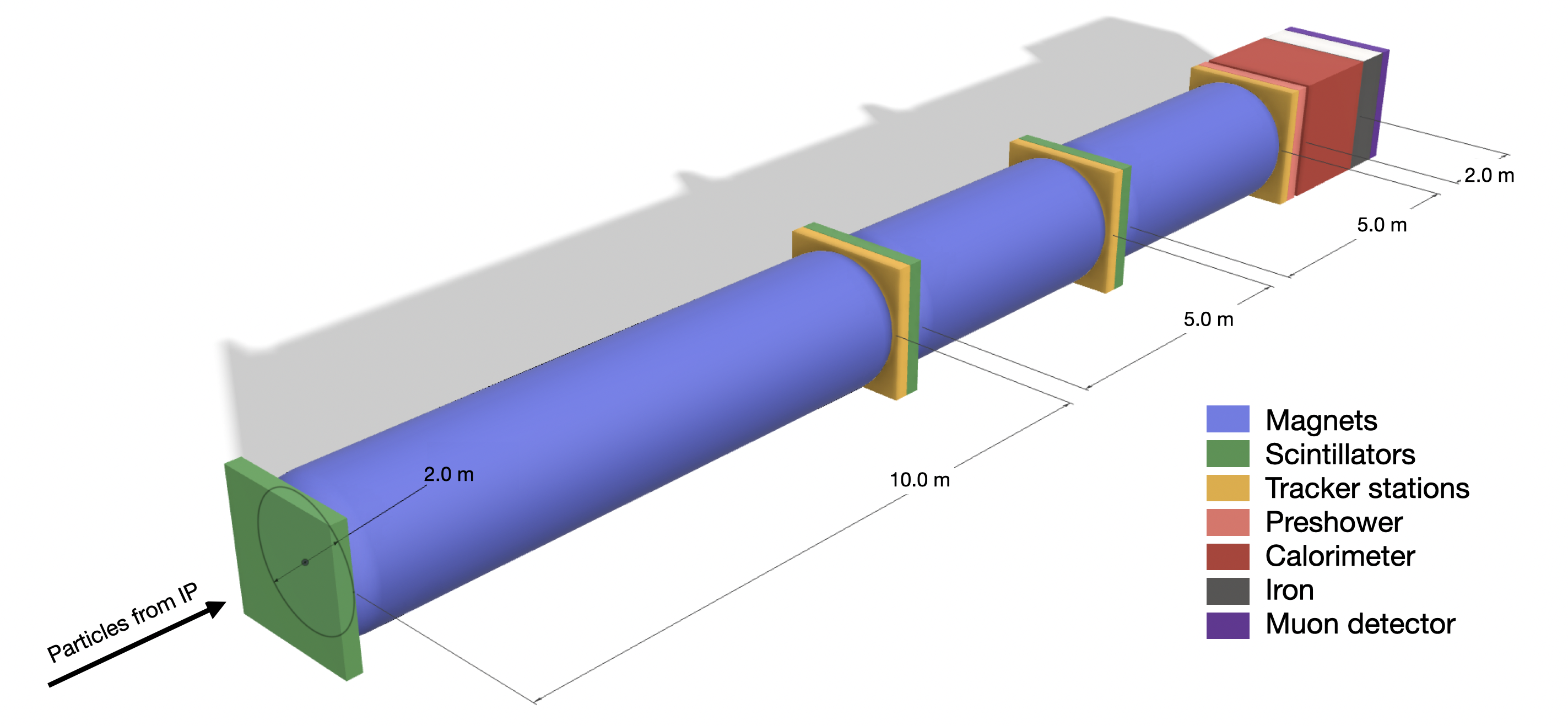}
  \caption{Schematic diagram of the proposed FASER2 detector.  The design shown has a cylindrical decay volume with 2 m diameter and 10 m length.  The total length is approximately 22 m, which could be accommodated in the purpose-built facility option for the FPF. }
  \label{fig:FASER2-Design}
\end{figure}

The FASER2 design can be optimized for either the UJ12 alcoves or purpose-built facility FPF options. The design is not yet strictly defined, but it will be similar to FASER in its general philosophy, modulo changes needed to ameliorate some of the additional challenges described above. A schematic layout of the FASER2 detector, assuming the purpose-built facility option, is given in \cref{fig:FASER2-Design}. The veto system will be scintillator-based, similar to FASER2. The significantly increased area of the active volume makes it impractical to use silicon tracker technology. A Silicon Photomultiplier and scintillating fiber tracker technology, such as LHCb's SciFi detector~\cite{Hopchev:2017tee}, is a strong candidate to replace the ATLAS SCT modules used in FASER. In addition, Monitored Drift Tube technology, similar to that used in the ATLAS New Small Wheel~\cite{Kawamoto:1552862}, is also being considered, although this option requires the use of gases in the LHC tunnel that could be problematic for the UJ12 alcoves option.  Superconducting magnet technology would be required to maintain sufficient field strength across the much larger aperture.  Suitable technology for this already exists and can be built for FASER2. There are several possibilities for the cooling of such magnets, and the use of cryocoolers and the possibility to share a single cryostat across several magnets are being considered. 

Searching for decays of new particles into neutral final states (such as $\text{ALP} \to \gamma \gamma$) motivates the need to be able to identify events with closely-spaced high-energy photons and to separate these from neutrino interactions in the calorimeter, for example, with a high granularity pre-shower in front of the calorimeter.  Such a detector is under consideration to be installed as an upgrade to the existing FASER experiment. In addition, the calorimeter needs to have good energy resolution; improved longitudinal separation with respect to FASER; and the capability to perform particle identification, separating, for example, electrons and pions. Dual readout calorimetry~\cite{Antonello:2018sna} is a good candidate to satisfy all these requirements. Finally, the ability to identify separately electrons and muons would be very important for signal characterization, background suppression, and for the interface with FASER$\nu$2. To achieve this, a mass of iron will be placed after the calorimeter, with sufficient depth to absorb pions and other hadrons, followed by a detector for muon identification.

To conclude, the physics potential of a larger-scale successor to FASER is clear. Possible scenarios for this larger detector are being explored, and initial studies strongly indicate a preference for the purpose-built facility option for the FPF.  

\subsection{FASER$\nu$2}
\label{sec:fasernu2}

FASER$\nu$~\cite{Abreu:2019yak} at the LHC was designed to directly detect collider neutrinos for the first time and study their properties at TeV energies. The FASER Collaboration has recently reported the first neutrino interaction candidates at the LHC in a small pilot detector exposed in 2018~\cite{FASER:2021mtu}. In LHC Run~3 beginning in 2022, FASER$\nu$ will measure $\sim$10,000 flavor-tagged, charged-current (CC) neutrino interactions, with a total tungsten target mass of 1.1 tonnes. However, only a handful of tau neutrino events are expected to be detected, which would be insufficient for sensitive physics studies.  

The FASER$\nu$2 detector is a much larger successor to FASER$\nu$. Following the example of FASER$\nu$, FASER$\nu$2 will be an emulsion-based detector able to identify heavy flavor particles produced in neutrino interactions, including $\tau$ leptons and charm and beauty particles. In the HL-LHC era, FASER$\nu$2 will be able to carry out precision tau neutrino measurements and heavy flavor physics studies, eventually testing lepton universality in neutrino scattering and new physics effects.  More generally, as discussed in the following physics sections, FASER$\nu$2 will provide extraordinary opportunities for a broad range of neutrino studies, with additional and important implications for QCD and astroparticle physics. 

\begin{figure}[tb]
\centering
\includegraphics[width=0.95\linewidth]{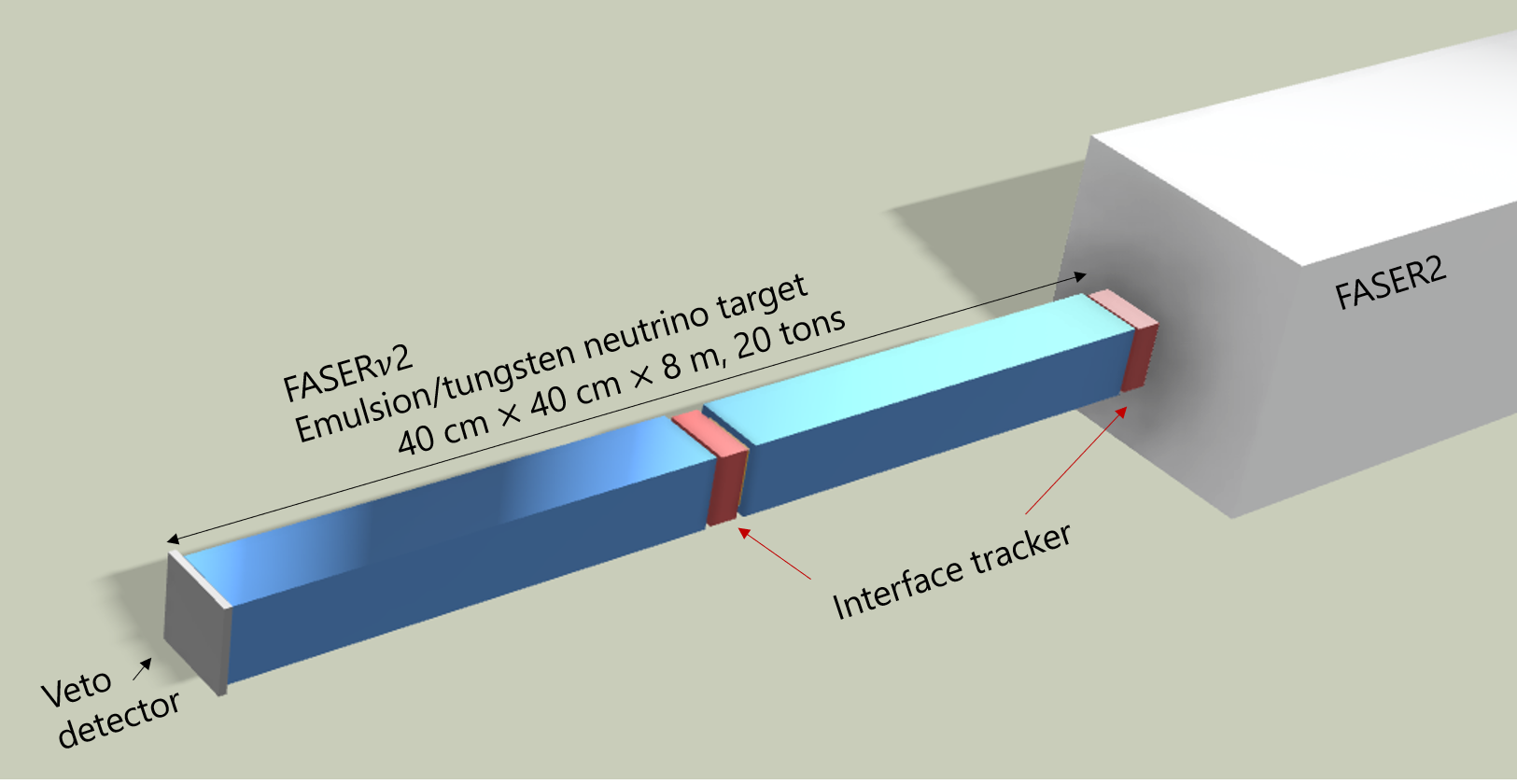}
\caption{Conceptual design of the FASER$\nu$2 detector.}
\label{fig:FASERnu2}
\end{figure}

\Figref{FASERnu2} shows a view of the FASER$\nu$2 detector. Its ideal location is in front of the FASER2 spectrometer along the beam collision axis to maximize the neutrino event rate per area for all three flavors. The FASER$\nu$2 detector is currently envisioned to be composed of 3300 emulsion layers~\cite{Ariga2020} interleaved with 2 mm-thick tungsten plates. It will also include a veto detector and interface detectors to the FASER2 spectrometer, with one detector in the middle of the emulsion modules and the other detector downstream of the emulsion modules to make the global analysis and muon charge measurement possible. Both the emulsion modules and interface detectors will be put in a cooling system. The total volume of the tungsten target is $40~\cm \times 40~\cm \times 6.6~\text{m}$, and the mass is 20 tonnes. The detector length, including the emulsion films and interface detectors, will be about 8 m. 

As described in Ref.~\cite{Abreu:2019yak}, analyses of the data collected in the emulsion modules will make possible the identification of muons, the measurement of muon and hadron momenta by the multiple Coulomb scattering coordinate method, and the energy measurement of electromagnetic showers. In addition, by conducting a global analysis that ties together information from FASER$\nu$2 with the FASER2 spectrometer via the interface detectors, the charges of muons will be identified.   Given 20 times the luminosity and 20 times the target mass of FASER$\nu$, FASER$\nu$2 will collect two orders of magnitude higher statistics than FASER$\nu$, allowing precision measurements of neutrino properties for all three flavors.

The high muon background in the LHC tunnel might be an experimental limitation. The possibility of sweeping away such muons with a magnetic field placed upstream of the detector is currently being explored, as described in \secref{sweepermagnet}. Considering the expected performance, emulsion films will be replaced every year during the winter stops. 

\subsection{Advanced SND@LHC \label{sec:AdvSND}}

\begin{figure}[tb]
\centering
\includegraphics[width=1.0\linewidth]{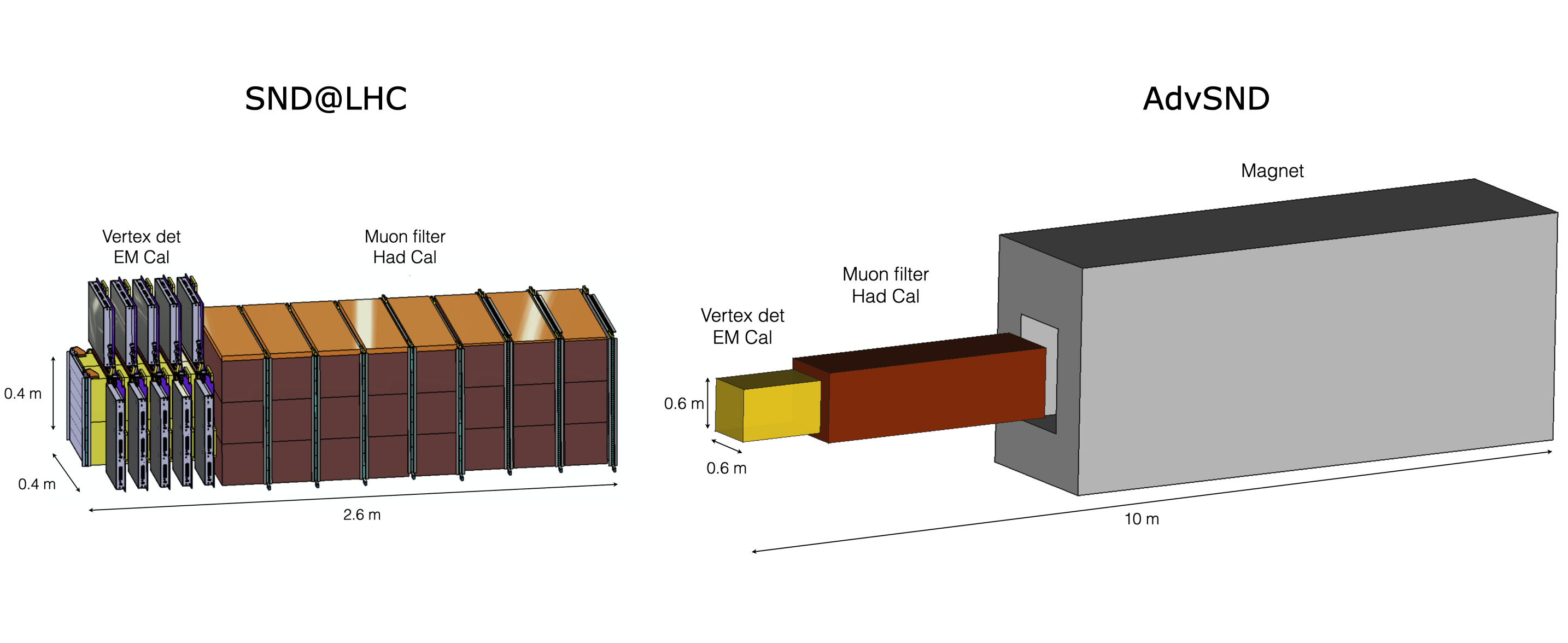}
\caption{Schematic layout of the SND@LHC detector (left) and conceptual design of Advanced SND@LHC (right).}
\label{fig:SND}
\end{figure}

The SND@LHC experiment~\cite{Ahdida:2750060} was designed to measure neutrinos and search for light DM at the LHC, via a scattering signature, in the pseudorapidity region  $7.2 < \eta < 8.6$, complementary to all other LHC experiments. The experiment is located 480 m downstream of IP1 in the TI18 tunnel. The detector is composed of a hybrid apparatus; see \cref{fig:SND}. The neutrino target region exploits the Emulsion Cloud Chamber (ECC) concept with tungsten plates interleaved with emulsion films acting as a micrometer-accuracy vertex detector. The target is distributed along five walls, each  interleaved with electronic trackers providing the time stamp for the interactions reconstructed in the emulsion and  the calorimetric measurement of the electromagnetic energy. Upstream of the target, a veto system identifies charged particles entering from outside. The target region is followed downstream by a hadronic calorimeter and a muon system. 

Neutrinos in the relevant pseudorapidity range come mostly from charmed hadron decays.  The experiment will therefore measure charm production in $pp$ collisions in this unexplored angular region, where the gluon parton distribution functions (PDFs) are unknown. The charm production measurement in this angular region is also relevant for understanding the atmospheric neutrino background in astrophysical neutrino searches, as discussed in \secref{atmosphericneutrinos}.  The experiment's sensitivity to feebly-interacting particles, including light DM, is reported in Ref.~\cite{Boyarsky:2021moj}.

An advanced version of the SND@LHC detector is envisaged for the HL-LHC. AdvSND will be made of two detectors, one in the FPF at $\eta \sim 8$ or roughly 50 cm off-axis (AdvSND1), and the other located closer to the ATLAS IP at $\eta \sim 5$ (AdvSND2).   AdvSND1 will have an angular acceptance similar to SND@LHC and will perform charm production measurements, tau neutrino studies, and lepton flavor universality tests with neutrinos at the percent level.  AdvSND2 will profit from the overlap with LHCb's pseudorapidity coverage to reduce systematic uncertainties. To increase the azimuthal angle coverage of the second detector, the idea is to search for a location in existing caverns that is closer to the IP. We consider this second module as a near detector meant for systematic uncertainty reduction, which would be placed outside of the FPF.  It should be noted that it might be sensitive also to neutrinos from $W/Z$ decays, thus extending the physics potential of the first detector. 

Each detector will be made of three elements, as shown in \cref{fig:SND}. The upstream one is the target region for  the vertex reconstruction and the electromagnetic energy measurement with a calorimetric approach. It will be followed downstream by a muon identification and hadronic calorimeter system. The third and most downstream element will be a magnet with two high-resolution tracking stations to enable muon charge and momentum measurement, thus allowing for neutrino/antineutrino separation for muon neutrinos and for tau neutrinos in the muonic decay channel of the $\tau$ lepton. 

The target will be made of thin sensitive layers interleaved with tungsten plates, for a total mass of at least 2 tonnes. The actual value of the target mass, ranging from 2 to about 10 tonnes, will be optimized by accounting for the systematic uncertainties in the measurements. Without a way to reduce the muon background, the use of nuclear emulsion at the HL-LHC is made difficult by the very high muon intensity that would make the replacement rate of the target incompatible with technical stops. The high muon background has motivated the possibility of a sweeper magnet, which could resolve this problem, as discussed in \secsref{sweepermagnet}{fasernu2}.  As an attractive independent solution, compact electronic trackers with high spatial resolution are being investigated to fulfill the tasks of both vertex reconstruction with micrometer accuracy and electromagnetic energy measurement. The hadronic calorimeter and the muon identification system will be about 10~$\lambda$, which will bring the average length of the hadronic calorimeter to about 11.5~$\lambda$, thus improving the muon identification efficiency and energy resolution. The magnetic field strength is assumed to be about 1.5 T over a $\sim$3 m length. The ancillary tracking stations, one upstream and the other one downstream of it, will be developed with the same technology adopted for the vertex detector, thus granting a high position accuracy that would turn into a high-resolution momentum measurement. 

The configuration of the detectors will efficiently distinguish all three neutrino flavors and measure their energy. AdvSND will open a unique opportunity to probe the physics of heavy flavor production at the LHC in a region inaccessible to other experiments.

\subsection{FLArE: Forward Liquid Argon Experiment \label{sec:FLArE}}

FLArE, a liquid argon time projection chamber (LArTPC), is considered for the suite of detectors for the FPF. Such a detector offers the possibility to precisely determine particle identification, track angle, and kinetic energy over a large dynamic range in energies. 

A LArTPC is well motivated by the requirements of the light DM search~\cite{Batell:2021blf,Batell:2021aja}, in which an energetic, isolated, forward-going electron must be identified and its energy measured. The most important background to the DM process of elastic scattering from electrons is neutrino-electron scattering. Isolated photons from muon scattering and other neutrino interactions will also contribute. To distinguish the DM signal process from background, we must identify the electron and measure it with angular precision less than $1^\circ$ and with excellent energy resolution from $\sim 10~\mev$ to a few GeV. The energy range for this performance depends on the mass of the DM candidate and backgrounds, both of which must be further studied.  

The same detector is also expected to measure millions of neutrino interactions, including tau neutrinos. The detector should have sufficient capability to measure these very high energy ($>100~\gev$) events, so that the cross section for each flavor can be measured. Identification of tau neutrinos with low backgrounds needs detailed simulations and reconstruction studies.

Table \ref{lartab} summarizes the main parameters of a LArTPC for the FPF. A detector with a fiducial mass of approximately 10~tonnes is envisioned. For $3~\iab$, such a detector will collect hundreds of thousands of muon neutrino/antineutrino CC events, about a hundred thousand electron neutrino events, and thousands of tau neutrino events. These numbers have large uncertainties due to the poorly understood production cross section in the forward region~\cite{Bai:2020ukz}. It is also important to note that this flux of events will have the same time structure as the LHC accelerator with a bunch spacing of 25 ns. At the same time, muons from interactions at the IP will produce a background muon flux of about $\sim 1~\cm^{-2}~\text{s}^{-1}$ at the nominal maximum luminosity of $5\times 10^{34}~\cm^{-2}~\text{s}^{-1}$ at the HL-LHC.

\begin{table} 
\begin{center}
\begin{tabular}{ l || l | l }
 \hline
 \hline
 & Value & Remarks \\
 \hline 
 Detector length &  7~m & Not including cryostat \\ 
 TPC drift length &  $0.75$ to $1.00$ ~m & 2 TPC volumes with HV cathode in center \\ 
 TPC height &  1.5~m &  \\ 
 Total LAr mass & $\sim 50$~tonnes & Volume in the cryostat \\ 
 Fiducial mass & $10-20$~tonnes & \\ 
 Background muon rate & $\sim 1/$cm$^2/$s & Maximum luminosity of $5\times10^{34}/$cm$^2/$s \\
 Neutrino event rate & $\sim 50/$tonne/fb$^{-1}$ & For all flavors of neutrinos \\ 
 \hline
 Stopping power (MIP) & 2.1~MeV/cm & \\
 Radiation length & 14~cm & \\
 Interaction length & 85~cm & \\
 Molière radius & 9~cm & \\
 Light yield & 50~ph/keV & at 0 V/cm \\
 Scintillation time & singlet 7~ns, triplet~1.6~$\mu$s & peaked at 128~nm\\
 Rayleigh scattering length & 90~cm & at 128~nm\\
 Ionization charge yield & 10~fC/cm & for MIP at 500~V/cm \\
 Electron drift velocity & 1.6~mm/$\mu$s & at 500~V/cm \\
 Electron diffusion coefficient &~7.2 cm$^2$/s & at 500~V/cm \\
 Achievable drifting electron lifetime & $> 10$~ms & $< 30$~ppt O$_2^{\text{eq}}$ contamination \\
 Demonstrated drift length & 3.6~m & $\approx 2.3$~ms drift time \\
 \hline
 \hline
\end{tabular}
\end{center}
\caption{Detector parameters for FLArE, a LArTPC for the FPF. The top part of the table shows the nominal geometric parameters for a detector to be considered for the FPF, and the bottom part shows the basic properties of a LArTPC. \label{lartab}}  
\end{table} 

The nominal configuration for FLArE would include a central cathode operating at a large high voltage and two anode planes on two sides of the detector parallel to the beam from the ATLAS IP. The electric field between the cathode and the anode will be at $\sim500$~V/cm, providing a drift field for ionization electrons; the drift time for a 1~m-long drift will be about 0.6~ms. For a detector with approximate cross section of $2$~m$^2$, we therefore expect about 12 muon tracks to be within a single drift time. Neutrino and DM events must be selected out of these overlaying background particle trajectories. For the TPC, a readout using wires or pixels is possible.   If the granularity of the readout is approximately $3-5$ mm in all dimensions, then an angular resolution of a few mrad for electromagnetic showers appears feasible. A readout of the scintillation light is crucial to allow the measurement of the distance along the drift. It is also important for the selection of events that originate in the detector (such as a neutrino or a DM event), as well as generating the trigger necessary for acquiring the data. 

There are two key technical issues that need to be resolved for the selection of the readout technique and the overall design of the LAr detector.  The measurement of isochronous tracks~\cite{Qian:2018qbv} or tracks that are parallel to the anode plane presents a particular difficulty for LHC events, given their extreme forward angles. Such tracks will produce a simultaneous signal on anode channels, making it difficult to measure their trajectory in the vertical plane. The second issue is the development of the scintillation photon detector. This detector system needs to have the capability to measure the time of the event precisely to isolate the several particle tracks that are within a single drift time, but it also needs to have basic fast pattern recognition capability to select interesting events at the trigger level.

The LArTPC is expected to be installed in a membrane cryostat with passive insulation and with inner dimensions of $2.1~\m \times 2.1~\m \times 8.2~\m$. Following the example of ProtoDUNE~\cite{DUNE:2017pqt,DUNE:2021hwx}, the membrane cryostat technology allows the cryostat to be constructed underground. The insulation, being passive, ensures reliable and safe long-term performance. The cryogenic system must re-condense the boil-off, keeping the ullage absolute pressure stable to better than 1 mbar, and purify the LAr bath. A standard approach is to re-condense the argon with a heat exchanger with liquid nitrogen. A LAr flow of 500~kg/h through the purification circuit is considered sufficient to reach and maintain the required LAr purity. 

The total heat input due to the cryostat and the cryogenics system is estimated to be of the order of 7~kW: 4~kW from the cryostat, 1~kW from the GAr circuit, 1~kW from the LAr purification, and 1~kW from other inefficiencies. To this, the detector electronics should be added. A Turbo-Brayton ($\sim 8~\m \times 1.6~\m \times 2.7~\m$) TBF-80 unit from Air Liquid installed in the vicinity of the cryostat provides approximately 10~kW cooling power from $\approx$100~kW electrical power and 5 kg/s of water at ambient temperature. Liquid argon and nitrogen storage tanks are required above ground and connected via piping to the underground cryogenics. Exhaust of gases will be done to the atmosphere on the surface. For safety reasons, the cryostat in the cavern will be placed in a trench 1.5~m deep, 6.9~m wide, and 12.6~m long, which collects the argon in case of a leak. Oxygen deficiency is the main risk associated with the LArTPC. A properly-dimensioned ventilation system will constantly extract air in the proximity of the cryostat/cryogenics. A detection of low oxygen content by the oxygen deficiency hazard system in the cavern and in the trench will trigger increased air extraction.

The FLArE detector could be an excellent choice for the detection of light DM scattering as well as neutrino events at the 10-tonne fiducial mass scale.  Further simulation work is needed to understand event reconstruction and background rejection.  For detector design, in particular, simulation work is needed to understand electromagnetic shower containment and energy resolution in a 7 m-long detector.  Study of kinematic resolution in the case of wire readout versus pixel readout is needed. And finally, the design and performance of the photon detector system needs to be investigated and demonstrated by R\&D.  In particular, the photon system has to serve three functions in order of increasing  difficulty: separation of beam-related muon and neutrino events, accurate timing performance to measure the location of the neutrino or DM events in the TPC, and the association of a neutrino or DM event with a bunch crossing in the collider detector.

\subsection{FORMOSA: FORward MicrOcharge SeArch \label{sec:FORMOSA}}

The FPF provides an ideal location for a next-generation experiment to search for BSM particles that have an electric charge that is a small fraction of that of the electron. Although the value of this  fraction can vary over several orders of magnitude, we generically refer to these new states as ``millicharged'' particles (mCPs). Since these new fermions are typically not charged under QCD, and because their electromagnetic interactions are suppressed by a factor of $(Q/e)^2$, they are ``feebly'' interacting and naturally arise in many BSM scenarios that invoke dark or otherwise hidden sectors. For the same reason, the experimental observation of mCPs requires a dedicated detector. Such a detector could also provide new sensitivity to other signatures, such as exotic heavy neutrinos with an electric dipole moment~\cite{Frank:2019pgk}. 

As proposed in Ref.~\cite{Foroughi-Abari:2020qar}, FORMOSA\footnote{The detector considered here corresponds to the FORMOSA-II setup introduced in Ref.~\cite{Foroughi-Abari:2020qar}. FORMOSA-I refers to a demonstrator prototype that could be installed in the UJ12/TI12 experimental areas near the current FASER experiment.} is an experiment to search for mCPs at the FPF which would consist of a milliQan-type detector~\cite{Haas:2014dda, Ball:2016zrp}.  This will be technically similar to what the milliQan Collaboration will install in the PX56 drainage gallery near LHC P5 near the CMS IP during Run~3~\cite{milliQan:2021lne}, but with a significantly larger active area and a more optimal location with respect to the expected mCP flux. 

\begin{figure}[tb]
  \centering
  \includegraphics[width=0.7\textwidth]{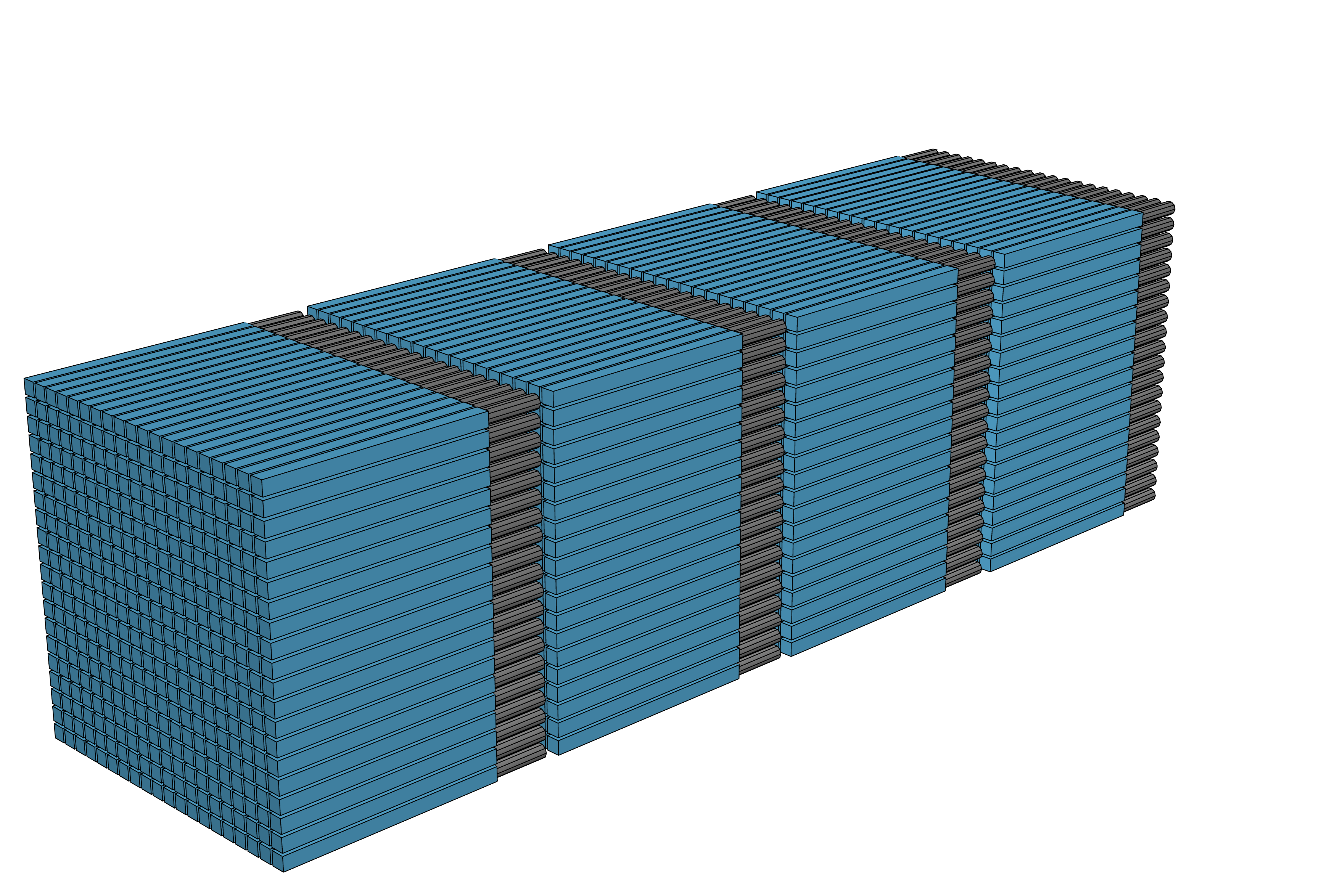}
  \caption{A diagram of the FORMOSA detector components. The scintillator bars are shown in blue connected to PMTs in black.}
  \label{fig:formosa-bars}
\end{figure}

To be sensitive to the small $dE/dx$ of a particle with $Q \lesssim 0.1e$, a mCP detector must contain a sufficient amount of sensitive material in the $x$ dimension, which in this case is chosen to be the longitudinal direction pointing to the IP. The optimal choice of scintillator material is under consideration. Currently, as in Ref.~\cite{Haas:2014dda}, plastic scintillator is chosen as the detection medium with the best known combination of photon yield per unit length, response time, and cost. Consequently, FORMOSA is planned to be a $1~\mathrm{m}~\times1~\mathrm{m}~\times5~\mathrm{m}$ array of suitable plastic scintillator (e.g., Eljen EJ-200~\cite{Eljen} or Saint-Gobain BC-408~\cite{SG}). The array will be oriented such that the long axis points at ATLAS IP1 and is located on the beam collision axis. The array contains four longitudinal ``layers'' arranged to facilitate a 4-fold coincident signal for feebly-interacting particles originating from the ATLAS IP. Each layer in turn contains one hundred $5~\mathrm{cm}\times5~\mathrm{cm}\times100~\mathrm{cm}$ scintillator ``bars'' in a $10\times10$ array. To maximize sensitivity to the smallest charges, each scintillator bar is coupled to a high-gain photomultiplier tube (PMT) capable of efficiently reconstructing the waveform produced by a single photoelectron (PE). To reduce random backgrounds, mCP signal candidates will be required to have a quadruple coincidence of hits with $\bar{N}_{\rm PE}\ge 1$ within a 20 ns time window. The PMTs must therefore measure the timing of the scintillator photon pulse with a resolution of $\le5$ ns. The bars will be held in place by a steel frame. A conceptual design of the FORMOSA detector is shown in \cref{fig:formosa-bars}. 

 Although omitted for clarity in \cref{fig:formosa-bars}, additional thin scintillator ``panels'' placed on each side of the detector will be used to actively veto cosmic muon shower and beam halo particles. Finally, thin scintillator panels will be placed on the front and back of the detector to aid in the identification of muons resulting from LHC proton collisions. During Run~2 of the LHC, a similar experimental apparatus (the milliQan ``demonstrator'') was deployed in the PX56 draining gallery at LHC P5 near the CMS IP. This device was used successfully to search for mCPs, proving the feasibility of such a detector~\cite{ball2020search}.

Even though the pointing, 4-layered, design will be very effective at reducing background processes, small residual contributions from sources of background that mimic the signal-like quadruple coincidence signature are expected. These include overlapping dark rate pulses, cosmic muon shower particles, and beam muon afterpulses. In Ref.~\cite{milliQan:2021lne}, data from the milliQan prototype were used to predict backgrounds from dark rate pulses and cosmic muon shower particles for a closely-related detector design and location. Based on these studies, such backgrounds are expected to be negligible for FORMOSA. Backgrounds from muon afterpulses are considered in Ref.~\cite{Foroughi-Abari:2020qar} and can be rejected by vetoing a 10 $\mu s$ time window in the detector following through-going beam muons. A prototype mCP detector in the FASER cavern would provide important insights into the optimal design of the FORMOSA detector and is being actively explored.

\section{Searches for New Physics}
\label{sec:bsm}




The majority of interactions at the LHC are soft, with GeV-scale momentum transfers between the colliding protons, and produce mesons at high rapidity, with typical angles of order $\text{GeV} / \text{TeV} \sim \text{mrad}$. The resulting fluxes of mesons in the region of the FPF are therefore extraordinary, despite its small solid angle coverage:~within 2~mrad of the beam axis, the HL-LHC will produce $4 \times 10^{17}~\pi^0$s, $6 \times 10^{16}~\eta$s, $2 \times 10^{15}~D$~mesons, and $10^{13}~B$ mesons. Due to these large rates, light BSM states could be copiously produced in meson decays even if the new decay channels have very small branching fractions. The general-purpose LHC detectors, with coverage only up to $|\eta| \approx 5$, are typically not sensitive to such new physics.

Light BSM fields feature prominently in solutions to many of the most significant outstanding questions in particle physics, including the nature of DM, neutrino masses, the strong CP problem, the hierarchy problem, and the matter-antimatter asymmetry of the universe. From a bottom-up perspective, they have also been invoked to explain several currently unresolved experimental anomalies. Here, we show the reach of a suite of FPF detectors to discover light, weakly-coupled new physics, considering three main categories of signatures. First, we consider particles that are produced and then decay to SM states in detectors for long-lived particles, such as the FASER2 experiment presented in \secref{FASER2}. Then, motivated by DM, we describe the ways in which light invisible states can scatter off dense forward detectors like FASER$\nu$2, AdvSND, and FLArE (see Secs.~\ref{sec:fasernu2}, \ref{sec:AdvSND}, and \ref{sec:FLArE}), complementing the neutrino studies that could be performed at these experiments. Last, we discuss the possibility of seeing new states with non-standard patterns of energy deposition in matter, focusing on mCPs at the FORMOSA detector described in \secref{FORMOSA}.

In presenting the FPF BSM physics case, we have employed many of the benchmarks of the BSM working group of the Physics Beyond Colliders initiative~\cite{Beacham:2019nyx}. For many of these scenarios, the FPF will test parameter regions that are otherwise inaccessible. There is also much well-motivated physics beyond these minimal benchmarks that remains to be studied at the FPF, such as new U(1) gauge groups and UV-complete models of light DM~\cite{Alexander:2016aln, Battaglieri:2017aum}. The results presented here should thus be considered as only a subset of possible BSM physics searches at the FPF, and we encourage the community to augment them with further studies.

\subsection{Long-Lived Particle Decays}
\label{sec:llps}

Long-lived particles are common in theories of hidden sectors~\cite{Alimena:2019zri}, where new physics is coupled to the SM through a mediator. The most minimal of these theories involves a single portal interaction between the mediator and the SM. The symmetries of the SM admit only three possibilities for this interaction at dimension-4: $F_{\mu\nu} F'^{\mu\nu}$, where $F$ and $F'$ are the field strengths associated with U(1)$_{\mathrm{EM}}$ and a new U(1), respectively (dark photon); $\mu \phi H^2$ or $\phi^2 H^2$, with $\phi$ being a new scalar and $H$ the SM Higgs doublet (dark scalar or dark Higgs); and $HLN$, where $L$ is one of the left-handed lepton doublets of the SM, and $N$ is a new singlet fermion (dark fermion, sterile neutrino, or heavy neutral lepton (HNL)). In addition to these minimal portals, the mediator could have non-renormalizable interactions with the SM, as in the case of ALPs $a$ coupling to one of the SM field strengths $F$ through $a F \widetilde{F}$. 

If the mediator's mass is at the GeV scale, its phenomenologically-allowed couplings to the SM are very small, and, in the absence of non-SM decays, the mediator is often long-lived. At the LHC, light mediators in hidden sector theories would be produced at high energy in the forward direction, and their decays to SM particles could be seen in FPF detectors. We focus here on a selection of minimal and non-minimal portals as an illustration of the reach of the FPF; many of these portals have been studied previously in the context of FASER. Below, we present the corresponding sensitivity reach for the FASER2 detector located in the purpose-built facility, assuming a location that is centered on the beam axis at a distance of $620~\m$ downstream from the IP and a cylindrical detector with a decay volume length of $5~\m$ and a radius of $1~\m$. The results have been obtained with FORESEE~\cite{Kling:2021fwx}, a package that can use the LLP production rates, lifetimes, and decay channels of any desired model to estimate the reach of forward detectors at $pp$ colliders. The results assume that LLP decays to the SM inside FASER2 can be seen with 100\% efficiency, underscoring the necessity of designing detectors that are sensitive to all possible final states in the highly collimated geometry typical of far-forward searches.

\begin{figure*}[t]
\centering
\includegraphics[width=0.49\textwidth]{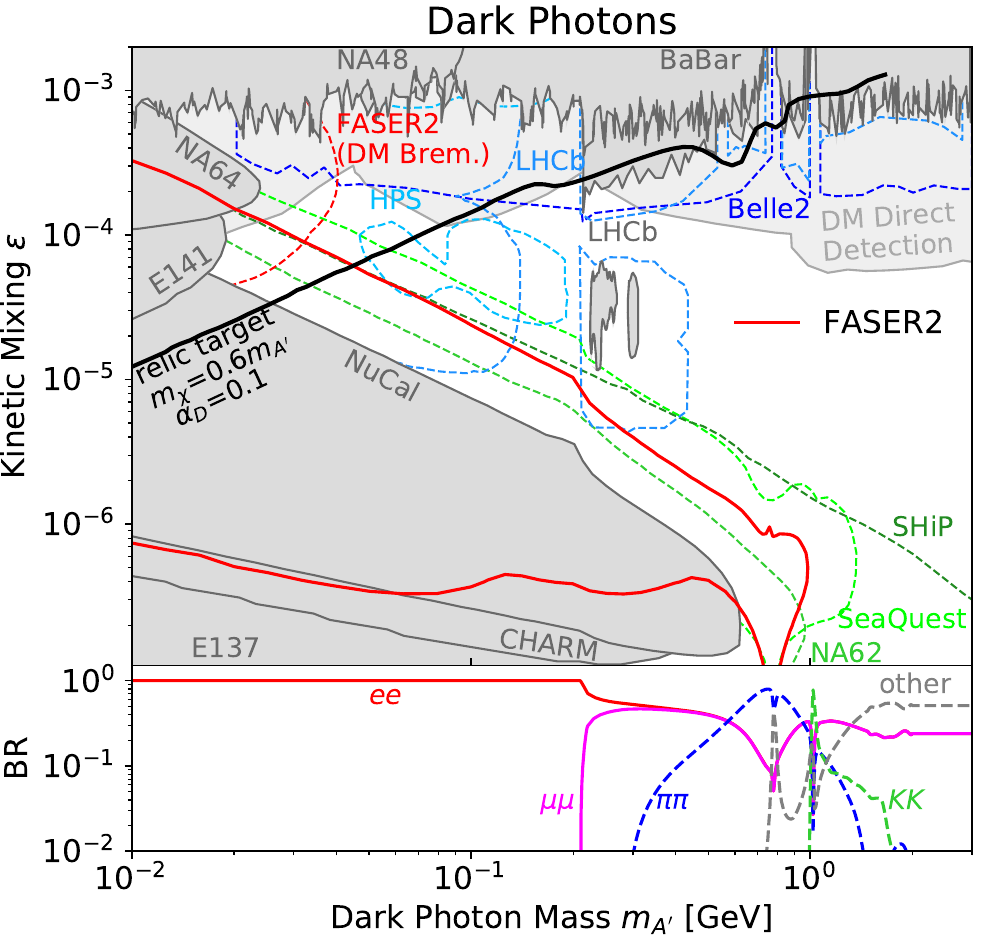}
\includegraphics[width=0.49\textwidth]{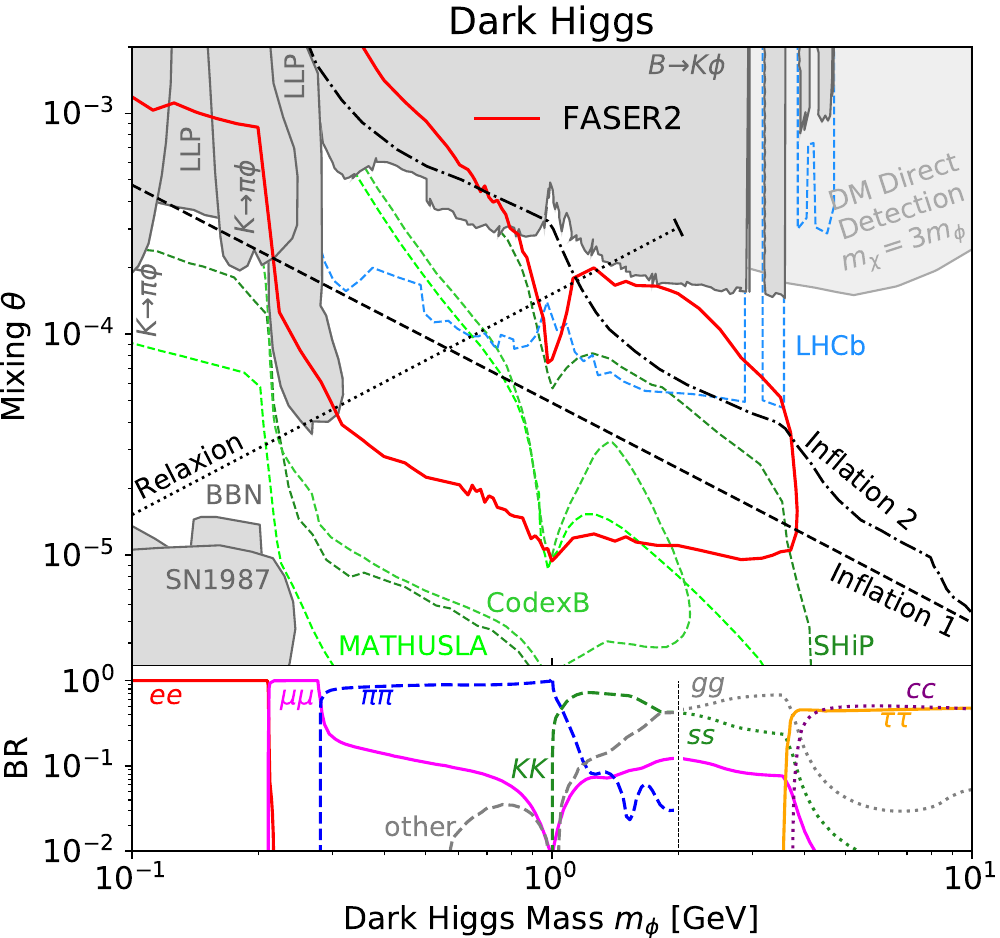}
\caption{Sensitivities for the dark photon (left) and  dark Higgs (right) in the $(\text{mass}, \text{coupling})$ plane. The sensitivity reaches of FASER2 are shown as solid red lines alongside existing constraints (dark gray-shaded regions) and projected sensitivities of other proposed searches and experiments (colorful dashed lines), as obtained in Refs.~\cite{Lees:2014xha, Gninenko:2012eq, Bjorken:1988as, Riordan:1987aw, Aaij:2019bvg, Batley:2015lha, Banerjee:2018vgk, Blumlein:1990ay, Kou:2018nap, Solt:2020zbi, Ilten:2015hya, Ilten:2016tkc, Dobrich:2018ezn, Berlin:2018pwi, Ahdida:2020new} for the dark photon and Refs.~\cite{Ruggiero:2020phq, Artamonov:2009sz, Aaij:2015tna, Aaij:2016qsm, Winkler:2018qyg, Foroughi-Abari:2020gju, MicroBooNE:2021ewq, Alekhin:2015byh, Gligorov:2017nwh, Evans:2017lvd, Curtin:2018mvb} for the dark Higgs; also see Ref.~\cite{Beacham:2019nyx} and references therein for further proposals on proposed searches and experiments. The solid black line shows the DM relic target for a complex scalar DM scenario with $m_\chi=0.6m_{A'}$ and $\alpha_D=0.1$, where the corresponding direct detection bounds are shown in light gray. The bottom panels show the LLP branching fractions, as obtained in Refs.~\cite{Ilten:2018crw, Winkler:2018qyg}.
}
\label{fig:bsm_llp1}
\end{figure*}

\textbf{Dark Photon:} Dark photons are motivated in a wide variety of theories with new U(1) symmetries~\cite{Battaglieri:2017aum}. At the LHC, they are dominantly produced through meson decays and bremsstrahlung in the region that would be covered by the FPF~\cite{Feng:2017uoz, Ariga:2018uku, Foroughi-Abari:2021zbm}. The left panel of \cref{fig:bsm_llp1} shows the FPF reach from dark photon decays in FASER2 as a function of the dark photon's mass $m_{A'}$ and kinetic mixing $\epsilon$ with U(1)$_{\text{EM}}$~\cite{Kling:2021fwx}. The upper limit of the projection is set by the requirement that the dark photon lifetime is sufficiently large to reach the detector. Despite having a longer baseline than existing experiments such as NuCal, FASER2 would achieve increased sensitivity because of the boost with which the dark photons are produced. The FPF would thus close a significant portion of the gap between searches for prompt dark photon decays and long-lived searches. 

The dark photon could also serve as a mediator for DM annihilation. We show a target line where the correct thermal relic density of a complex scalar DM particle $\chi$ is obtained for $m_\chi = 0.6 m_{A'}$ and dark coupling $\alpha_D = 0.1$. At low masses, direct detection does not constrain this model. In this scenario, the dark photon could also be produced via scattering of DM in the material before FASER2, $\chi N \to \chi N A'$, and we show the potential additional FASER2 reach under the same mass and coupling assumptions~\cite{Jodlowski:2019ycu}. The case of resonant DM annihilation through a dark photon has also been studied~\cite{Feng:2017drg,Bernreuther:2020koj}. Finally, new gauge symmetries could lead to vector mediators whose productions and decays offer further phenomenology beyond the dark photon case. Those for which the FASER2 reach has been investigated include a $B\!-\!L$ gauge boson~\cite{Ariga:2018uku, Bauer:2018onh, Mohapatra:2019ysk, Okada:2020cue}, lepton-specific $L_i\!-\!L_j$ and $B\!-\!3L_i$ gauge bosons~\cite{Bauer:2018onh, Bauer:2020itv}, gauge bosons decaying to tau neutrinos~\cite{Bahraminasr:2020ssz, Kling:2020iar}, and a $T_{3R}$ gauge boson~\cite{Dutta:2020enk}.

\textbf{Dark Higgs:} New hidden sector scalars mixing with the Higgs arise in theories addressing the hierarchy problem~\cite{Winkler:2018qyg, Flacke:2016szy}, DM~\cite{Burgess:2000yq}, inflation~\cite{Bezrukov:2009yw, Bramante:2016yju, Okada:2019opp}, and the cosmological constant problem~\cite{Foot:2011et}. Unlike the dark photon, a dark scalar $\phi$ would mostly be produced at the LHC through the decay $B \to X_s \phi$, given its preferential couplings to heavy quarks~\cite{Feng:2017vli, Ariga:2018uku}. In this regard, the FPF offers an advantage over many accelerator probes at lower energies, where heavy-flavor meson production is more kinematically suppressed. The right panel of \cref{fig:bsm_llp1} shows the projected FASER2 reach for a minimal dark scalar $\phi$ with mixing angle $\theta$ with the SM Higgs. The minimum mixing angles probed, as low as $10^{-5}$, are a substantial improvement over existing limits. In addition, compared with other proposed LHC LLP detectors, FASER2 would probe larger (though as yet unconstrained) mixing angles because its forward position allows for observation of LHC dark scalars with relatively high energies, leading to extended reach at shorter lifetimes. B-factories could also probe dark scalars with small mixing angles by searching for displaced vertices~\cite{BaBar:2015jvu, Filimonova:2019tuy}.

If the dark Higgs couples to DM $\chi$ with $m_\chi > m_\phi$, secluded annihilation $\chi \chi \to \phi \phi$ can lead to the correct relic abundance~\cite{Feng:2017vli, Winkler:2018qyg} through thermal freeze-out independent of the mixing angle. We show the current direct detection limits in this scenario assuming $m_\chi = 3 m_\phi$~\cite{DarkSide:2018kuk, CRESST:2019jnq, XENON:2018voc}. An alternative DM scenario involving freeze-in was discussed in Ref.~\cite{Hryczuk:2021qtz}. \Figref{bsm_llp1} also shows theory targets motivated by the hierarchy problem and cosmology. The dotted target line indicates the preferred parameters for the relaxion providing a dynamical solution to the electroweak hierarchy problem (with the QCD' scale $\Lambda=2~\gev$), as discussed in Ref.~\cite{Winkler:2018qyg}.  In addition, the dark Higgs can play the role of an inflaton driving  cosmological inflation in the early universe, as illustrated by additional lines in the figure. A first scenario, labeled Inflation 1, corresponds to a theory where the inflation potential exhibits classical conformal invariance, which is broken radiatively via the Coleman-Weinberg mechanism~\cite{Okada:2019opp}. In this case, the inflationary predictions, in particular the tensor-to-scalar ratio $r$, are uniquely determined by $m_\phi$ and $\theta$, and we show the lower bound on $\theta$ arising from Planck 2018 measurements $r<0.064$ as a dashed line. A second scenario, labeled Inflation 2, considers a viable low-scale inflaton-curvaton model that could be discovered in the FPF search space~\cite{Bramante:2016yju}. The dot-dashed line illustrates the mass and mixing angle for an inflaton that decays when the universe reaches a density around the electroweak scale of $\rho \sim (100~\gev)^4$, corresponding to a lower coupling limit motivated by the incorporation of electroweak baryogenesis.  In addition, scalars with couplings different from the Higgs mixing expectation have been considered at forward LHC detectors in the case of dominant couplings to muons~\cite{Batell:2017kty} or up quarks~\cite{Kling:2021fwx, Batell:2021xsi}, an additional coupling to the Higgs boson~\cite{Boiarska:2019vid}, as well as the dilaton of Ref.~\cite{Csaki:2020zqz} employed to address the hierarchy problem.

\begin{figure*}[t]
\centering
\includegraphics[width=0.49\textwidth]{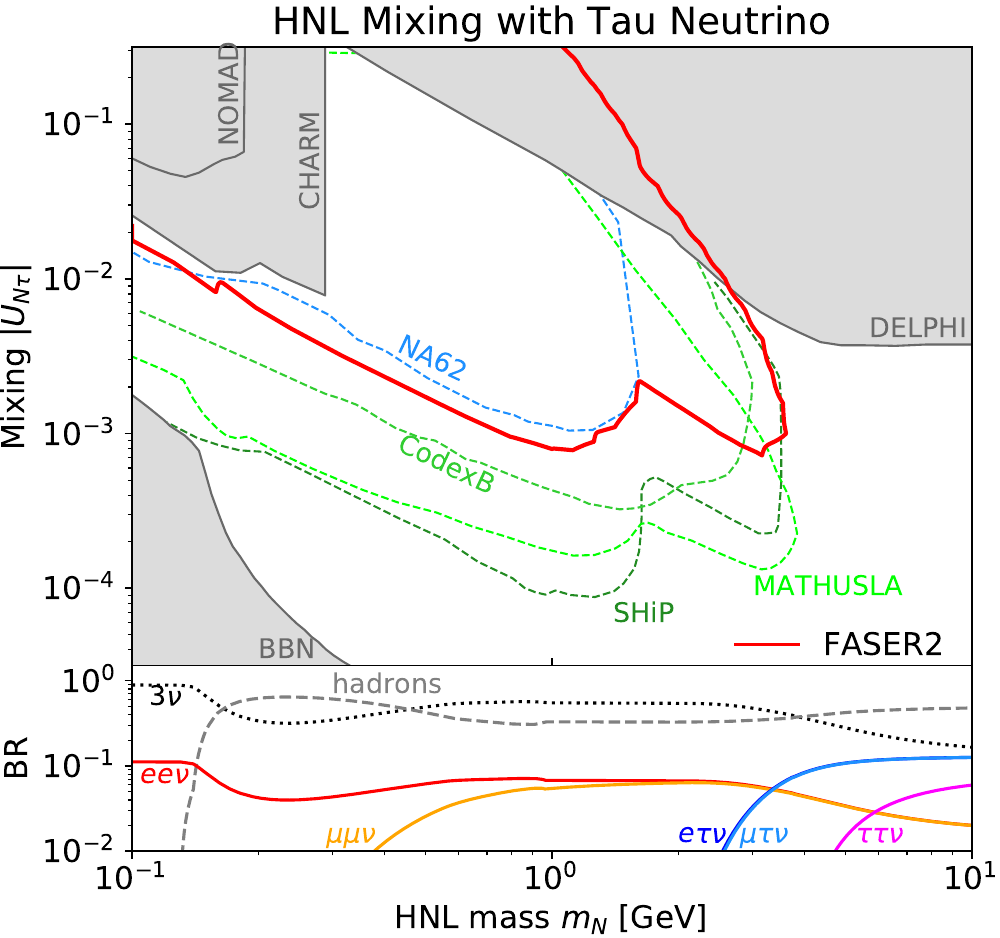}
\includegraphics[width=0.49\textwidth]{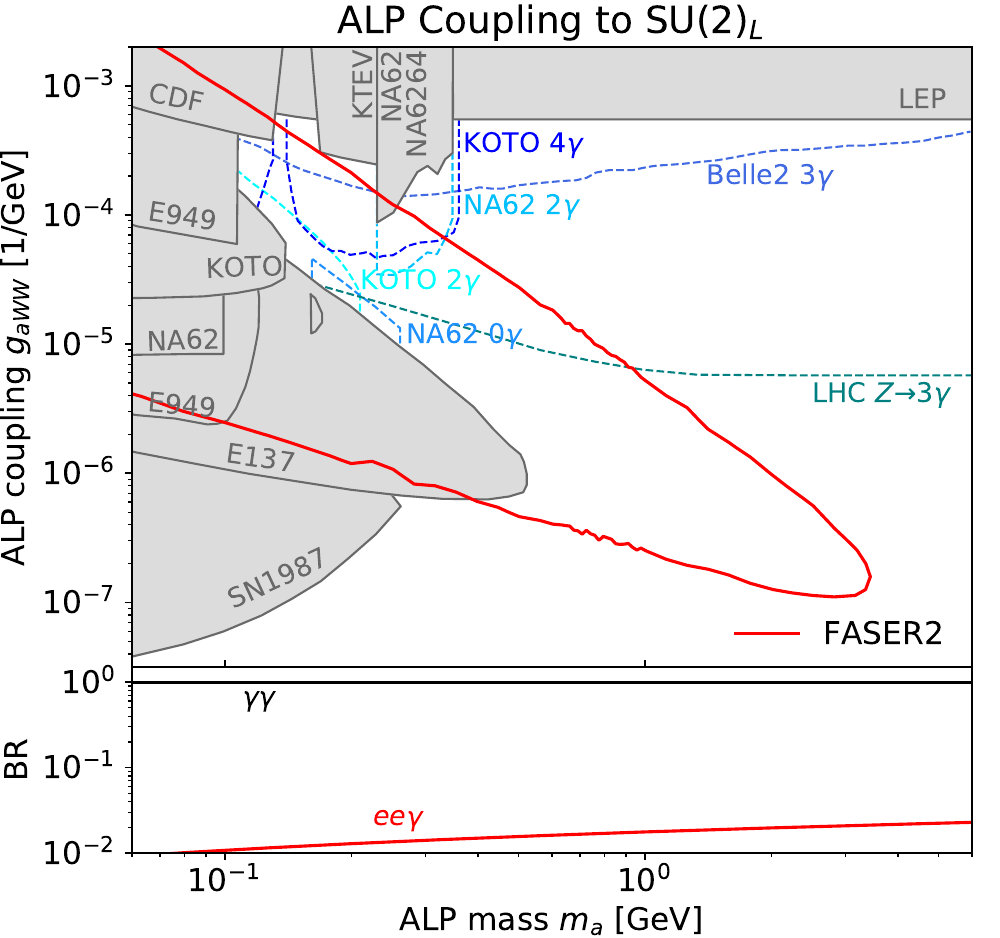}
\caption{Sensitivities for a HNL mixing with the tau neutrino (left) and an ALP coupling to $SU(2)_L$ gauge bosons (right) in the $(\text{mass}, \text{coupling})$ plane. The sensitivity reaches of FASER2 are shown as solid red lines alongside existing constraints (dark gray-shaded regions) and projected sensitivities of other proposed searches and experiments (colorful dashed lines), as obtained in Refs.~\cite{Orloff:2002de, NOMAD:2001eyx, DELPHI:1996qcc, Ruchayskiy:2012si, Beacham:2019nyx} for the HNL and Refs.~\cite{Izaguirre:2016dfi, Gori:2020xvq} for the ALP. The bottom panels show the LLP's branching fractions, as obtained in Ref.~\cite{Ariga:2018uku}.}
\label{fig:bsm_llp2} 
\end{figure*}

\textbf{HNLs:} Heavy neutral leptons mixing with the SM neutrinos offer an explanation for neutrino masses through the seesaw mechanism, can be responsible for the matter-antimatter asymmetry of the universe through leptogenesis, and can provide a compelling DM candidate~\cite{Asaka:2005pn, Shaposhnikov:2006nn}. In the GeV range, HNLs could be produced at the LHC through meson and $\tau$ decays, and are typically long-lived. At the FPF, their decays to hadrons and/or charged leptons would be visible in FASER2. LHC production of HNLs mixing with all three neutrino flavors have been studied in Refs.~\cite{Ariga:2018uku, Kling:2018wct, Helo:2018qej, Cline:2020mdt}. Although a HNL giving neutrino masses through the type I seesaw would decay to all flavors, given the mixing angles required by neutrino oscillations, in less minimal models, such as the linear seesaw or inverse seesaw, there is more freedom to fit the neutrino data. In particular, scenarios can be built where the decay of the HNL is predominantly to taus. We consider this benchmark in the left panel of \cref{fig:bsm_llp2}, showing the expected FASER2 reach. Because the $\nu_\tau$ has significant production from heavy meson decays, the large rates for forward $D$ and $B$ production at the LHC benefit the FPF, allowing for sensitivity to HNL masses up to several GeV. Besides the minimal $HLN$ mixing case, HNLs decaying through higher-dimensional operators have been studied in Refs.~\cite{deVries:2020qns, Cottin:2021lzz}, and dipole interactions have been considered in Ref.~\cite{Jodlowski:2020vhr}. In addition, other portals between fermionic BSM states and the SM have also been investigated, including in supersymmetric models with light neutralinos~\cite{Helo:2018qej, Dercks:2018eua}, which are a promising solution to the hierarchy problem, and in effective field theory~\cite{Darme:2020ral}. 

\textbf{ALPs:} In addition to the well-known QCD axion that addresses the strong CP problem, light new pseudoscalars appear generically in string theory~\cite{Jaeckel:2010ni}. These ALPs can have feeble couplings to the SM through dimension-5 interactions, leading to long lifetimes. The production and subsequent decays of ALPs in forward LHC detectors have been considered for the cases of dominant couplings to photons~\cite{Ariga:2018uku, Feng:2018pew}, weak gauge bosons~\cite{Kling:2020mch}, gluons~\cite{Ariga:2018uku}, and fermions~\cite{Ariga:2018uku}. In the right panel of \cref{fig:bsm_llp2}, we consider the benchmark of an ALP coupling to $W$ bosons through the interaction $a W^{\mu\nu} \widetilde{W}_{\mu\nu}$, as discussed in Ref.~\cite{Izaguirre:2016dfi, Gori:2020xvq, Kling:2020mch}. Such ALPs are mainly produced at the LHC through decays of $B$ mesons, as well as through the Primakoff process occurring in the TAN, and the primary decay mode is to diphotons. Similarly to the case of the dark scalar, the reach of FASER2 goes significantly beyond existing limits because of the substantial forward $B$ production at the LHC. In particular, in the GeV mass range, much lower couplings can be probed than in searches for photons in prompt ALP decays at the LHC and $B$ factories. ALPs with other couplings can have different phenomenology. For instance, in the charming ALP scenario of Ref.~\cite{Carmona:2021seb}, in which the only tree-level ALP couplings are to right-handed up-type quarks, the main production mode is through the decay of $D$ mesons. Such ALPs decay primarily to photons and, if kinematically allowed, to pions; depending on the exact coupling structure, decays to leptons can also be dominant. For ALP production through $D$ meson decays, FASER2 can extend the current limits from CHARM significantly. 

\textbf{Non-Minimal Models:} While we have described some extensions of the benchmark portals in the preceding discussion, more models with additional portal fields and/or interactions exist, often motivated by considerations of UV completeness or experimental anomalies. These theories have the advantage of decoupling the LLP's production rate from its decay lifetime, giving rise to rich phenomenology. They can also lead to qualitatively new sources of LLPs, as in the case mentioned above of secondary production for a dark photon coupling to DM. A scenario that has been well studied in the literature consists of a $Z'$ and a HNL~\cite{Jho:2020jfz, Das:2021nqj, Chauhan:2020mgv}, which could also explain the MiniBooNE anomaly~\cite{MiniBooNE:2010idf}. In addition to its usual neutrino-mixing based production modes, the HNL could be produced through $Z' \!\to\! N N$ decays~\cite{Deppisch:2019kvs}, or via the upscattering of SM neutrinos, $\nu p \!\to\! N p$~\cite{Jodlowski:2020vhr}. If the HNL is produced in neutrino scattering in the upstream rock, it could also be detected via a multiple coincident muon signature traversing the FPF experiments~\cite{Bakhti:2020szu}. Other non-minimal models that have been studied for FASER2 include models with both a dark photon and a dark Higgs~\cite{Jodlowski:2019ycu, Araki:2020wkq, Bertuzzo:2020rzo}, models with both a dark photon and an axion~\cite{deNiverville:2019xsx}, and models with an ALP decaying into dark sector states~\cite{Bakhti:2020vfq}. Finally, models of inelastic DM have been studied in Refs.~\cite{Berlin:2018jbm, Jodlowski:2019ycu}.

\subsection{Dark Matter Scattering and Production}

\begin{figure}[tbp]
    \centering
    \vspace{-0.5cm}
    \includegraphics[trim={.5cm 0.2cm 0.7cm 0.55cm 0},clip,width=0.6\textwidth]{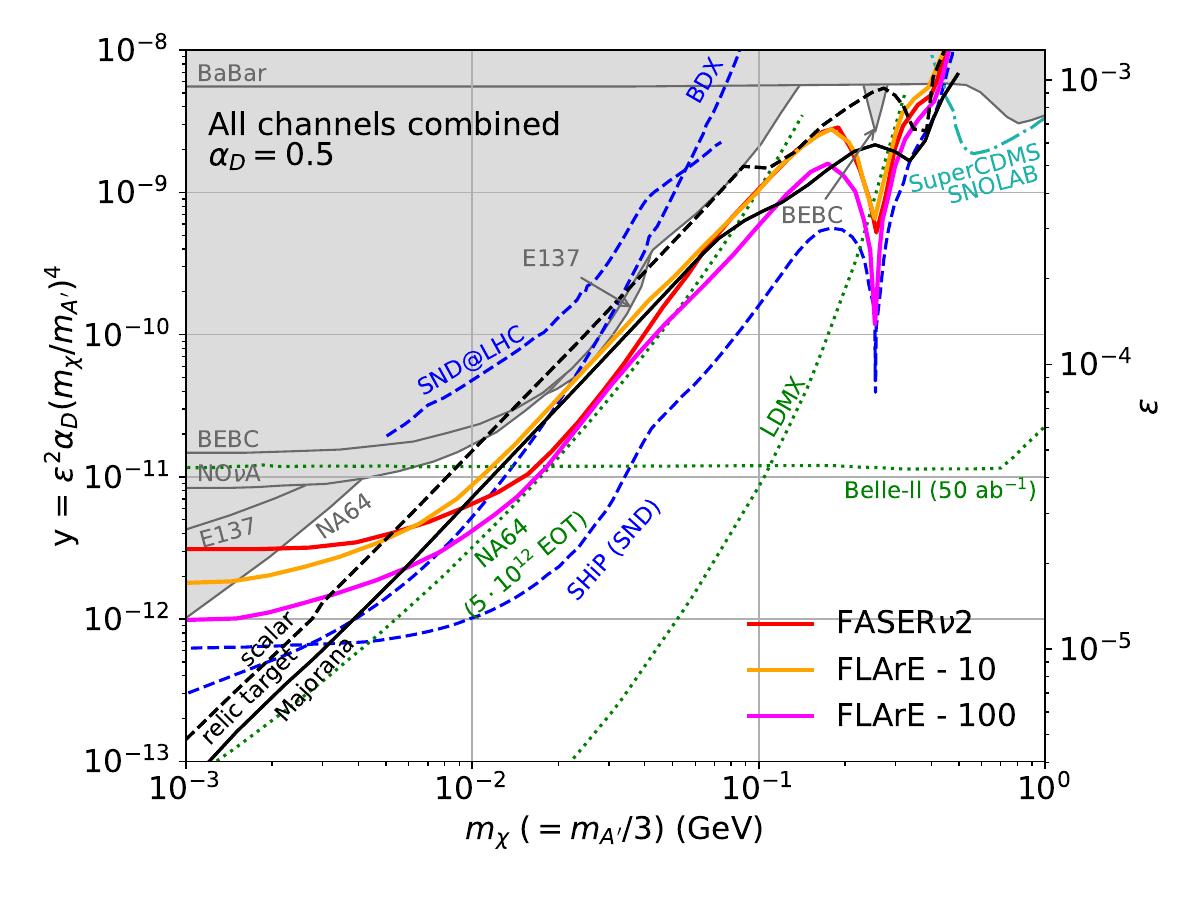}    
    \caption{Parameter space of dark photon-mediated DM models shown in the $(m_{\chi}, y)$ plane, where $y= \epsilon^2 \alpha_D (m_\chi/m_{A'})^4$, for $m_{A'} \!=\! 3 m_\chi$ and $\alpha_D \!=\! 0.5$. Shown are 90\% CL projected exclusion bounds for the proposed FPF scattering detectors, including the emulsion detector FASER$\nu2$ (red) and the 10- and 100-tonne LArTPC detectors FLArE-10 (orange) and FLArE-100 (magenta). Model parameters predicting the correct thermal DM relic abundance are shown for Majorana fermion DM (solid black) and complex scalar DM (dashed black).
    Parameter space excluded by previous experiments is shown in gray, and  projections from several other proposed experiments are also displayed.  From Ref.~\cite{Batell:2021aja}.} 
    \label{fig:dark-photon-DM}
\end{figure}

Identifying DM and discerning its fundamental properties is one of the main drivers in particle physics today. Particle colliders such as the LHC have an important role to play in this effort. As is well known, heavy DM with masses near the weak scale lead to the characteristic signature of missing transverse energy at colliders, and a mature program is underway at the LHC to search for DM of this kind. In contrast, in simple models, light DM with MeV to GeV masses is dominantly produced in the forward direction at the LHC, rendering traditional searches ineffective. Such DM can scatter at FPF neutrino detectors. Additionally, in models where the muons and neutrinos reaching the FPF can produce DM, the production of DM can also be inferred by searches for missing energy in the neutrino detectors. We briefly discuss the possibilities for each of these signatures.

\textbf{Scattering of Light Dark Matter:} One promising approach to probe light DM is to search for its scattering with ordinary matter using suitable detectors housed in the FPF. As a similar search strategy is called for to directly detect collider-produced neutrinos, emulsion detectors such as FASER$\nu$ and SND@LHC, and their potential HL-LHC era upgrades, as well as the proposed LArTPC detector FLArE, would be well-suited to study the scattering signatures of light DM produced in the forward direction at the LHC.

The prospects for detecting light DM in this way have been examined in several recent studies~\cite{Batell:2021blf, Boyarsky:2021moj, Batell:2021aja}. In simple, well-motivated models with a kinetically-mixed dark photon mediator, scattering detectors in the FPF can probe both the electronic and nuclear interactions of DM through a variety of reactions, including elastic DM-electron scattering, elastic DM-nucleon scattering, inelastic resonant pion production, and DM DIS. With reasonable assumptions for detector energy thresholds and energy and spatial resolutions, one can employ suitable topological and kinematic cuts to separate the DM signature from neutrino-induced background processes.  In \cref{fig:dark-photon-DM}, the 90\% CL projected exclusion bounds for the proposed emulsion experiment FASER$\nu$2, as well as for the 10- and 100-tonne LArTPC detectors FLArE-10 and FLArE-100, are displayed in the dark photon model parameter space. In the plot, we have assumed that muon-induced backgrounds can be suppressed to negligible levels with the use of timing information. Crucially, with the full HL-LHC dataset, the FPF experiments can probe regions of parameter space in these models that explain the observed DM abundance through simple thermal freeze-out, shown by black lines in \cref{fig:dark-photon-DM} for Majorana fermion and complex scalar DM.

It is worth emphasizing that the search for light DM scattering in the FPF detectors at the LHC will probe DM interactions in the relativistic regime, and as such is rather insensitive to  the particular DM particle type and interaction structure. This is in contrast to traditional DM direct detection experiments searching for the scattering of non-relativistic halo DM, where event rates can be substantially suppressed in certain models (e.g., spin- or momentum-dependent scattering).  In addition, the DM scattering rate in the LHC far-forward detectors is not sensitive to the precise DM abundance obtained from its thermal production. We then expect the DM signal to grow with increasing values of the coupling constants. This is not the case for direct detection searches, since larger couplings lead to larger annihilation cross sections and smaller values of the thermal relic density. The complementarity between the two types of searches could then become very instructive when probing subdominant components of DM. Finally, it is worth highlighting the complementarity with experiments utilizing missing energy/momentum techniques with lepton beams. Unlike the FPF detectors at the LHC, such experiments do not detect the re-scattering of DM and also have diminished sensitivity to the hadronic interactions of DM. 

\textbf{Production of Light Dark Matter in Neutrino Scattering:} Another promising approach to probe for light dark sector states is to search for an associated missing energy signature. At the FPF, such a signature could occur in models with a neutrino-philic mediator $X$, which could be produced in neutrino interactions $\nu q \to \ell q' X$. Many well-motivated theories for BSM physics provide new mediators that exclusively couple to neutrinos. A well-known example is the Majoron~\cite{Chikashige:1980ui}. Another scenario with a mediator mass below the weak scale has been proposed to address the relic density of the sterile neutrino DM~\cite{deGouvea:2019phk, Kelly:2020pcy, Kelly:2020aks}, which otherwise is in severe tension with existing constraints~\cite{Abazajian:2017tcc}. The thermal freeze-out mechanism for DM via the neutrino-philic mediator has also been explored in Ref.~\cite{Kelly:2019wow}. An additional well-motivated scenario is a minimal U(1)$_{L_\mu-L_\tau}$ gauge boson with a vector-like DM candidate, where the  mediator mass is below the muon mass~\cite{Foldenauer:2018zrz, Holst:2021lzm}. Neutrino-philic mediators provide motivated targets for current and upcoming experiments, and the search for these particles across a large range of masses is an important task for the future of particle physics. 

Constraints on neutrino-philic mediators arise from cosmology~\cite{Heurtier:2016otg, Kelly:2020aks, Escudero:2020ped}, precision decay width measurements~\cite{Pasquini:2015fjv, Blinov:2019gcj, Brdar:2020nbj}, and neutrino-less double beta decay~\cite{Agostini:2015nwa, Blum:2018ljv}, but are typically significantly weaker than those on force carriers coupled to the SM charged fermions. In the future, planned accelerator neutrino experiments, such as DUNE, will be able to search for the production of such mediators in neutrino scattering and probe the sub-GeV mass range~\cite{Berryman:2018ogk, Kelly:2019wow}. The FPF offers the potential of using the LHC's existing TeV-energy neutrino beam to search for neutrino-philic scalars with even higher masses in a so-far unconstrained parameter space~\cite{Kelly:2021mcd}. Relevant scalar mediator masses are in the range of 0.5 GeV to 100 GeV, and the range of coupling to neutrinos between 0.1 and 1 are relevant to FPF experimental constraints.

\subsection{Millicharged Particles}

The search for mCPs is related to several deep mysteries of the Universe, including charge quantization~\cite{Dirac:1931kp} and DM~\cite{Brahm:1989jh,Feng:2009mn,Cline:2012is}. A mCP with electric fractional charge $Q_{\chi}$ can be modeled simply as $\mathcal{L}_{\rm mCP} = \bar{\chi}(i \cancel{\partial} - \epsilon e\cancel{A}- m_{\rm mCP})\chi$, where $\epsilon \equiv Q_{\chi}/e$. A mCP can be a low-energy consequence of a particle coupled to an kinetically-mixed U(1)$^{\prime}$ dark photon in the massless phase~\cite{Holdom:1985ag}, but it can also just be a particle with a small hypercharge, without the existence of a dark photon.

\begin{figure}
\centering
\includegraphics[trim={.5cm 0 0.4cm 0},clip,width=0.6\textwidth]{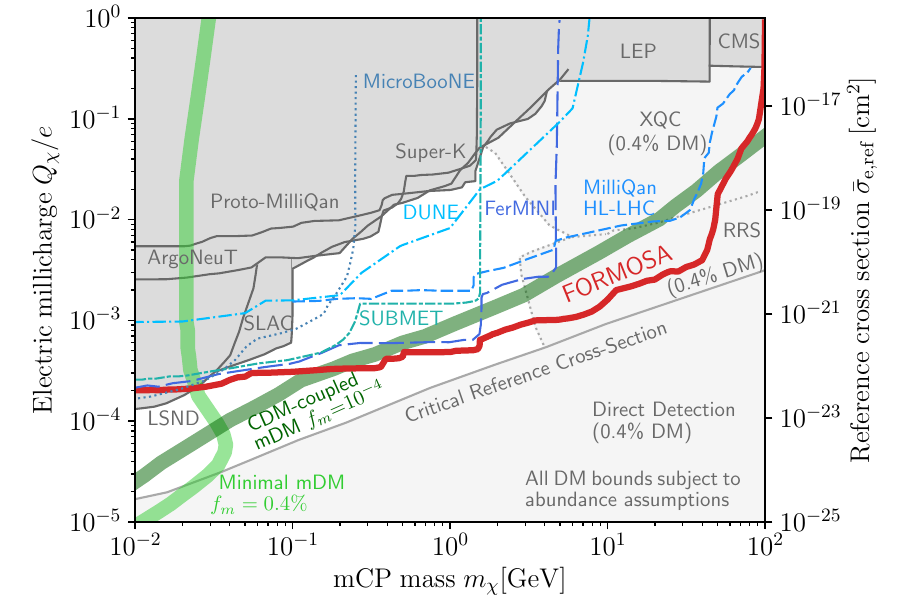}
\caption{Parameter space for mCPs shown in the $(m_{\chi}, Q_{\chi}/e)$ plane. The sensitivity reach of FORMOSA is shown as a solid red curve assuming an on-axis location, a $1\, \m \times 1\, \m$ cross-sectional area, and an integrated luminosity of $3~\iab$. Exclusions from previous collider searches are shown in dark gray~\cite{Prinz:1998ua, Akers:1995az, Davidson:2000hf, CMS:2012xi, Jaeckel:2012yz, Magill:2018tbb, Acciarri:2019jly, Ball:2020dnx, Plestid:2020kdm}, and projections for proposed searches are shown as blue dashed lines~\cite{milliQan:2021lne, Kelly:2018brz, Choi:2020mbk, Acciarri:2019jly, Magill:2018jla,Marocco:2020dqu}. The millicharged SIDM window is also presented in terms of the reference cross section $\bar{\sigma}_{\rm e,ref}$ (right vertical axis). In this case, additional constraints arise from underground direct-detection, balloon, and satellite experiments~\cite{Mahdawi:2018euy, Emken:2019tni, Erickcek:2007jv, Rich:1987st}, which are presented assuming a mSIDM contributes $0.4\%$ to the DM abundance. The green bands correspond to the millicharged dark matter explaining the EDGES anomaly~\cite{Liu:2019knx}. 
}
\label{fig:mCP}
\end{figure}

The FPF provides one of the best opportunities to search for mCPs, since a large flux of high energy mCPs can be produced, independent of the DM assumptions, and detected by a detector located in the forward region. The FORMOSA experiment, discussed in \secref{FORMOSA}, is a dedicated proposal to exploit this opportunity. Its projected sensitivity, alongside existing accelerator constraints and other proposed searches, is presented in \cref{fig:mCP}. FORMOSA will be the most sensitive experiment to study mCPs in the 10 MeV to 100 GeV mass window. In addition, a liquid argon detector at the FPF, such as FLArE, might also have the ability to study mCPs, although the sensitivity is not yet properly studied.

Such a mCP can also account for a fraction of the observed DM abundance, making it an example of milli-charged strongly interacting dark matter (mSIDM) with a large ``reference cross section'' $\bar{\sigma}_{\rm e,ref}$. Such a scenario cannot be probed by conventional underground direct-detection detectors, since the mSIDM particle flux would be attenuated through interactions in the Earth's atmosphere and crust, and they would lose too much energy to meet the threshold for traditional underground direct-detection experiments~\cite{Emken:2019tni, Mahdawi:2018euy,Plestid:2020kdm}. Accelerator experiments, including FORMOSA, have advantages in probing  mSIDM since they directly produce high-energy mCPs that would not be attenuated before reaching the detector. The probe of mSIDM is a clear demonstration of the power of accelerator searches in new physics scenarios where DM interacts strongly with SM particles. 

Another interesting physics motivation related to the parameter space that FORMOSA can probe is related to the result reported by the Experiment to Detect the Global EoR Signature (EDGES) Collaboration~\cite{Bowman:2018yin, Barkana:2018lgd, Slatyer:2018aqg, Liu:2019knx}. To explain the absorption spectrum reported by EDGES, a mCP can cool the gas to allow a stronger 21 cm hydrogen absorption signal at redshift $z \sim 17$. As again shown in \cref{fig:mCP}, a large parameter space of millicharged dark matter models motivated by the EDGES anomaly can be probed by FORMOSA.

In addition to mCPs, there are a variety of other models that could lead to an anomalous energy deposition signature that could be probed at FPF experiments, such as models of DM with electromagnetic form factors~\cite{Chu:2018qrm, Chu:2020ysb, Kling:2022ykt}.

\section{Neutrino Physics}
\label{sec:neutrinos}




As the particle accelerator with the highest energy built thus far, the LHC is also the source of the most energetic human-made neutrinos~\cite{DeRujula:1984pg, DeRujula:1992sn, Beni:2019pyp, Bai:2020ukz, Abreu:2019yak, Abreu:2020ddv, Abreu:2021hol, Ahdida:2020evc, Ahdida:2750060, Foldenauer:2021gkm}. The $pp$ collisions occurring at the LHC IPs produce a large number  of hadrons along the beam direction, which can inherit an $\mathcal{O}(1)$ fraction of the proton energy.  The decays of these particles then lead to an intense and strongly collimated beam of highly energetic neutrinos of all three flavors in the far-forward direction. In this section, we discuss expected neutrino fluxes at the FPF experiments, explore the potential neutrino cross section measurements, and lay out some examples of the BSM neutrino physics that can be explored at FPF experiments. The potential neutrino physics discussed here by no means covers all the possible scenarios that can be explored at the FPF experiments. 

\subsection{Neutrino Fluxes} 
\label{sec:nu-fluxes}

A crucial ingredient for the FPF's neutrino physics program will be reliable estimates of the LHC's forward neutrino fluxes and their uncertainties.  The neutrinos at the FPF originate from weak decay of forward-going hadrons, in particular pions, kaons, hyperons, and charmed hadrons. While the forward production of light hadrons typically relies on non-perturbative hadronic interaction models tuned to data, the production of forward charm can be calculated using perturbative QCD methods. More details of forward neutrino production are also discussed in \secsref{qcd}{astro}. 

\begin{figure*}[t]
    \centering
    \includegraphics[width=0.48\textwidth]{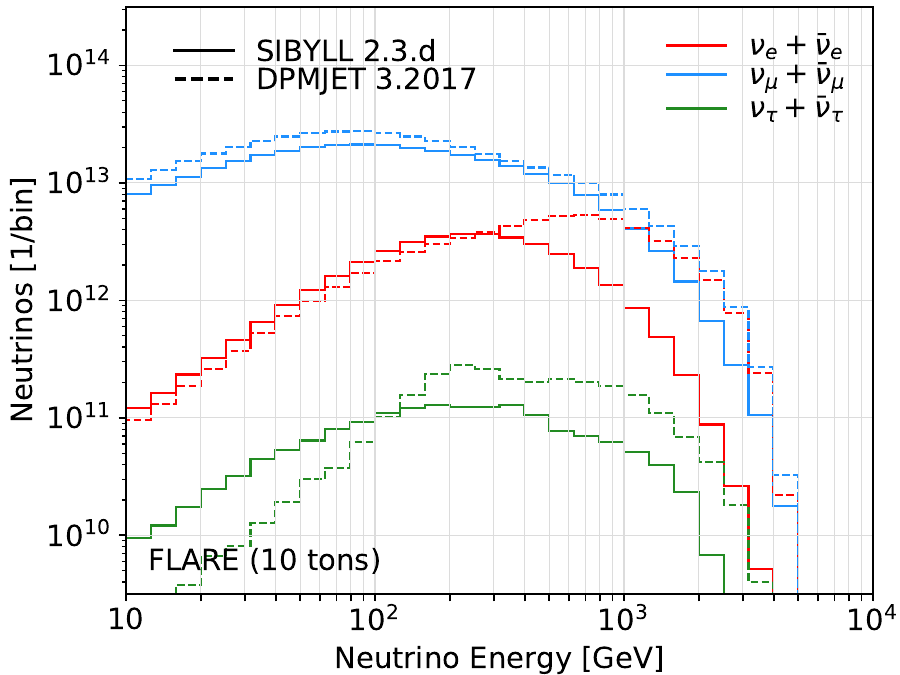}
    \includegraphics[width=0.48\textwidth]{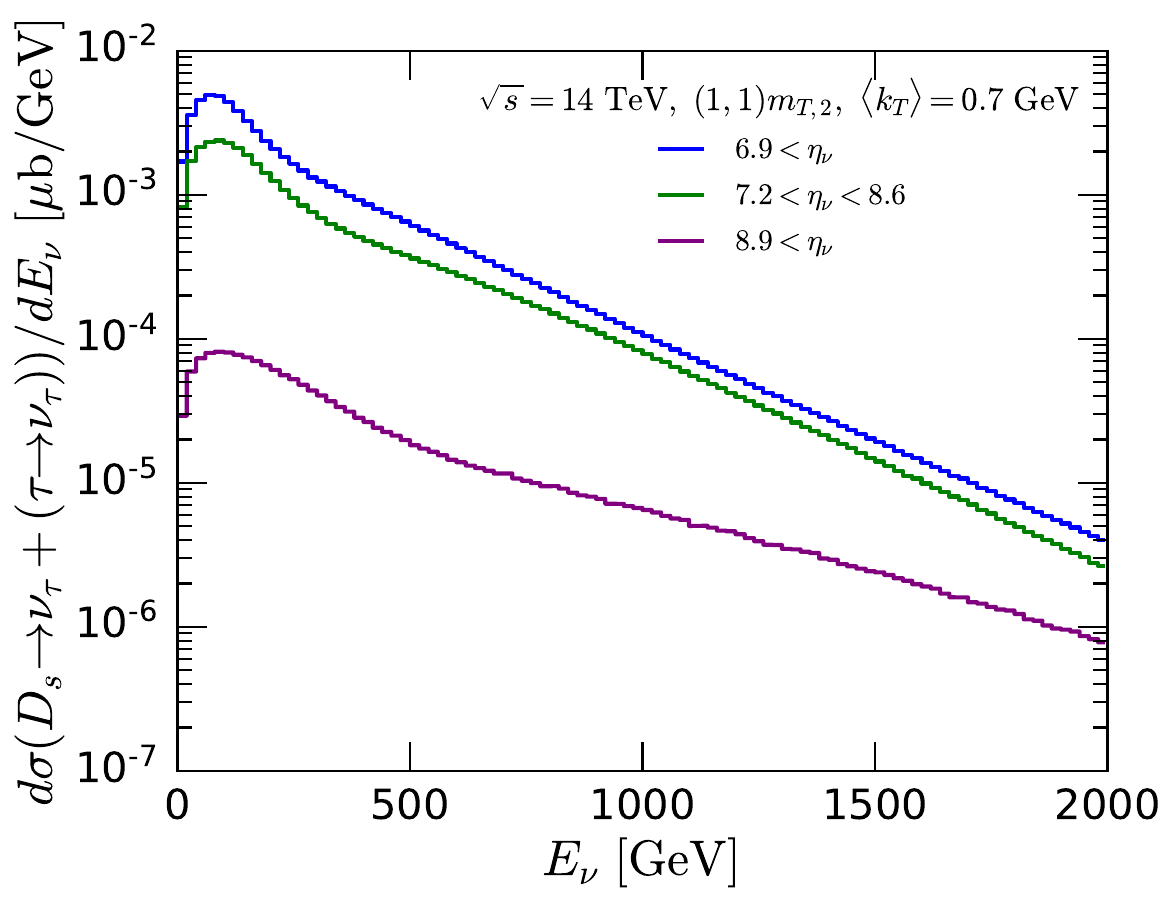}
    \caption{Left: The number of neutrinos per energy bin passing through the FLArE detector with an assumed cross-sectional area of $1~\m \times 1~\m$ ($\eta_\nu\gtrsim 7.2$). The flux is estimated with two different event generators, \texttt{Sibyll 2.3d} and \texttt{DPMJET 3.2017}, and uses the fast neutrino flux simulation of Ref.~\cite{Kling:2021gos}. We further assume the detector to be located $620~\m$ downstream of the ATLAS IP at the HL-LHC with 14 TeV $pp$ collisions and an integrated luminosity of $\mathcal{L}=3~\iab$. Right: The differential cross section for the $\nu_\tau+\bar{\nu}_\tau$ energy distribution for different rapidity ranges for $2\pi$ azimuthal coverage. The results are evaluated using NLO perturbative QCD with the PROSA PDFs~\cite{Zenaiev:2019ktw} and scale proportional to $m_{T,2}=\sqrt{4 m_c^2+p_T^2}$ for $D_s$ production and decay~\cite{Bai:2020ukz,Bai:2021ira}.}
    \label{fig:neutrinoflux}
\end{figure*}

In the left panel of \cref{fig:neutrinoflux}, the estimated neutrino flux for all three neutrino flavors is shown. We show the energy spectrum of neutrinos going through a $1~\m \times 1~\m$ cross-sectional area, corresponding to the FLArE detector, evaluated using two different event generators, \texttt{Sibyll 2.3d}~\cite{Ahn:2009wx, Riehn:2015oba, Riehn:2017mfm, Riehn:2019jet} and \texttt{DPMJET 3.2017}~\cite{Roesler:2000he, Fedynitch:2015kcn}, as implemented in the CRMC simulation package~\cite{CRMC}. The fast neutrino flux simulation introduced in Ref.~\cite{Kling:2021gos} is used to propagate SM hadrons through the LHC beam pipe and magnets and to simulate their decays into neutrinos. The electron and muon neutrino fluxes have both light and heavy flavor hadron contributions, where the highest energy neutrinos at the FPF come predominantly from charm hadron decays. In contrast, the charm decay $D_s^ \pm \to \tau^\pm\nu_\tau$ and the subsequent tau decays dominate the tau neutrino flux over the full energy range. Table~\ref{tab:interactingnu} shows the numbers of CC and neutral-current (NC) DIS neutrino interactions in the detectors estimated using the fluxes from \texttt{Sibyll 2.3d} and \texttt{DPMJET 3.2017}. While we only show the combined results for $\nu + \bar\nu$, it is worth noting that the LHC's neutrino beam consists of a similar number of neutrinos and antineutrinos. We also note that there are currently large differences between the predictions for the event rate of these two event generators, which are mainly related to the modeling of the charm component. NLO perturbative evaluations of charm production improve the predicted high-energy neutrino fluxes by decreasing the huge LO uncertainties.  The right panel of \cref{fig:neutrinoflux} shows a NLO perturbative QCD evaluation of the energy distribution of $\nu_\tau+\bar{\nu}_\tau$ from $D_s^\pm$~\cite{Bai:2020ukz, Bai:2021ira} using the PROSA PDFs~\cite{Zenaiev:2019ktw}, for several rapidity ranges and full azimuthal coverage. 

\begin{table}[tbp]
\setlength{\tabcolsep}{5.2pt}
    \centering
    \begin{tabular}{c|c|c||c|c|c|c}
    \hline\hline
      \multicolumn{3}{c||}{Detector} & 
      \multicolumn{4}{c}{Interactions at FPF} \\
      \hline
      Name &  Mass & Coverage
      & CC $\nu_e\!\!+\!\bar{\nu}_e$ 
      & CC $\nu_\mu\!\!+\!\bar{\nu}_\mu$
      & CC $\nu_\tau\!\!+\!\bar{\nu}_\tau$
      & NC\\
       \hline\hline
       FASER$\nu$2  
       & 20 tonnes & $\eta \gtrsim 8.5$
       & 178k / 668k & 943k / 1.4M & 2.3k / 20k & 408k / 857k \\
       \hline
       FLArE 
       & 10 tonnes & $\eta \gtrsim 7.5$
       & 36k / 113k & 203k / 268k & 1.5k / 4k & 89k / 157k \\
       \hline
       AdvSND1  
       & 2 tonnes & $7.2 \lesssim \eta \lesssim 9.2$
       & 6.5k / 20k & 41k / 53k & 190 / 754 & 17k / 29k \\
       \hline
       AdvSND2 
       & 2 tonnes & $\eta \sim 5$
       & 29 / 14 & 48 / 29 & 2.6  / 0.9 & 32 / 17 \\
       \hline\hline
    \end{tabular}
    \caption{The estimated number of neutrino interactions as obtained using two different event generators, \texttt{Sibyll 2.3d} and \texttt{DPMJET 3.2017}, for FPF experiments located $620~\m$ downstream of the ATLAS IP at the HL-LHC with 14 TeV $pp$ collisions and an integrated luminosity of $\mathcal{L}=3~\iab$.}
    \label{tab:interactingnu}
\end{table}

As is clear from the discussion and results presented above, there are sizable uncertainties associated with the predictions of the neutrino fluxes. On the one hand, this makes the measurement of neutrino fluxes an interesting physics goal that can help us better understand forward particle production. This will be discussed in more detail in \secsref{qcd}{astro}. On the other hand, it is also a source of systematic uncertainties for many measurements, for example, the neutrino interaction cross section discussed in the next section. For these applications, it is essential to have reliable flux estimates and quantify their uncertainties.

Since different approaches are used to describe forward particle production, different strategies are required to quantify their uncertainties. The inclusive production of heavy-flavored hadrons, relevant for the description of the $\nu_\tau$ flux and of the $\nu_e$ flux at high neutrino energies, can be described by calculations with a perturbative QCD core. For fixed-order computations of the hard scattering, associated perturbative QCD uncertainties are given by renormalization and factorization scale variations. As the accuracy of these calculations is limited, related uncertainties turn out to be significant. Additionally, these calculations have a non-perturbative component. A complete assessment of uncertainties therefore requires including uncertainties associated with non-perturbative ingredients, such as PDFs and fragmentation functions (FFs). In contrast, the inclusive production of light hadrons, which is most relevant for the evaluation of the $\nu_\mu$ fluxes and of the $\nu_e$ fluxes at low neutrino energies, is simulated with hadronic interaction models. They provide a sophisticated description of microscopic physics at the expense of a sizable number of phenomenological parameters tuned to data. One approach often used in astroparticle physics is to consider the spread of generator predictions as a first estimate of the uncertainties. While this approach captures some differences due to both tuning and underlying modeling, it is unclear how to interpret its results statistically. An alternative approach to address this problem is computing tuning uncertainties~\cite{Abreu:2020ddv, Buckley:2018wdv, Krishnamoorthy:2021nwv}. Here, multiple additional tunes are obtained that deviate from the central tune so that they represent uncertainties at a given confidence level. Work in this direction is in progress.

\subsection{Neutrino Interactions and Cross Sections}

As we have seen above, the FPF neutrino experiments can detect many neutrino interactions at the highest human-made energies. In the following, we will first discuss the different types of neutrino interactions observed at the FPF and how the FPF can help constrain the associated neutrino interaction cross sections at TeV energies. Lastly, we comment on the final state hadronic physics effects in these high-energy neutrino interactions.

{\bf Deep-inelastic scattering (DIS):} Due to the large neutrino beam energy, and hence a large energy transfer $Q^2$, the majority of neutrino interaction events can be described by DIS. These events are characterized by about $5 - 10$ energetic hadronic particles, carrying an $\mathcal{O}(1)$ fraction of the incoming neutrino energy, which form a collimated jet in the detector~\cite{Abreu:2019yak}.

\begin{figure*}[t]
\centering
\includegraphics[width=0.99\textwidth]{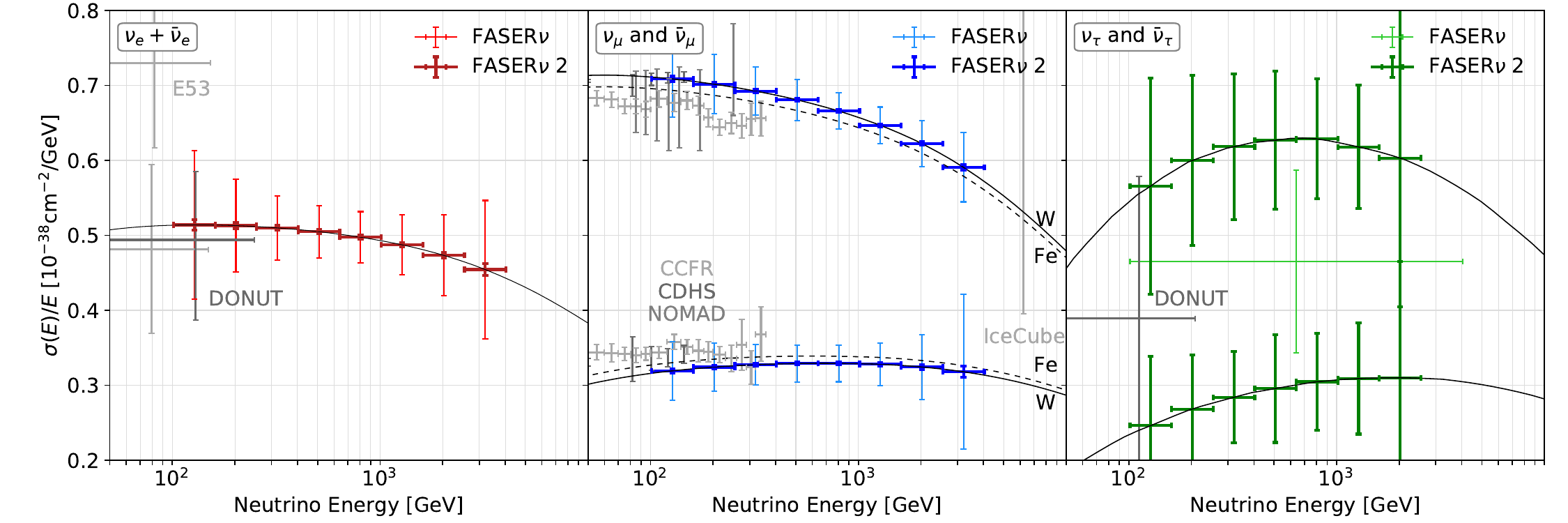}
\caption{Illustration of the estimated statistical uncertainties for FPF experimental measurements of the $\nu$-nucleon CC cross section for electron (left), muon (center), and tau (right) neutrinos. Here we present results for FASER$\nu$ at the LHC and FASER$\nu$2 at the HL-LHC with integrated luminosities of $150~\ifb$ and $3~\iab$, respectively. For muon and tau neutrinos, we show separate results for the neutrino and antineutrino cross section, where we assume perfect charge identification. The sizes of systematic uncertainties are under study and not included in this figure. Existing constraints are shown in gray. The black curves are the theoretical predictions for the average DIS cross section at LO per tungsten-weighted (solid) and iron-weighted (dashed) nucleon. }
\label{fig:xs_sensitivity}
\end{figure*}

DIS neutrino interaction cross sections have been measured by beam dump experiments at low energies $E_\nu < 350~\gev$~\cite{ParticleDataGroup:2020ssz} and by IceCube at high energies, $E_\nu>6.3~\tev$, for muon neutrinos~\cite{Aartsen:2017kpd}. The first cross section measurements at TeV energies will be performed by the FASER$\nu$ and SND@LHC detectors during Run~3 of the LHC~\cite{Abreu:2019yak, Abreu:2020ddv, Ahdida:2020evc, Ahdida:2750060}. These cross section measurements can be further improved by the FPF neutrino detectors FLArE, FASER$\nu$2, and AdvSND in the HL-LHC phase with significantly larger event statistics for all three neutrino flavors. This is illustrated in \cref{fig:xs_sensitivity}. The black solid curve is the theoretical prediction for the average DIS CC cross section per tungsten-weighted nucleon. It is evaluated at LO using \texttt{nCTEQ15} PDFs~\cite{Kovarik:2015cma}, where we have included a suppression factor for the tau neutrino cross section obtained in Ref.~\cite{Kretzer:2002fr}. The gray error bars correspond to existing neutrino cross section measurements. Note that these plots compare CC cross sections for different target nuclei, introducing small differences between the different scattering cross sections. As an example of this effect, we also show the cross section for iron, which is the target material of CCFR and CDHS, as a dashed black line in the central panel. To illustrate the capabilities of experiments in the FPF to measure neutrino cross sections, we show the potential statistical uncertainties in cross section measurements, here illustrated with the tungsten-based detector FASER$\nu$2 at the HL-LHC, with improvements over projected measurements of FASER$\nu$ during Run~3 of the LHC. Similar results are expected for the other detectors. Systematic uncertainties, included neutrino flux uncertainties as well as experimental uncertainties, are currently under study and not included in \cref{fig:xs_sensitivity}. 

Magnetized detector components would allow identification of the charge of muons produced in neutrino interactions, and hence allow measurements of the neutrino and antineutrino CC cross sections separately. This applies to muon neutrinos, but also to tau neutrino interactions, where the produced tau lepton decays into a muon. The FPF neutrino experiments would therefore be able to differentiate tau neutrinos and tau antineutrinos for the first time. This is illustrated in \cref{fig:xs_sensitivity}, where we show separate cross section measurements for neutrinos and antineutrinos in the center and right panel, assuming perfect charge identification. For tau neutrinos, we include only tau decay into muons, which occurs with a branching fraction of about $17\%$. Further information about the outgoing muon could be obtained by associating the event in the neutrino detector with the activity in subsequent detectors placed in the FPF, including FASER2.

In addition to CC cross sections, NC scattering can also be measured by neutrino detectors at the FPF. Although neutral hadron backgrounds need to be identified and removed~\cite{Ismail:2020yqc}, the timing capabilities of the FPF detectors will allow for efficient vetoing of neutral hadron events. The ratio of the NC and CC cross sections can be interpreted as a measurement of the weak mixing angle, which could have sensitivity comparable to existing high energy neutrino datasets from CHARM~\cite{CHARM:1987pwr} and NuTeV~\cite{NuTeV:2001whx}, if uncertainties can be brought below the percent level. This ratio can also be used to limit neutrino non-standard interactions~\cite{Ismail:2020yqc}.

{\bf Quasi-elastic (QE) and resonant (RES) scattering:} In addition to the total inclusive scattering cross section, one can also study specific exclusive neutrino interaction processes. In the $10$-tonne detectors operating during the HL-LHC era, we expect $\mathcal{O}(10^3)$ such CC quasi-elastic (CCQE) events and a similar number of processes in which only the soft pions are produced (CCRES)~\cite{Batell:2021aja}. The dominant contributions to both types of processes are related to the interactions of the muon (anti)neutrinos with the mean neutrino energy for the interactions typically between $200$ and $300~\gev$. Notably, the up-to-date CCQE cross section measurements have been performed for $E_\nu\lesssim 100~\gev$; see Ref.~\cite{Formaggio:2013kya} for a review. The relevant measurements in the FPF could then extend these results to larger values of $E_\nu$. In addition, the measurement of non-DIS CC events, in which the outgoing lepton energy is a very good estimator of the incoming neutrino energy, would provide independent information about the incident neutrino spectrum, which has important implications for QCD and neutrino oscillation studies.

{\bf Shallow inelastic scattering (SIS) and the SIS-DIS transition region:} Beyond the resonance states is a region of continuum non-resonant $\pi$ production that starts at hadronic invariant mass $W = M_N + M_\pi$, where $M_N$ is the nucleon mass. The non-resonant $\pi$ production in the $ 1.4~\gev \leq W \leq 2~\gev$ and $Q^2 \leq 1~\gev^2$ kinematic range is defined as the SIS region, which transitions into the DIS region ($W \geq 2~\gev$ and $Q^2 \geq 1~\gev^2$). The boundary between the SIS and DIS regions is not well defined, and it lies in the transition region between where the interactions are described in terms of hadronic degrees of freedom and where the interactions are described with quark and gluon degrees of freedom. In the electromagnetic sector, these can be described by the quark-hadron duality phenomenon that provides a connection between the average value of interaction strengths in the quark-gluon description of the DIS formalism at high $Q^2$, and the average value of interaction strengths in the pion-nucleon description in the region of resonance excitation at low $Q^2$~\cite{Bloom:1970xb,Bloom:1971ye}. Although the duality has been extensively studied both experimentally and theoretically with electromagnetic-induced processes, it is only poorly known in the weak sector~\cite{SajjadAthar:2020nvy}. The experiments at FPF are expected to have $\mathcal{O}(10^3)$ neutrino events in this kinematic region and would provide a rare opportunity to study quark-hadron duality in the weak sector. Additionally, the so-called Soft-DIS region (kinematically defined as $W \geq 2~\gev$ and $Q^2 \leq 1~\gev^2$), as well as the SIS region above the well-studied $\Delta$-resonance region are only minimally studied both experimentally and theoretically.  Data from FPF experiments would provide unique insights in studying these processes. 

{\bf Neutrino-electron scattering:} The exceptionally hard neutrino spectrum in the forward region of the LHC allows one to observe neutrino scatterings off electrons, which are highly suppressed at low energies. This relies on the NC scatterings of muon neutrinos, as well as on both the CC and NC interactions of electron neutrinos. During the HL-LHC era in the $10$-tonne detectors placed in the FPF, we expect about $50$ such scatterings leading to final-state electrons and $\sim 200$ events with an outgoing muon. The latter are mostly due to $\nu_\mu e^-\to \nu_e \mu^-$ and can therefore provide an independent measurement of the total muon neutrino flux. Importantly, the scatterings off electrons typically lead to far-forward outgoing leptons, which allows for discriminating between them and the CCQE scatterings of $\nu_\mu$ discussed above.

{\bf Test of lepton universality:} An intense beam of neutrinos of all three flavors in the far-forward direction coupled with the high expected CC events statistics (see Table~\ref{tab:interactingnu}) at FPF experiments will provide a unique opportunity to test lepton universality in neutrino scattering by comparing the CC cross section of all three neutrino flavors.

{\bf Final state hadronic physics effects in neutrino interactions:} The neutrino experiments at the FPF will observe high-energy neutrino interactions with various nuclear targets, e.g., tungsten and argon. On the one hand, an accurate description of these interactions and its uncertainties is important for many of the considered SM and BSM applications. On the other hand, this high-energy neutrino-nucleus collider setup also provides additional opportunities to improve our understanding of the associated hadronic and nuclear physics. 

Most of the detected events can be described by deep inelastic neutrino scattering, for which the cross sections are available at NNLO and beyond~\cite{Gao:2021fle}. Hadronic and nuclear effects associated with the initial state are included through (nuclear) PDFs, and are discussed in more detail in \cref{sec:CCDIS}. In addition, there are a variety of final state hadronic and nuclear effects that need to be considered when modeling high energy neutrino interactions. This includes the parton shower, the hadronization process, and the interactions of both partons and hadrons when passing through the dense matter of the target nucleus.  

All these effects have an impact on the kinematics of the final state, such as the multiplicity and energies of hadrons. The FPF neutrino detectors, in particular an emulsion detector with its high spatial resolution, will be able to measure the shape of the neutrino events.  Analogous to electron-nucleus DIS at the EIC~(see Sec.~3.3.2 in Ref.~\cite{Accardi:2012qut}), neutrino-nucleus DIS at the FPF could then be used to obtain complementary information on, for example, (i) the response of nuclear matter to fast moving light and heavy quarks, (ii) medium-induced energy losses and their impact on FFs, (iii) color transparency, and (iv) final state interactions of hadrons in nuclear matter. In addition, these event shapes could also be used as valuable input to tune neutrino event generators, such as GENIE~\cite{Andreopoulos:2009rq, Andreopoulos:2015wxa} or GiBUU~\cite{Buss:2011mx}. 

\subsection{BSM Neutrino Physics: Examples}

If the neutrino fluxes and interactions are sufficiently well understood, the energy spectrum and total rate of interacting neutrinos at FPF experiments can also be used to probe BSM effects. In the following we consider three such scenarios: non-standard interactions, neutrino dipole moments, and sterile neutrino oscillations. 

{\bf Non-standard interactions and effective field theories:} One of the main goals of the FPF neutrino experiments is to measure neutrino interaction cross sections at TeV energies. These measurements can be used to probe new physics associated with the interactions between neutrinos and hadronic matter. Given the excellent agreement between experiments and SM predictions up to LHC energies, one motivated approach is to parametrize the effects of new physics in terms of the standard model effective field theory (SMEFT). The SMEFT has the same particle content and respects the same local gauge symmetries as the SM, and effects of new physics are included via higher dimensional non-renormalizable operators that are added to the Lagrangian.  In the top-down approach, the SMEFT can be obtained by integrating out the heavy particles from some UV models above the weak scale~\cite{Skiba:2010xn, Gaillard:1985uh, Cheyette:1987qz, Henning:2014wua}. After further integrating out the top quark and weak bosons, one obtains new 4-fermion interactions between leptons and leptons/quarks, which are also referred to as neutrino non-standard interactions (NSIs)~\cite{Farzan:2017xzy, Grossman:1995wx, GonzalezGarcia:2001mp}. Constraints obtained at different energies can then be translated by considering the matching and running between different EFTs~\cite{Jenkins:2017jig, Alonso:2013hga, Alonso:2014zka, Grojean:2013kd, Jenkins:2013zja, Jenkins:2013wua, Jenkins:2017dyc, Aebischer:2017gaw, Aebischer:2015fzz}.

NSIs can give observable effects in the production, propagation, and detection of neutrinos, and so the relevant coefficients can be probed at neutrino detectors~\cite{Bischer:2018zcz, Bischer:2019ttk, Falkowski:2019xoe, Falkowski:2019kfn, Falkowski:2021bkq, Du:2020dwr, Du:2021rdg, Du:2021idh, Du:2021nyb, Luo:2020sho}. Using the approach introduced in Refs.~\cite{Falkowski:2019xoe, Falkowski:2019kfn}, the sensitivity of the FASER$\nu$ detectors was investigated taking into account SMEFT coefficients that modify either neutrino production through meson decays or neutrino detection through DIS~\cite{Falkowski:2021bkq}. In this study, it was shown that forward LHC neutrino experiments can constrain interactions that are in principle weaker than the SM weak interactions (by two to three orders of magnitude), corresponding to new physics at the multi-TeV scale.

\begin{figure*}[t]
  \centering
  \includegraphics[width=0.51\textwidth]{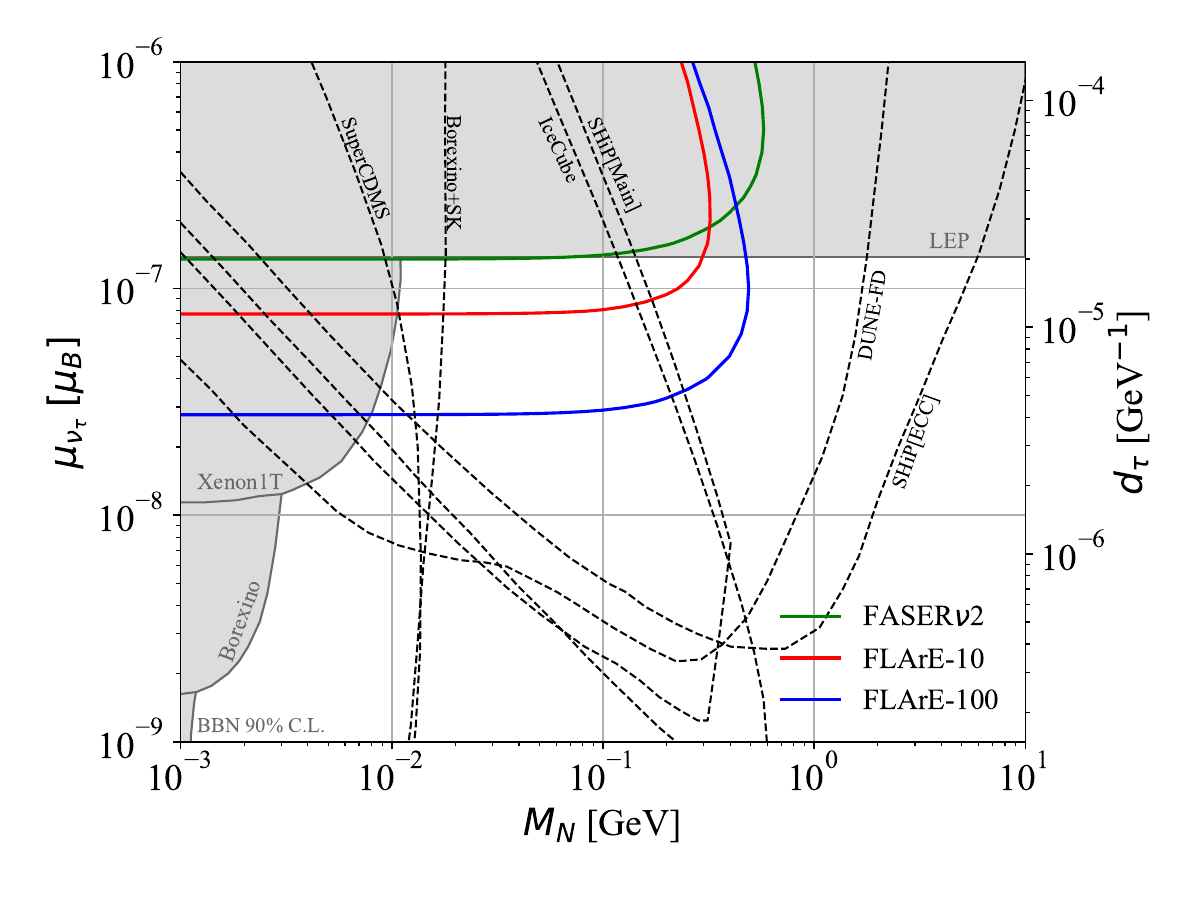}
  \includegraphics[width=0.48\textwidth]{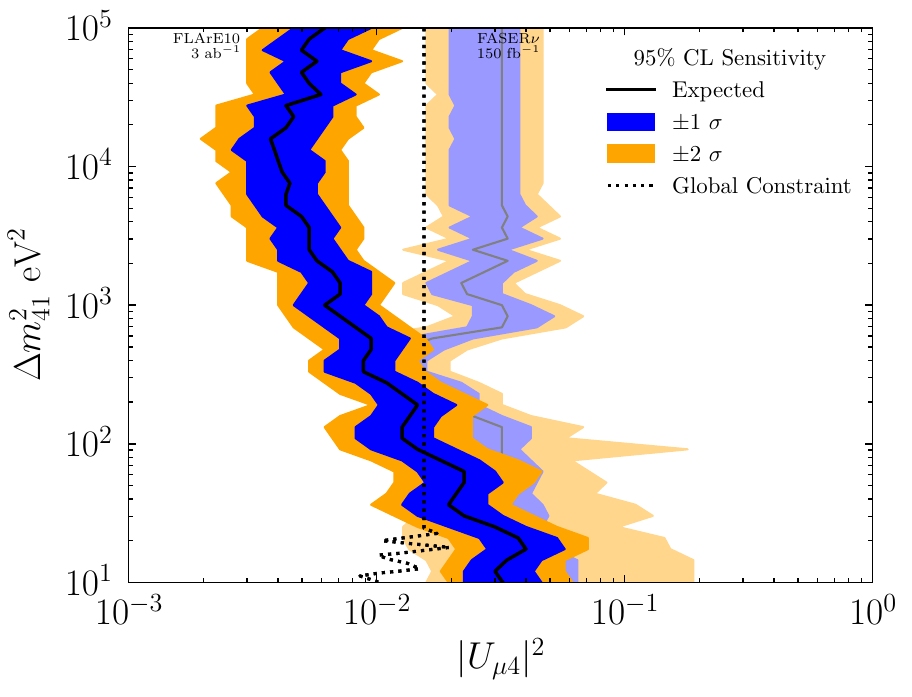}
  \caption{Left: Projected 90\% CL exclusion bounds at FASER$\nu$2, FLArE-10, and FLArE-100 for the tau neutrino's magnetic moment $\mu_{\nu_\tau}$. The grey shaded regions are current constraints from terrestrial experiments, and the black dashed lines are projected sensitivities. Right: Constraints on the mixing parameter of a sterile neutrino, $|U_{\mu4}|^2$, that can be probed by FASER$\nu$ at LHC Run~3 with $150~\ifb$ (light shaded regions) and by FLArE-10 at the HL-LHC with $3~\iab$ (dark shaded regions) at 95\% CL. }
  \label{fig:sensitivity-um4_dipoleNMM_tau}
\end{figure*}

{\bf Neutrino magnetic moments:} The electromagnetic properties of the SM neutrinos are of much interest recently, as they could explain various anomalies, such as the XENON1T excess~\cite{XENON:2020rca}, the observation of black holes in the mass gap region~\cite{Sakstein:2020axg}, and the MiniBooNE excesses~\cite{MiniBooNE:2007uho}. In addition, an observation of a neutrino magnetic moment could shed light on the origin of neutrino masses and allow to distinguish a Dirac or Majorana nature~\cite{Fujikawa:1980yx, Pal:1981rm, Bell:2005kz, Bell:2006wi}. 

Although the SM prediction for the neutrino magnetic dipole moment is very low, $\leq 10^{-19}~\mu_B$~\cite{Giunti:2014ixa}, it can be substantially larger in some BSM theories~\cite{Babu:2020ivd}. The unprecedented flux of neutrinos at the FPF, particularly the tau neutrino flux, and the controlled backgrounds allow one to place stringent constraints on neutrino magnetic moments by looking at neutrino-electron scattering. In the absence of a right-handed neutrino, one can generate a substantial neutrino magnetic moment with the addition of the operator, $\overline{\nu} \sigma_{\mu \nu} \nu F^{\mu \nu}$. FLArE-10 can put an upper limit on the $\nu_\tau$ magnetic moment of a few $10^{-8}\mu_B$~\cite{AbariTsaiAbraham}, which is an order of magnitude lower than the current direct bounds from DONUT~\cite{DONUT:2001zvi}. In theories with a right-handed neutrino, one can include the so-called dipole portal~\cite{Magill:2018jla, Shoemaker:2018vii, Brdar:2020quo}, $\mathcal{L} \supset \frac{1}{2} \mu_{\nu}^{\alpha} \bar{\nu}_{L}^{\alpha} \sigma^{\mu \nu} N_{R} F_{\mu \nu}$, which can lead to excess events at low electron recoil energies~\cite{Jodlowski:2020vhr}. Bounds derived for $\nu_\tau$ are shown in \cref{fig:sensitivity-um4_dipoleNMM_tau} (left) in the $(M_N, \mu_{\nu_\tau})$ plane~\cite{Ismail:2021dyp}. 

{\bf Oscillations to sterile neutrinos:} The large flux of neutrinos and the capability of FPF experiments to detect and identify their flavors will provide an opportunity to probe sterile neutrinos via their oscillations. Given that the baseline is $L \simeq 600$ m and the neutrino energies are typically $E_\nu \sim \text{few}~100$ GeV, the sterile neutrino masses in the sensitivity range of the FPF will be of the order of tens of eV, i.e.,~$\Delta m_{41}^2 \sim 1000 {\ \rm eV^2}$. The possibility of exploring such sterile neutrino oscillations at LHC forward experiments has been investigated in Refs.~\cite{Abreu:2019yak, Bai:2020ukz}. These studies can be performed with muon and electron neutrinos at FASER$\nu$ and SND@LHC during Run~3.  During the HL-LHC era, however, the FPF would enable similar studies with tau neutrinos, which would be impossible otherwise, given the low number of tau neutrinos. 

In \cref{fig:sensitivity-um4_dipoleNMM_tau} (right) we show the $(|U_{\mu4}|^2, \Delta m_{41}^2)$ parameter space that can be constrained at 95\% CL by FASER$\nu$ in Run~3 and FLArE-10 in the HL-LHC era.  These results use the Feldman-Cousins procedure~\cite{Feldman:1997qc} and the current global oscillation constraints~\cite{Dentler:2018sju} are shown for reference indicating that FLArE-10 will have leading sterile neutrino constraints in the $m_4\gtrsim10$ eV mass region \cite{Bolton:2019pcu}. The neutrino flux uncertainty is modeled with a single nuisance parameter scaling the flux between the predictions from \texttt{Sibyll 2.3d} and \texttt{DPMJET 3.2017}. Cross section uncertainties are not included.

\section{QCD}
\label{sec:qcd}

      
        

Quantum Chromodynamics (QCD) is unanimously accepted as the theory of the strong interac\-tions. Yet, there are kinematic regimes in which QCD has not been stringently tested. The FPF offers a number of unique opportunities for testing and studying QCD in some of these regimes, as can be inferred from the fact that predictions for fluxes and cross sections at the FPF introduce unique challenges for QCD theory. We expect that the neutrinos reaching the FPF will be mostly emitted in the decays of various hadrons produced in collisions at the LHC ATLAS IP. In particular, as explained in \secref{neutrinos}, muon neutrinos will be produced mostly in the decays of light mesons and, to a lesser extent, the light baryons. Tau neutrinos will be produced by decays of heavy-flavored hadrons, especially $D_s^\pm$ mesons. Electron neutrinos will be produced in decays of both light and heavy-flavored hadrons, with the latter dominating at the largest neutrino energies. 

Therefore, the FPF, with its capability of distinguishing neutrinos and antineutrinos of different flavors, will provide versatile experimental data on both light- and heavy-flavor production. Interpretation of these data will require diverse theoretical approaches. When describing heavy-meson production, charm and bottom quark masses above 1 GeV allow one to apply perturbative QCD (pQCD) methods down to $p_T$ = 0. However, the smallness of the $c$ and $b$ masses compared to the other physical scales, notably the LHC center-of-mass energy $\sqrt{s}$, introduces typical pQCD challenges associated with so-called multi-scale processes. Additionally, non-perturbative QCD effects are expected to be enhanced in forward heavy-flavor production. On the other hand, low-$p_T$ light-flavor production is dominated by non-perturbative QCD effects and multiple parton interactions, compensating for the long-distance pQCD divergences in hard-scattering contributions. Production of all these hadrons can be described either by dedicated calculations, with different levels of accuracy and approximations employed, or by general-purpose event generators.

As we discuss in the following, QCD opportunities can be enhanced by covering a wide rapidity range either by placing the FPF detectors at different radial distances from the beam collision axis or by making the FPF detectors work in coincidence with the ATLAS detector. Deployment of diverse detection techniques, with several detectors having partial overlap in their rapidity ranges, will allow one to cross-check the consistency and robustness of independent measurements. The use of a range of nuclear targets with mass numbers varying in a wide range will fundamentally enhance the FPF potential for constraining nuclear PDFs. 

\begin{figure}[tbp]
\includegraphics[width=0.48\textwidth]{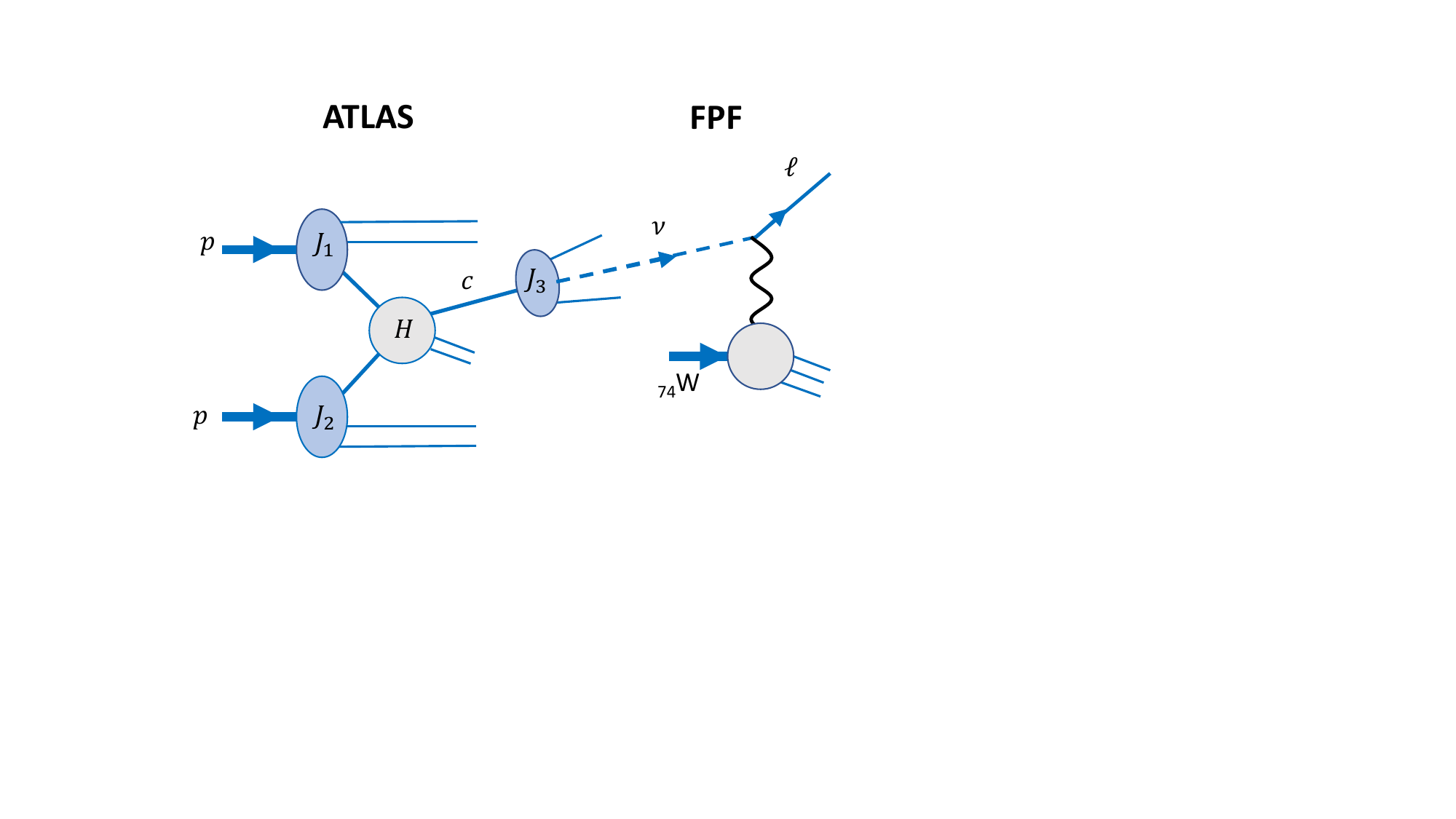}
\hspace{1.2cm}
\includegraphics[width=0.31\textwidth]{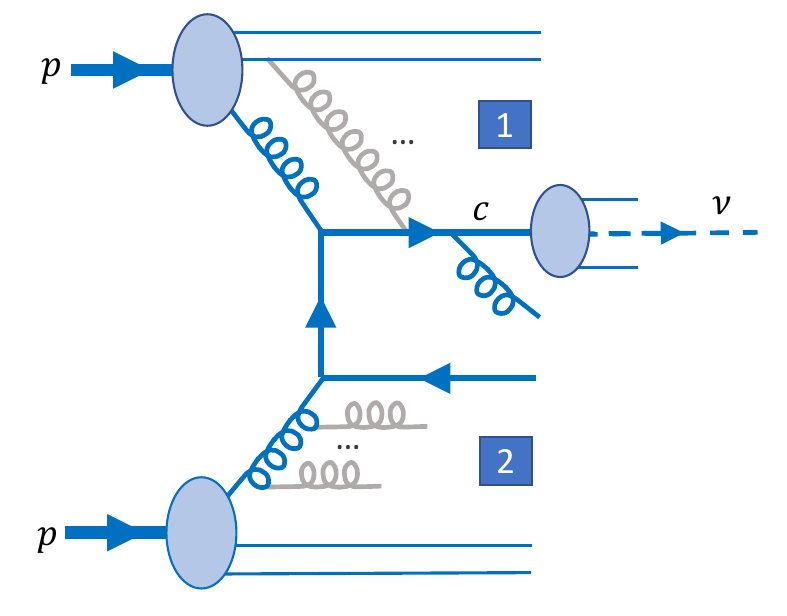}
\caption{Left:~Production of a forward neutrino at the ATLAS IP, and its detection via CC DIS on a tungsten target in the FPF. Right:~The Feynman graph shown in dark blue represents a typical radiative contribution included in the collinear factorization framework at central rapidities.  In far-forward $pp\to c\bar c$ production, additional enhanced corrections from higher orders of $\alpha_s$ are expected, such as those obtained by including the partonic interactions and emissions indicated in grey color, as discussed in the main text.}
\label{fig:FeynmanGraph1}
\end{figure} 

The novelty of the QCD regime probed at the remote FPF site can be illustrated by the example of forward neutrinos from charmed meson decays in ATLAS, as shown in the left panel of Fig.~\ref{fig:FeynmanGraph1}. For the production of charmed hadrons observed inside ATLAS at central rapidity, the standard collinear factorization framework presents the $pp$ cross section as the factorized convolution of the hard scattering cross section $H$, jet functions $J_1$ and $J_2$ describing the breakup of initial protons 1 and 2, and a jet function $J_3$ describing the fragmentation of the charm quark into a charmed hadron whose decay, in turn, may generate a neutrino. The initial-state jet functions $J_1$ and $J_2$ depend on PDFs. The final-state jet function $J_3$ depends on the FFs into charmed mesons and the branching ratios for meson decays into neutrinos. The aforementioned jet functions are essentially independent of one another and of the hard cross section $H$. Indeed, when the charm quark escapes at a large angle to the beam axis, these jet functions, accounting for collinear and soft-collinear emissions, describe the development of parton showers in the directions of the initial-state protons and in the direction of the fragmenting charm quark, respectively. These kinematic sectors are spacelike-separated, and so the dominant dynamical contributions in these sectors can be calculated independently (in this approximation, we are neglecting wide-angle, soft-gluon interactions producing color correlations between different sectors and driving hadronization at the end of the perturbative phase of the scattering event.).

On the other hand, to produce a neutrino in the direction of the FPF, the charm quark escapes close to the beam axis in nearly the same direction as the comoving remnants of proton 1. For pseudorapidities of at least 6 and possibly as high as 9, the incident parton 1 carries nearly all the energy/longitudinal momentum of proton 1, while the incident parton 2 carries a vanishing fraction of the energy/longitudinal momentum of proton 2.  In this configuration, we expect important QCD contributions that are not included in the standard collinear factorization. In the right panel of Fig.~\ref{fig:FeynmanGraph1}, the dark-blue Feynman graph is a typical QCD contribution included in computations of charm production at central rapidities. These contributions are known at least to the next-to-leading order (NLO) accuracy in the QCD strong coupling constant $\alpha_s$.  In the far-forward region, there are at least two more categories of important contributions. First, the forward charm quark now travels for an extended duration of time in the mean gluon field created by the remnants of proton 1. Contributions to jet functions $J_1$ and $J_3$ are no longer cleanly separated. While the gluon PDF in the leading-power contribution falls off rapidly in the $x_1\to 1$ limit, one may encounter enhanced power-suppressed contributions with the charm quark connected to jet 1 by two or more gluon propagators, as illustrated by the grey gluon propagator in the upper half of the right panel of Fig.~\ref{fig:FeynmanGraph1}.  Some of these effects can be estimated by introducing a non-perturbative, or ``intrinsic,'' charm PDF. Also, the final-state fragmentation may be affected by interactions with the proton remnants. Second, accounting for the multiple gluon emissions responsible for the immense longitudinal energy loss from the initial-state partons with small $x$ (see emissions from the incident parton 2 in the right panel of Fig.~\ref{fig:FeynmanGraph1}) requires one of the theoretical approaches that introduces alternative forms of factorized cross sections applicable for such kinematics, as summarized in Sec.~\ref{sec:High-s}. While the full theory for describing both kinds of novel QCD effects is still to be developed, Secs.~\ref{sec:ForwardCharmHybrid} and \ref{sec:CollinearPDFs} review some estimates that can already be made by the extensions of available techniques.

We note that these measurements ultimately shed light on fundamental aspects of QCD factorization for hadron-hadron collisions at the highest energy $\sqrt{s}$ reached at a collider. Systematic proofs of all-order factorization have been made only for the simplest hadron-hadron observables, such as the inclusive cross section for the lepton pair production process (see Refs.~\cite{Bodwin:1984hc,Collins:1985ue,Collins:1988ig} and Chapter 14 of Ref.~\cite{Collins:2011zzd}). These proofs are complicated to a large extent by multiperipheral scattering contributions in precisely the forward directions relevant for the FPF. Clean measurements of forward cross sections may shed light on the emergence of various types of QCD factorization or the situations when factorization is absent. For example, measurements of neutrino production rates in $pp$ collisions can shed light on QCD factorization and saturation and probe PDFs in so-far unconstrained regions; see Sec.~\ref{sec:CollinearPDFs}.

On the detection side, the primary method to observe forward neutrinos in the FPF is via CC DIS on a nuclear target, such as argon or tungsten (see the left panel of Fig.~\ref{fig:FeynmanGraph1}). Successful measurements must be able to control the nuclear dynamics in CC DIS, and in turn they will provide an opportunity to test poorly known aspects of nuclear PDFs, such as those related to strange quarks and PDF ratios for sea quarks of different flavors. Section~\ref{sec:CCDIS} addresses the physics issues and opportunities for CC DIS on heavy nuclei in the FPF detectors.

In addition to observing forward neutrinos, the FPF rapidity reach may be extended to detect processes involving the associated production of one of these neutrinos with a less forward hadron/jet/lepton, by employing the timing coincidence technique reviewed in Sec.~\ref{sec:HadronFPF}, which discusses some of the physics opportunities presented by this FPF configuration.
Finally Sec.~\ref{sec:MC} describes the FPF potential for constraining and further developing the phenomenological models of non-perturbative QCD embedded in Monte Carlo event generators currently used for the LHC and astroparticle physics. Given the rich panoply of the barely explored QCD effects that will impact FPF measurements, it is clear that the successful interpretation of FPF measurements requires a coordinated program to study the relevant QCD effects at the FPF and other facilities, notably forward production at LHCb, large-$x$ CC DIS at the EIC~\cite{AbdulKhalek:2021gbh}, and small-$x$ dynamics at the LHC and in future DIS experiments.

\subsection{QCD Theory for High-Energy Particle Production \label{sec:High-s}}

The minimal longitudinal momentum fraction $x$ accessed in a hard-scattering process is $Q^2/s$, where $Q$ is the invariant mass of the hadronic final state produced in the hard scattering $H$ (see Fig.~\ref{fig:FeynmanGraph1}). With $\sqrt{s}=14$ TeV at the LHC and $Q$ no higher than a few GeV in most events contributing to forward neutrino production, the hierarchy of energy scales, $s \gg Q^2 \gg  \Lambda_{\mathrm{QCD}}^2$, puts the FPF squarely in the kinematic regime where collinear factorization possibly needs to be augmented or replaced by an alternative theoretical approach.  Fixed-order calculations in pQCD within the collinear factorization framework have a long record of successes in describing experimental data and have been consistently at the core of QCD validation in lepton-lepton, lepton-hadron, and hadron-hadron collisions. However, there are  kinematic regimes in experiments involving~$ep$,~$eA$, $pp$, $pA$, and $AA$ collisions, where fixed-order QCD calculations might not be enough. The high-energy or \emph{Regge} limit belongs to this category and corresponds to the kinematic limit where $s \gg |t|$,~with $s$ and $t$ being the usual Mandelstam variables. It is related to the presence of large logarithms~of the form $\ln(s/Q^2) \approx \ln(1/x)$.  These large logarithms originate in specific Feynman diagrams (so-called ladder diagrams) of arbitrarily high order. Leading logarithmic terms at order $n$ have the form $(\alpha_s \ln s)^n$, whereas the next-to-leading ones have the form $\alpha_s(\alpha_s \ln s)^n$. The all-order resummation of the aforementioned towers of logarithms is provided by the Balitsky--Fadin--Kuraev--Lipatov (BFKL) framework~\cite{Fadin:1975cb,Kuraev:1976ge,Kuraev:1977fs,Balitsky:1978ic}. An important addition to the BFKL resummation program is the inclusion of non-linear effects. These are associated with high partonic densities, and they also restore unitarity at very high center-of-mass energies~\cite{Balitsky:1995ub,Kovchegov:1999yj,Balitsky:2007feb,Jalilian-Marian:1997jhx,Jalilian-Marian:1997qno,Weigert:2000gi,Iancu:2000hn,Ferreiro:2001qy,Gelis:2010nm}.

Typically, for observables in hadron colliders, a hybrid approach is adopted where one incorporates the BFKL resummation inside the standard collinear description. By ``hybrid'' here we mean cases where either (i) the cross section is a convolution of the PDFs of the two colliding partons, the FFs of the outgoing partons, and the partonic cross section, which itself is built as a convolution between the partonic impact factors and the BFKL Green's function, or (ii) both the PDF of one colliding parton and the $k_T$-dependent unintegrated gluon distribution (UGD) for the other one appear in the factorization formula. While approach (i) is suitable for processes with two hard final states with a large rapidity separation, as in, e.g., forward-central production processes (see Sec.~\ref{sec:HadronFPF}), approach (ii) is appropriate for  single-inclusive forward heavy-flavor production (see Sec.~\ref{sec:ForwardCharmHybrid}), in which the hadronic impact factor for the emission of the identified final-state particle from the initial-state large-$x$ parton is taken in collinear factorization, i.e., built from a standard collinear PDF, and in turn is convoluted with a $k_T$-dependent UGD~\cite{Catani:1990eg} associated with the small-$x$ initial-state parton and given by the solution of the BFKL equation.

In the following we will show examples of the application of the hybrid formalism to make predictions for the FPF. A possible first sign of the onset of BFKL dynamics at LHC energies emerged in the Mueller--Navelet dijet channel~\cite{Mueller:1986ey}, where next-to-leading logarithmic approximation (NLA) BFKL predictions~\cite{Ducloue:2013bva,Caporale:2014gpa} and CMS data~\cite{Khachatryan:2016udy} were compared among each other for azimuthal-angle distributions at 7~TeV, with a satisfactory level of agreement. Quite recently, it was shown how certain kinematic configurations probed in the production of jets~\cite{Celiberto:2015yba,Celiberto:2015mpa} and identified hadron final states~\cite{Celiberto:2020wpk, Bolognino:2018oth} might allow for a clear discrimination between BFKL-driven predictions and the standard collinear factorization ones, when data will be available. As discussed in the following, we expect that measurements at the FPF will also help to discriminate between these two approaches.

\subsection{Forward Charm Production in the Hybrid Formalism \label{sec:ForwardCharmHybrid}}

In recent years the LHCb Collaboration has performed several analyses of heavy meson production at high energies and rapidities $y$ up to 4.5 (see e.g.~\cite{LHCb:2013xam, LHCb:2015swx, LHCb:2016ikn}). IceCube has probed the astrophysical and atmospheric neutrino fluxes at even higher energies~\cite{IceCube:2020wum}. Such distinct sets of data are intrinsically related, since a robust description of heavy meson production at high-energy colliders is indispensable for reliable predictions of the prompt neutrino flux, which is expected to dominate the atmospheric neutrino flux for large-enough neutrino energies. In particular, the results presented in Ref.~\cite{Goncalves:2017lvq} and in \secref{astro} indicate that the behavior of the prompt atmospheric neutrino flux at the highest energies accessible at IceCube and at future neutrino telescopes is determined by the production of charmed mesons at very forward rapidities, beyond those probed by the current LHC main detectors. The relevant kinematic region may be accessible at the FPF, provided it is built to cover a wide-enough rapidity range. According to the present layout, described in Sec.~\ref{sec:facility}, the FPF neutrino rapidity coverage will extend down to at least 6.5. Smaller rapidities, down to $\sim$ 5.8, will be possible only for processes with particularly high cross sections, considering that for an off-axis detector with a typical area of 1 m$^2$, only a very few percent of the geometric acceptance can be seen. The FPF will provide information complementary to that from the central and mid-rapidity detectors (ALICE, ATLAS, CMS, LHCb), which are not capable of detecting individual neutrinos, but only of providing the whole missing energy associated with a collision event. 

Charm production was first computed in QCD by using collinear factorization. The core of these calculations are hard-scattering partonic cross sections for the production of $c\bar{c}$ pairs, which are currently known up to NLO accuracy. Already long ago, however, first calculations in the $k_T$ factorization framework~\cite{Catani:1990eg}, involving two UGDs associated to the initial-state partons, started to appear. Currently the hybrid approach which combines elements of collinear and $k_T$-factorization is in use for BFKL-related applications in the kinematic regime of the FPF, as discussed in Sec.~\ref{sec:High-s}. At present, the accuracy of the calculations of short-distance cross sections/impact factors for heavy-quark hadroproduction in this alternative framework is, however, still limited to the leading order.

\begin{figure}[tbp]
\begin{center}
  \includegraphics[trim={0.8cm 0 1.1cm 0},clip,width=0.325\textwidth]{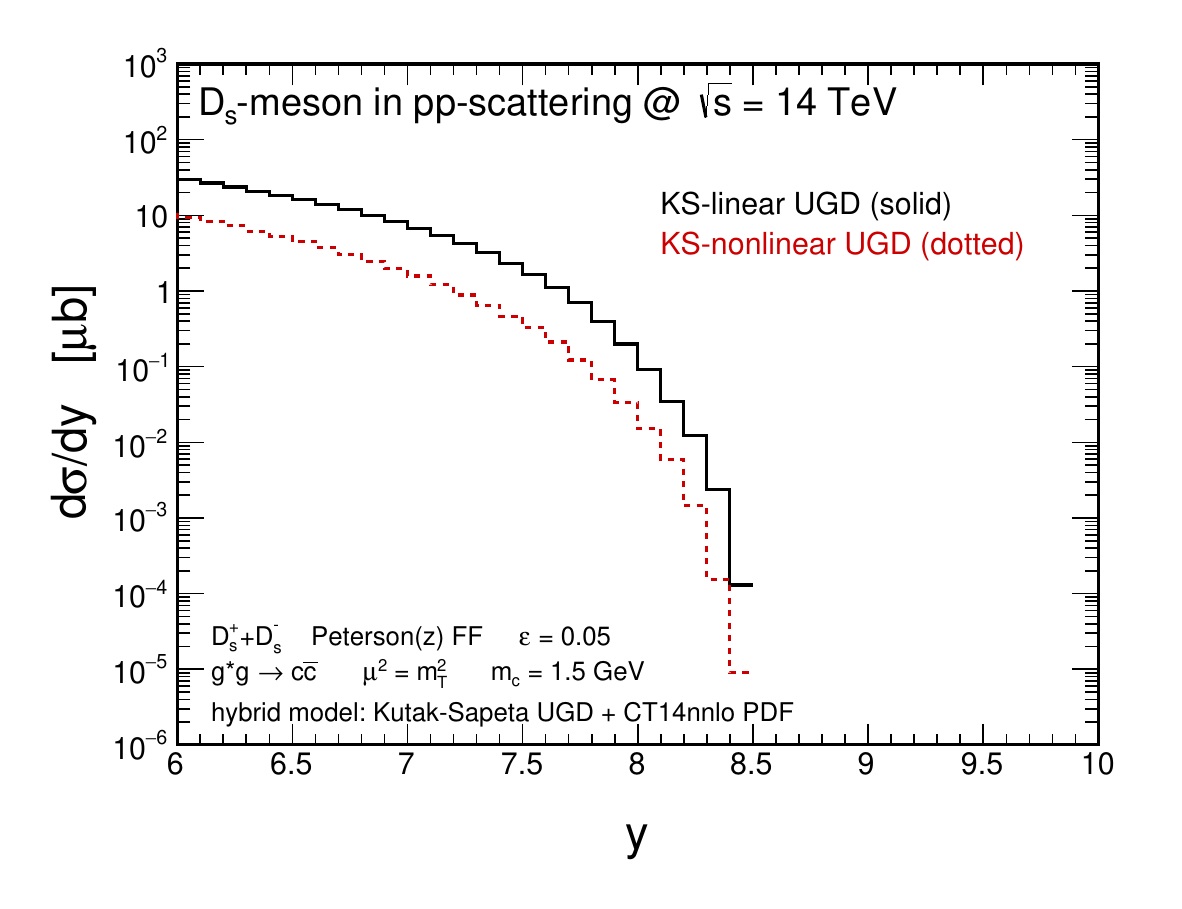}
  \includegraphics[trim={0.8cm 0 1.1cm 0},clip,width=0.325\textwidth]{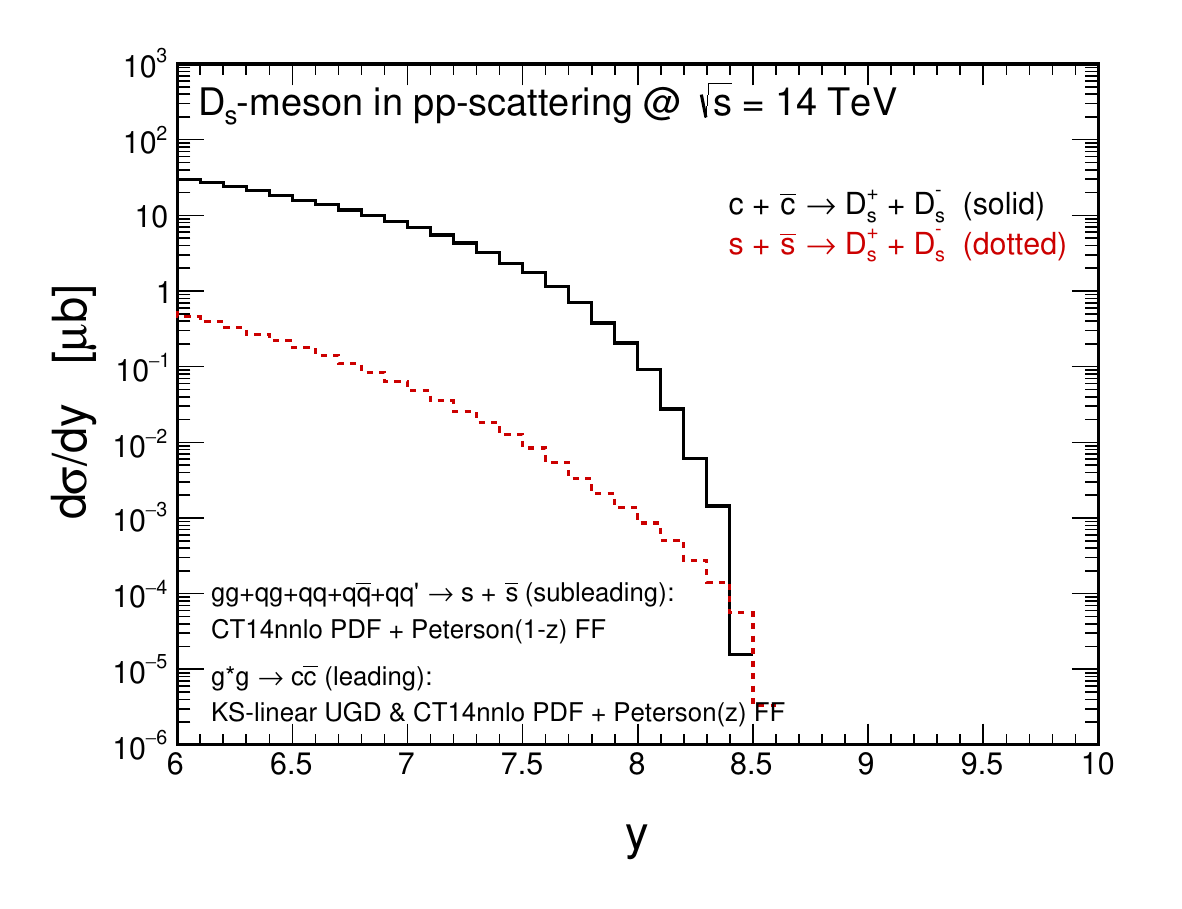} 
  \includegraphics[trim={0.8cm 0 1.1cm 0},clip,width=0.325\textwidth]{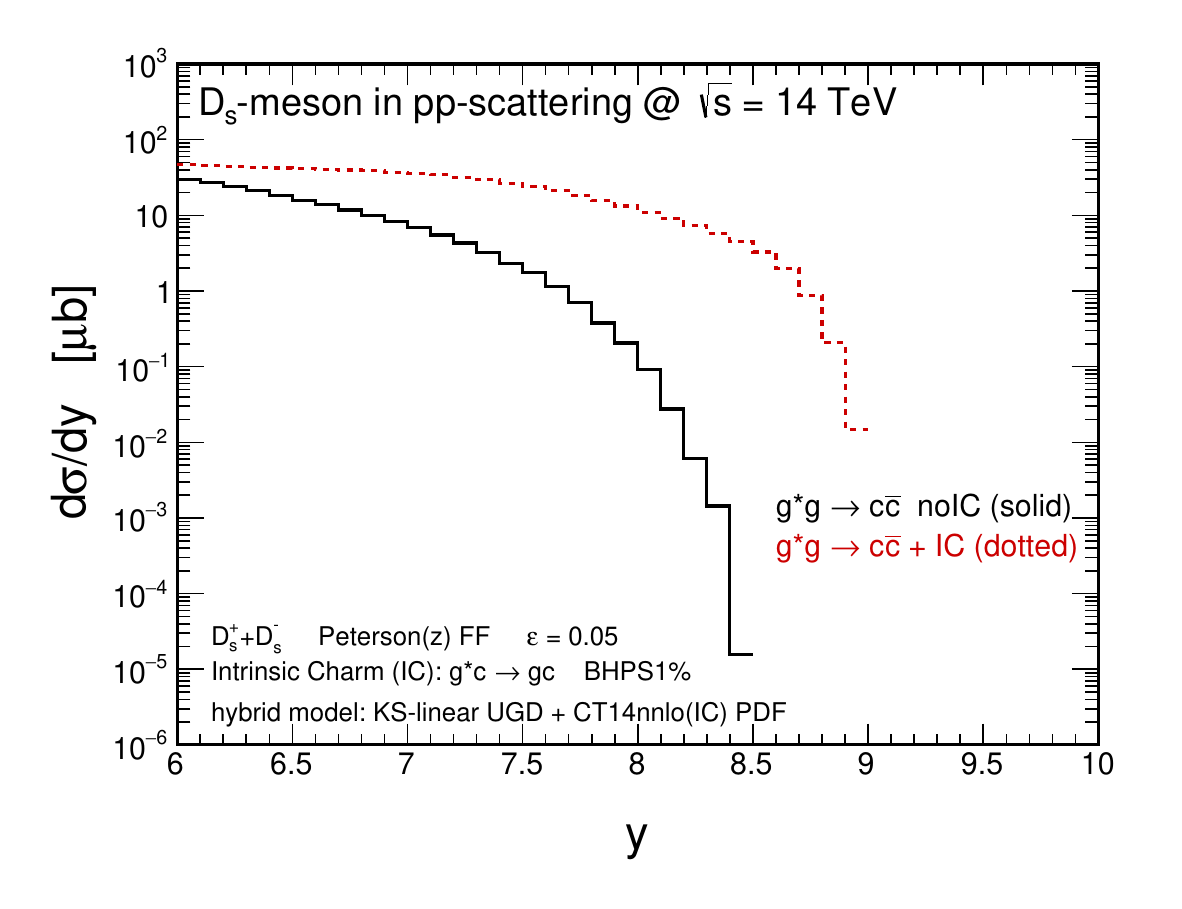}
\end{center}
\vspace*{-.3in}
\caption{The impact of nonlinear effects (left), subleading fragmentation (center), and the intrinsic charm component (right) on the rapidity distribution for $D_s$ meson production in $pp$ collisions at $\sqrt{s} = 14$ TeV.
  }
\label{fig:rapidity}
\end{figure} 

Future measurements at the FPF will advance our understanding of several aspects of forward charm production that are currently topics of intense debate, including:
\vspace*{-0.5em}
\begin{itemize}
\setlength\itemsep{-0.4em}

\item {\bf Discrimination between the purely collinear and the hybrid factorization formalism.} Forward charm production at the LHC mostly proceeds through the collision of one large-$x$ and one small-$x$ parton. Present predictions in the standard collinear framework are in agreement with the available LHC experimental data, at least when considering the large uncertainty bands affecting the calculations. However, the calculations in collinear factorization may not take into account all the relevant small-$x$ corrections. The FPF, probing more extreme $x$ values than LHCb, may clarify if and up to which extent the resummation of the latter corrections is important at LHC energies. To fully realize this program in practice, however, it is critical to upgrade the calculations in both formalisms to higher orders with the aim of reducing the present uncertainties, which are large in collinear factorization and even larger in the hybrid formalism.  
  
\item {\bf The evidence  (or not) of  nonlinear saturation effects resulting from extreme partonic densities in the QCD dynamics at high energies}~\cite{Gelis:2010nm}{\bf .} Charm production at very forward rapidities ($y \gtrsim 5$)  probes the hadronic wave function at very small partonic momentum fractions ($x \lesssim 10^{-5}$), where the recombination process $gg \rightarrow g$ may become important, especially at low hadronic scales $Q \sim 1$ GeV. The typical outcome of saturation is to tame the growth of the gluon PDF to protect the unitarity of the cross sections. The results presented in Refs.~\cite{Giannini:2018utr, Goncalves:2021yvw} indicate that the nonlinear (saturation) effects strongly modify the magnitude of the $x_F$ distribution, suppressing charm production in the kinematic region that will be probed by FPF experiments. The impact of the nonlinear effects on $D_s^\pm$ production in the kinematic range of an FPF experiment can be estimated using the hybrid approach described in Ref.~\cite{Goncalves:2021yvw}, and the results are presented in \figref{rapidity}, left panel. It turns out that, in comparison to the linear predictions, denoted as ``KS-linear UGD,'' the inclusion of the nonlinear effects suppresses the rapidity distribution by a factor of approximately 3.
 
\item {\bf The role of subleading fragmentation channels in charmed meson production.} Additional scattering channels may become competitive in cases where the channel with the dominant fragmentation mode, such as $c\to D$, is suppressed.  In recent years, it has become clear that the description of LHC measurements of charge production asymmetries in heavy meson production~\cite{LHCb:2012fb, LHCb:2018elu} is still a challenge for many theoretical approaches. In the FPF case, the studies performed in Refs.~\cite{Maciula:2017wov, Goncalves:2018zzf}  indicate that the channels with subleading fragmentation may become dominant at very forward rapidities and even enhance the overall rate. Indeed, the main charm production channel $g+g\to c+\bar c$ with the leading $c\to D_s$ FF is kinematically suppressed at the highest rapidity by the rapidly falling gluon PDF, while the $gg+gq+q\bar q+qq' \to s+ \bar s$ channel with the subleading fragmentation $s\to D_s$ may win on balance because of the much larger quark PDFs. The possible impact of subleading fragmentation on $D_s^\pm$ meson production in the kinematic range probed by FPF experiments is presented in \figref{rapidity}, central panel. This contribution becomes non-negligible at very forward rapidities, implying the enhancement of the associated $\nu_{\tau}$ flux probed by FPF experiments. We note that both the leading and subleading curves in the figure have very large PDF uncertainties at the edge of phase space at $y \sim 8.5$. At most, we can conclude at this point that the subleading mode can become competitive. Constraints on the large-$x$ PDFs from other experiments, such as the EIC, will help reduce these uncertainties.

\item {\bf The evidence (or not) for intrinsic heavy quarks}. A significant volume of research suggests that production of charm quarks may receive substantial enhancements, as compared to the gluon-fusion $g+g\to c+\bar c$ channel, when an initial-state charm quark/antiquark component carrying a large longitudinal momentum fraction of the parent nucleon is included. Differently from the extrinsic  heavy quarks/antiquarks that are generated by perturbative gluon splittings at the leading power, this non-perturbative, or intrinsic, component arises from charm-quark subgraphs with multiple connections to the light partons in the proton~\cite{Brodsky:1980pb, Navarra:1995rq}. At the diagrammatic level, the intrinsic component of the charm quark PDF arises as the leading part of ``type-1'' scattering contributions in the right panel of Fig.~\ref{fig:FeynmanGraph1} with two or more gluon connections between the hard charm quark and the proton jet function~\cite{Hou:2017khm}. As explained below in Sec.~\ref{sec:ForwardCharmSACOTMPS}, such an intrinsic component generically enhances charm PDFs at large $x$  ($> 0.2$) as a result of kinematic conditions satisfied by a stable proton bound state~\cite{Brodsky:1980pb,Blumlein:2015qcn}. In recent years, a nonzero initial condition for the evolution of the charm PDFs from the initial scale $Q_0$ has been included in several global QCD analyses~\cite{Ball:2016neh,Hou:2017khm,Ball:2021leu}. The resulting intrinsic-charm (IC) PDFs are compatible with the world experimental data, while the magnitude of the proton's momentum carried by the IC at scale $Q_0$ is constrained to be below 1\%, with the specific constraints varying between the different analyses~\cite{Jimenez-Delgado:2014zga,Brodsky:2015uwa,Jimenez-Delgado:2015tma}. The hybrid framework computations in Refs.~\cite{Giannini:2018utr,Goncalves:2021yvw} indicate that if an intrinsic component is present in the hadronic wave function, the atmospheric neutrino flux for $E_\nu > 5 \times 10^5$ GeV will be enhanced by a factor $\gtrsim 2$, depending of the probability of finding an intrinsic heavy quark component in the nucleon. The right panel of \figref{rapidity} shows predictions for the rapidity distribution of the $D_s^\pm$ production in $pp$ collisions at $\sqrt{s} = 14$ TeV in the rapidity range probed by FPF experiments, with and without including an intrinsic charm component accounted for by the use of the CT14NNLO IC PDFs (BHPS 1.0\%) in the hybrid factorization approach described in Ref.~\cite{Goncalves:2021yvw}. We can see that, in the rapidity range of interest for the FPF, for increasing rapidities, the IC component generically enhances the cross section by a multiplicative factor that can be large in the allowed models. 
\end{itemize}  

\subsection{PDFs and Forward Charm Production According to Collinear Factorization \label{sec:CollinearPDFs}}
  
We now turn back to collinear factorization in the $\overline{\rm MS}$ factorization scheme, which serves as the backbone QCD formalism for making the most of LHC predictions and obtaining the non-perturbative PDFs and FFs needed for these predictions. These PDFs and FFs not only play a key role in collinear factorization, but are also an essential ingredient for predictions in 
the hybrid factorization framework, as discussed in Secs.~\ref{sec:High-s}, \ref{sec:ForwardCharmHybrid}, and~\ref{sec:HadronFPF}. The collinear factorization framework may accommodate to some extent both the large-$x$ and small-$x$ QCD effects present in forward charm hadroproduction at the FPF. This is done in part by judiciously choosing the factorization scales and other auxiliary parameters in the QCD cross sections, implementing logarithmic expansions of all-order resummed cross sections, and constraining the currently uncertain PDFs in the relevant $x$ regions using measurements either at the FPF itself~\cite{Bai:2020ukz} or in other experiments.

In a typical very forward kinematic configuration accessible to the FPF experiments, neutrinos are produced from decays of charmed mesons with large rapidity values. In particular, charm $y$ values up to $\sim 9$ correspond to QCD scattering contributions with disparate partonic momentum fractions as high as $x_1\sim 0.5$ in one proton beam and as low as $x_2 \sim 5\times 10^{-8}$ in the other. As we have already pointed out, the standard QCD framework is modified in both limits of $x\to 1$ and $x\to 0$, where little or no experimental measurements currently exist. 
  
\subsubsection{Constraints on Small-$x$ PDFs}

As discussed in Sec.~\ref{sec:High-s}, for $x \to 0$, such very forward measurements are likely sensitive to BFKL phenomena or saturation~\cite{Golec-Biernat:1998zce} effects, the onset of which may have already been observed in the inclusive HERA data~\cite{Ball:2017otu,Abdolmaleki:2018jln,Hou:2019efy}. For small $x$ below $10^{-4}$, higher-order QCD terms with $\ln(1/x)$ dependence grow quickly at factorization scales of order 1 GeV. Nevertheless, collinear factorization employing precisely known PDFs can provide useful small-$x$ extrapolations for applications in astroparticle physics, such as the calculation of the ultra-high-energy neutrino-nucleus cross sections~\cite{Bertone:2018dse}, the attenuation rates of astrophysical neutrinos as they cross the Earth on their way to the detector~\cite{Garcia:2020jwr}, and the flux of prompt neutrinos arising from charm production in cosmic rays collisions in the atmosphere~\cite{Gauld:2015kvh, Garzelli:2016xmx}.
  
\begin{figure}[tbp]
\begin{center}
\includegraphics[width=0.47\textwidth]{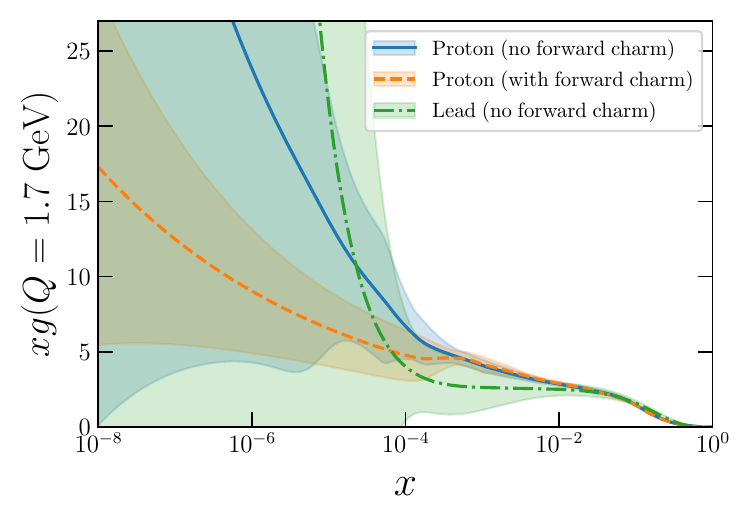}
\hfil
\includegraphics[width=0.47\textwidth]{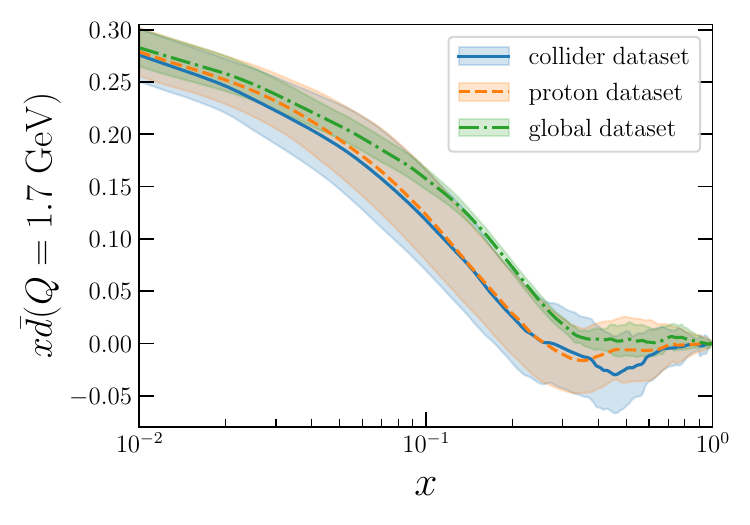}
\vspace*{-.2in}
\caption{Left: The small-$x$ gluon PDF at $Q=1.7$ GeV in the NNPDF3.1 proton fits without and with forward $D$ meson data from LHCb, together with the lead nuclear PDF in the nNNPDF2.0 analysis. Right: The $\bar{d}(x,Q)$ PDF at large $x$, discussed in Sec.~\ref{sec:CCDIS}, here comparing the NNPDF3.1 global proton fit with variants where all fixed-target DIS data are removed and where only data associated with proton targets are retained. \label{fig:PDFs-FPF} }
\end{center}
\end{figure}

The impact of forward charm production data on the small-$x$ PDFs~\cite{PROSA:2015yid} is quantified in Fig.~\ref{fig:PDFs-FPF}, where NNPDF3.1 global fits without and with the LHCb $D$ meson data~\cite{Gauld:2016kpd} at 5, 7, and 13 TeV are compared. One can observe how forward charm measurements constrain markedly the small-$x$ PDFs, as also pointed out in other analyses using open heavy-flavour data~\cite{Zenaiev:2019ktw, Garzelli:2020fmd}.
Similar or even stronger constraints could be expected from the corresponding forward FPF measurements, considering the aforementioned $x$ coverage. It will be interesting to compare the new constraints with those by complementary studies with forthcoming data on charmonium exclusive production in $pp$ collisions and DIS (for first studies in this direction, see Ref.~\cite{Flett:2019pux, Flett:2020duk}), processes that cannot be measured at the FPF, at least according to the present setup.

It would be appealing, but challenging, to extend 
the QCD analyses  
to neutrinos from heavy-flavor production and decay in proton-nucleus collisions, with either the proton or nucleus beam traveling toward the FPF, which offers the possibility of probing different physics mechanisms.  Provided that these measurements are technically feasible and the integrated luminosity will be sufficient to accumulate enough statistics, in the simplest interpretation, such measurements would constrain nuclear PDFs in regions where they are currently even less constrained than proton PDFs, as shown in Fig.~\ref{fig:PDFs-FPF}, where the lead PDF from the nNNPDF2.0 fit~\cite{AbdulKhalek:2020yuc} is compared to the proton one. Given much higher parton densities in the nuclei, the onset of saturation is expected to happen at much higher $x_2$ in proton-lead collisions than in $pp$, facilitating the study of this mechanism. The FPF would then provide access to information on non-linear dynamics in a nuclear environment, complementing the constraints on large-$x$ nuclear PDFs expected from the study of lepton-nucleus DIS at the EIC~\cite{AbdulKhalek:2021gbh,Khalek:2021ulf}.
  
\subsubsection{FPF Forward Charm Production at NLO with Massive Quarks and the Intrinsic Component \label{sec:ForwardCharmSACOTMPS}}

\begin{figure}[tbp] 
\includegraphics[width=0.47\textwidth]{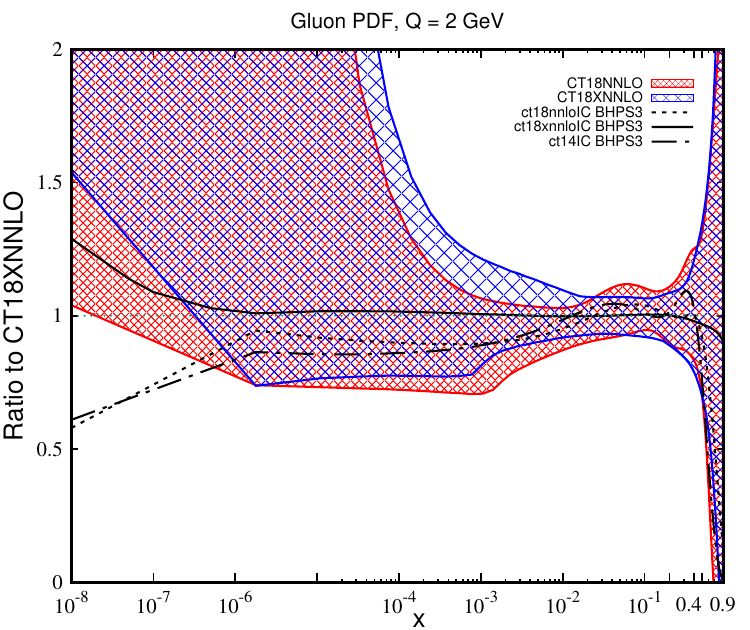}\quad\quad 
\includegraphics[width=0.47\textwidth]{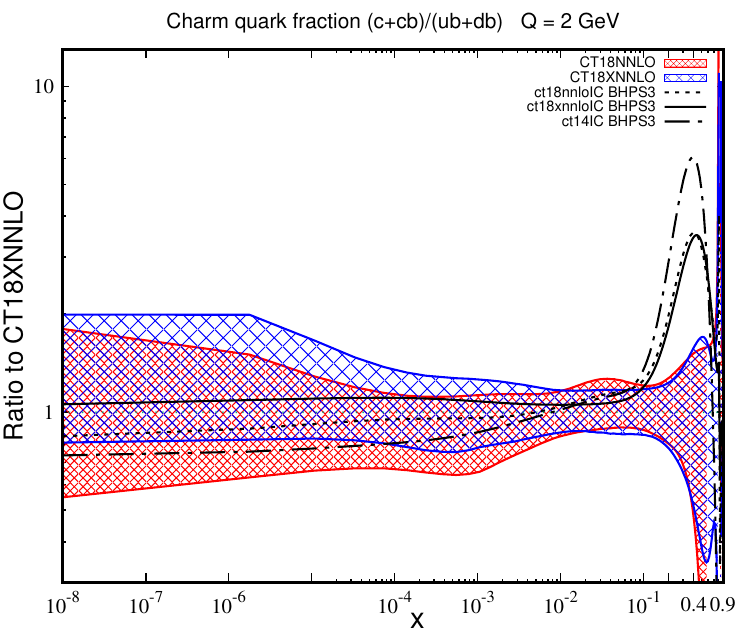}
\caption{Left: Gluon PDF ratios. Right: Charm-quark fraction ratios. The error bands represent the CT18NNLO (red) and CT18XNNLO (blue) PDF uncertainties at 90\% C.L.~\cite{Hou:2019efy}.
The solid, dashed, and dot-dashed lines correspond to CT18XNNLO-IC/CT18XNNLO, CT18NNLO-IC/CT18XNNLO, and CT14NNLO-IC/CT18XNNLO, respectively.}
\label{fig:CT18PDFs}
\end{figure}

At $x\to 1$, intrinsic charm (IC) production contributions arising from power-suppressed (higher-twist) scattering processes may strongly enhance the event rate prediction based on the leading-power (twist-2, or perturbative) calculation. The earliest parton-model formulation~\cite{Brodsky:1980pb, Brodsky:1981se, Pumplin:2005yf, Chang:2011vx, Brodsky:2015fna, Blumlein:2015qcn} introduces IC as a component of the charm PDF that arises from excited $| uudc\bar c\rangle $ Fock states of the proton wave function rather than from $g\to c\bar c$ perturbative splittings. ``Fitted charm'' is a phenomenological parametrization of the IC that is determined in a global QCD analysis as an independent PDF functional form~\cite{Jimenez-Delgado:2014zga,Dulat:2013hea,Hou:2017khm,Ball:2016neh}. In the context of QCD collinear factorization, IC is best understood in DIS, where it contributes via convolutions of universal twist-4 non-perturbative correlator functions with process-dependent hard cross sections~\cite{Hou:2017khm}. Therefore, the IC contributions parametrized by the fitted charm PDF need not coincide in reactions like $ep\rightarrow ec X$ and $pp\rightarrow cX$. Charm hadroproduction and $Z+c$ production at the LHC can constrain the IC contributions in $pp$ collisions~\cite{Hou:2017khm}. Some of these measurements can be extended to the FPF case and could further benefit from the possibility of coordination between the FPF and ATLAS detectors. This is particularly interesting for events with at least two identified final-state objects, with one at small rapidity seen by ATLAS and the other one emitting a large-rapidity neutrino seen by a properly configured  FPF detector. 

\begin{figure}[!ht] 
\includegraphics[width=0.47\textwidth]{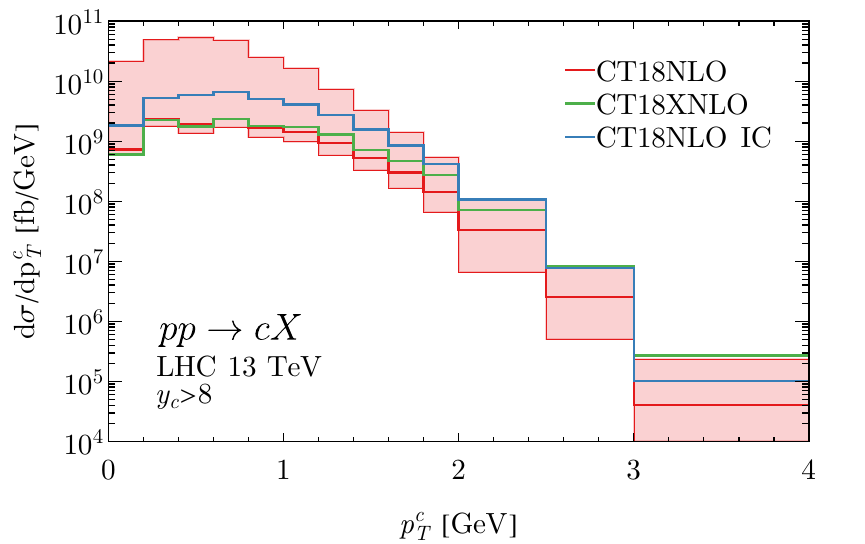}
\includegraphics[width=0.47\textwidth]{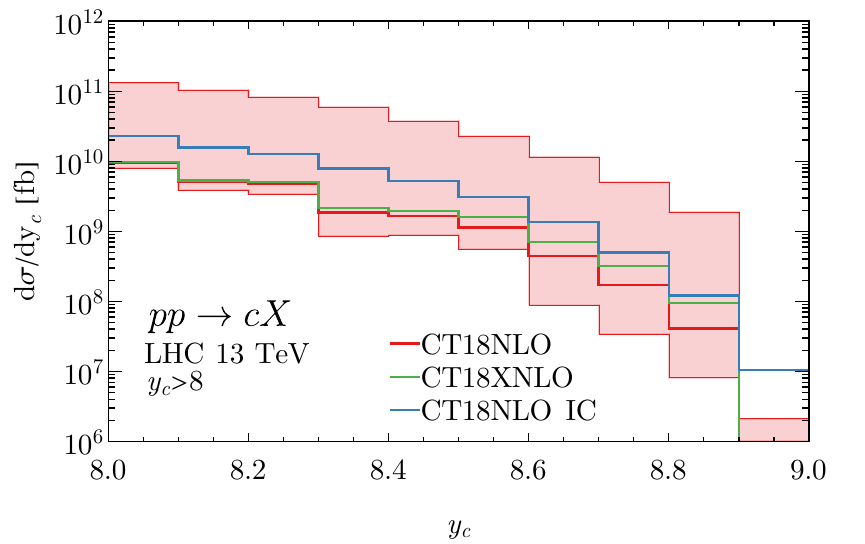}\\
\includegraphics[width=0.47\textwidth]{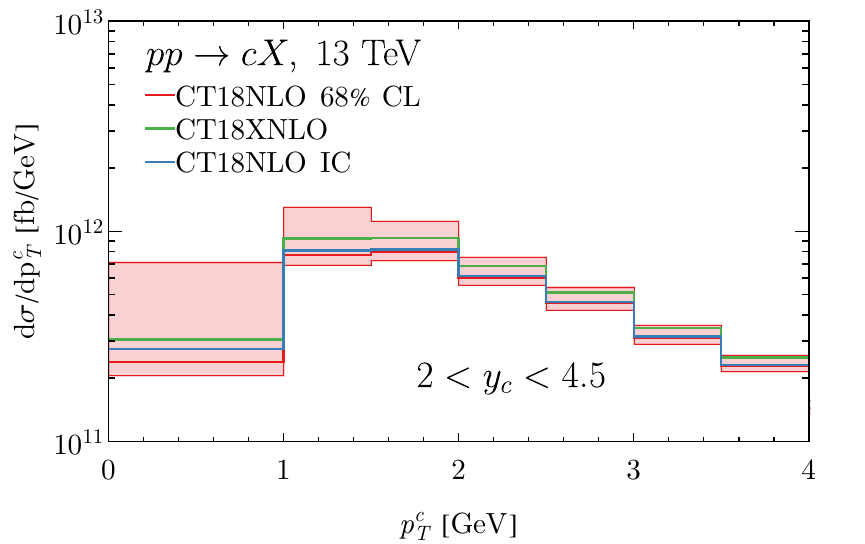}
\includegraphics[width=0.47\textwidth]{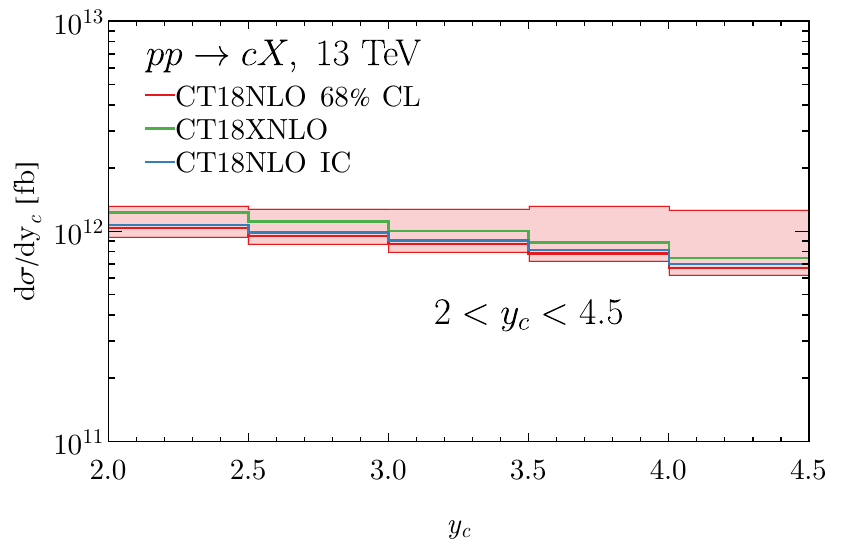}
\caption{\label{fig:charm} Transverse momentum and rapidity distributions of prompt charm produced in $pp$ collisions at the LHC for $\sqrt{s} = 13$ TeV in the very forward region ($y_c>8$) (upper panels) and at mid-rapidity ($2 < y_c < 4.5$) (lower panels). Central predictions refer to the CT18NLO, CT18XNLO and CT18NLO IC PDF sets. The red error band represents the CT18NLO-induced PDF uncertainty at 68\% C.L.}	
\end{figure}

\Figref{CT18PDFs} shows the standard CT18 NNLO PDFs~\cite{Hou:2019efy} that include only extrinsically produced charm, together with the alternative CT18``BHPS3'' PDFs~\cite{Hou:2017khm,Guzzi:2021fre} with the large-$x$ intrinsic sea PDFs included according to the model from Ref.~\cite{Blumlein:2015qcn}, as well as the CT18X PDFs~\cite{Hou:2019efy} determined with a scale choice in DIS that mimics the impact of low-$x$ resummation. The differences among the PDF models propagate into predictions for prompt charm production shown in Fig.~\ref{fig:charm} for the transverse momentum and rapidity distributions of charm quarks with rapidity $y_c>8$ at the NLO in QCD. The theory predictions are obtained in the recently developed S-ACOT general-mass factorization scheme with massive phase space (S-ACOT-MPS)~\cite{Xie:2019eoe}, an amended version of the S-ACOT scheme~\cite{Aivazis:1993kh,Aivazis:1993pi,Collins:1998rz,Kramer:2000hn,Tung:2001mv,Guzzi:2011ew} applied to the case of $pp$ collisions. CT18, CT18X, and CT18 IC NLO PDFs were used in this figure. The error bands indicate the 68\% C.L. PDF uncertainties. 

The differences among the predictions in Fig.~\ref{fig:charm} originate from the underlying gluon and charm PDFs in the relevant $x$ intervals in Fig.~\ref{fig:CT18PDFs}. On the one hand, the IC contribution substantially enhances the electron and tau neutrino interaction rates at the FPF by an amount that varies among the IC models~\cite{FASER:2019dxq,Bai:2020ukz}. On the other hand, the bulk of the PDF uncertainty for the charm kinematic distributions in Fig.~\ref{fig:charm} arises from the small-$x$ region, where the gluon PDF is hardly known and differs when comparing the CT18 and CT18X fits. The CT18X prediction gives a significantly larger cross section in the large $p_{T,c}$ and $y_c$ tails. The tails of the $p_{T,c}$ and $y_c$ distributions fall off rapidly in this very forward region. The IC prediction mainly enhances the overall rate. The CT18 PDF uncertainty covers all central predictions, except in the largest $y_c$ bin. 

The analysis of Figs.~\ref{fig:PDFs-FPF}-\ref{fig:charm} thus suggests that FPF experiments will be most successful as a part of a larger physics program that includes LHCb, the EIC, and possibly the LHeC~\cite{LHeC:2020van} to shed light on the appropriate QCD theoretical formalism(s) and constrain the PDFs in the currently unconstrained regimes of extreme $x$ values. NC $ep$ DIS at the EIC with tagged charmed final states is highly discriminating among the various IC models~\cite{GuzziIC2011,Hobbs:2017fom}. Forward and light-state production processes at the LHC probe the novel small-$x$ dynamics. A coordinated effort will in turn dramatically reduce the PDF uncertainty in the predictions for neutrino fluxes in FPF experiments. The FPF will allow us to better investigate the relevance of small-$x$ resummation, the transition from collinear to more general factorization frameworks, including the effects of initial-state partonic transverse momentum, and, as we will see in \secref{CCDIS}, CC DIS on heavy nuclei. 

\subsection{Neutrino-Induced Deep Inelastic Scattering \label{sec:CCDIS}}

The high-energy neutrino beam reaching the FPF can be used to measure CC DIS events upon interaction with FPF nuclear targets (liquid argon for LAr experiments and tungsten or other materials for emulsion experiments). These observations will constrain the nuclear PDFs for the target. Using different nuclear targets would probe the nuclear modification effects as a function of the nucleus mass number, which would be valuable, given that the scattering measurements for determination of the nuclear PDFs remain limited to a restricted set of mass numbers. Analogous information from the previous neutrino-induced DIS measurements, such as CHORUS and NuTeV, plays a prominent role in many global PDF fits of nucleon and nuclear PDFs (with the two related via nuclear corrections). The reason is that inclusive CC DIS and especially semi-inclusive charm production in CC DIS are the primary channels to probe the PDFs for strange quarks and anti-quarks. Strangeness PDFs offer insights about the nonperturbative proton structure~\cite{Chang:2014jba}, while they are also responsible for a large part of the PDF uncertainty in weak boson mass measurements at the LHC~\cite{Nadolsky:2008zw}. On the experimental side, determination of the (anti-)strangeness PDF has been one of the hottest topics for the PDF community.  This is because, depending on whether one uses lepton-hadron or hadron-hadron data, or whether one uses results from emulsion experiments or those using calorimetric techniques, the fits prefer somewhat different shapes for the strangeness PDFs for reasons that are not fully understood~\cite{Alekhin:2014sya, Alekhin:2017olj,Hou:2019efy,Faura:2020oom}. The elevated PDF uncertainty from fitting such inconsistent experiments propagates into various pQCD predictions (see e.g. those recently presented in Ref.~\cite{Bevilacqua:2021ovq}). 

Dimuon production in CC DIS with various (anti)neutrino flavors thus offers a window on the differences between the PDFs for $s$, $\bar s$, $\bar u$, and $\bar d$ (anti)quarks. The right panel of Fig.~\ref{fig:PDFs-FPF} shows an example of the role played by neutrino-nucleus DIS data in global fits. The figure compares the down antiquark PDF from the NNPDF3.1 global proton PDF fit~\cite{Ball:2017nwa}, which includes, among others, neutrino DIS data obtained with deuteron and heavy nuclear targets, with two fits from which the neutrino DIS data have been removed: one without any fixed-target DIS data, and another where only data measured on proton targets are retained. One sees that the PDF uncertainties on $\bar{d}(x,Q)$ increase significantly for $x\ge 0.05$  if neutrino DIS data are excluded. Analogous considerations would apply to the nuclear PDF case. Hence an important benefit of the FPF would be to provide measurements analogous to existing neutrino DIS experiments, but now at a higher energy, where measurements have fewer theoretical uncertainties.  Experiments at the FPF, being able to measure charm production in DIS, while distinguishing neutrino and antineutrino induced events, will help to solve the strangeness puzzle introduced by the tensions between the already existing data mentioned above. A peculiar aspect will be the possibility of using different techniques for charm tagging, considering the fact that at the FPF not only experiments allowing to tag charm through dimuon events, but even emulsion experiments, allowing to tag various kinds of charmed mesons and baryons by reconstructing in detail the topology of their decays, will be present. In this respect, emulsion experiments are expected to be particularly powerful. 

As the incoming neutrino beam will be quite broad, one must measure observables in which uncertainties associated with the incoming flux partially cancel out, such as in the ratio between charm-production and inclusive events. Also, it has not been excluded that the nuclear medium effects on the nuclear PDFs are different at some level in NC and CC DIS~\cite{Kovarik:2010uv,Paukkunen:2013grz,Paukkunen:2014nqa}, and that final-state nuclear medium effects may affect the extraction of PDFs from DIS events~\cite{Accardi:2009qv,Accardi:2015lcq}.
  
\subsection{Single-inclusive Forward and Forward-Central Events at the FPF + ATLAS \label{sec:HadronFPF}}

In this subsection we consider the detection of far-forward hadrons at the FPF using the time-coincidence method, which offers an unprecedented opportunity for deeper tests of QCD in the high-energy regime, opening the road to the possibility that the FPF might complement the reach of the ATLAS detector. 

The possibility of combining information from ATLAS and a forward detector at the FPF relies on the ability to use an FPF event to trigger ATLAS.  This requires very precise timing and has consequences for the design of the forward detector. The forward detector should at least be capable of separating the various bunch crossings, which implies  that, at the very least, one needs a resolution better than 25 ns if one wants to associate the forward signal in the FPF with a particular bunch crossing in ATLAS. For a detector located at $\sim$ 620 meters from the IP, taking into account a trigger latency of $\sim$ 10 $\mu$s for the ATLAS Level-0 system and the time needed for the neutrino to reach an FPF detector and for the trigger signal (traveling in air-core cables with speed $\beta$ of about 0.8) to reach the Central Trigger Processor, the trigger decision should be taken within 5-6 $\mu$s to be able to be used by the ATLAS Level-1 trigger system. 

An additional issue is pileup. Under the extreme values of the pileup parameter $\langle \mu \rangle \sim 150-200$ expected at the HL-LHC, related to the average number of $pp$ collisions in a bunch crossing, events with multiple hard scattering processes within the same bunch crossing will be more common than in previous Runs. With very precise timing measurements, it would be possible not only to assign the forward signal to a specific bunch crossing, but also to a subset of the luminous region. Since a time resolution of 100~ps, quite challenging to obtain by a large detector, corresponds to a position resolution of about 3 cm, a similar precision would allow one to identify the area of the luminous region from which the particles detected in the forward detector are coming, and thereby reduce the background due to the pileup. The effect of pileup can indeed be mitigated by requiring events with very hard objects, which would in turn allow one to work with coarser time resolutions. In any case, considering the present status of developments of timing techniques in association with LAr detectors, time resolutions of $\sim$ 1 ns are within reach. On the other hand, obtaining time resolutions of $\sim$~100~ps will require some R\&D, which will also be useful in view of possible applications to other experiments (see, e.g., the ICARUS case~\cite{ICARUS:2020wmd}).  

A first class of reactions that can be investigated at the FPF includes \emph{single-inclusive forward} emissions, where a neutrino with rapidity $y \gtrsim 6.5$ is identified. Both inclusive and exclusive processes can be measured and predicted using the hybrid small-$x$ formalism discussed in Sec.~\ref{sec:High-s}. Tests of this formalism and the UGD evolution can be done at the FPF by considering the emission of several kinds of final states, such as charged light hadrons, vector mesons (extensively studied at HERA~\cite{Anikin:2011sa,Besse:2013muy,Bolognino:2018rhb,Celiberto:2019slj,Bolognino:2019pba,Bautista:2016xnp,Garcia:2019tne,Hentschinski:2020yfm}), and mesons with open charm/beauty, accompanied by their decay producing at least a forward neutrino. This kind of study can be performed by the FPF detectors alone.  Requiring coincidence with ATLAS may allow identification of states with large invariant masses, whose decay products are not entirely captured by the FPF, but fall partly into the FPF's and partly into ATLAS's coverage areas. Additionally, combining the data on \emph{single-inclusive forward} emissions at ATLAS and at the FPF will allow one to investigate in which rapidities and kinematic configurations the aforementioned hybrid formalism 
provides a better physics description than
the standard collinear formalism for single-inclusive particle production and in which ones, instead, it does not, with the long-term aim of filling the gap between the two descriptions. 

\begin{figure}
\includegraphics[width=0.30\textwidth]{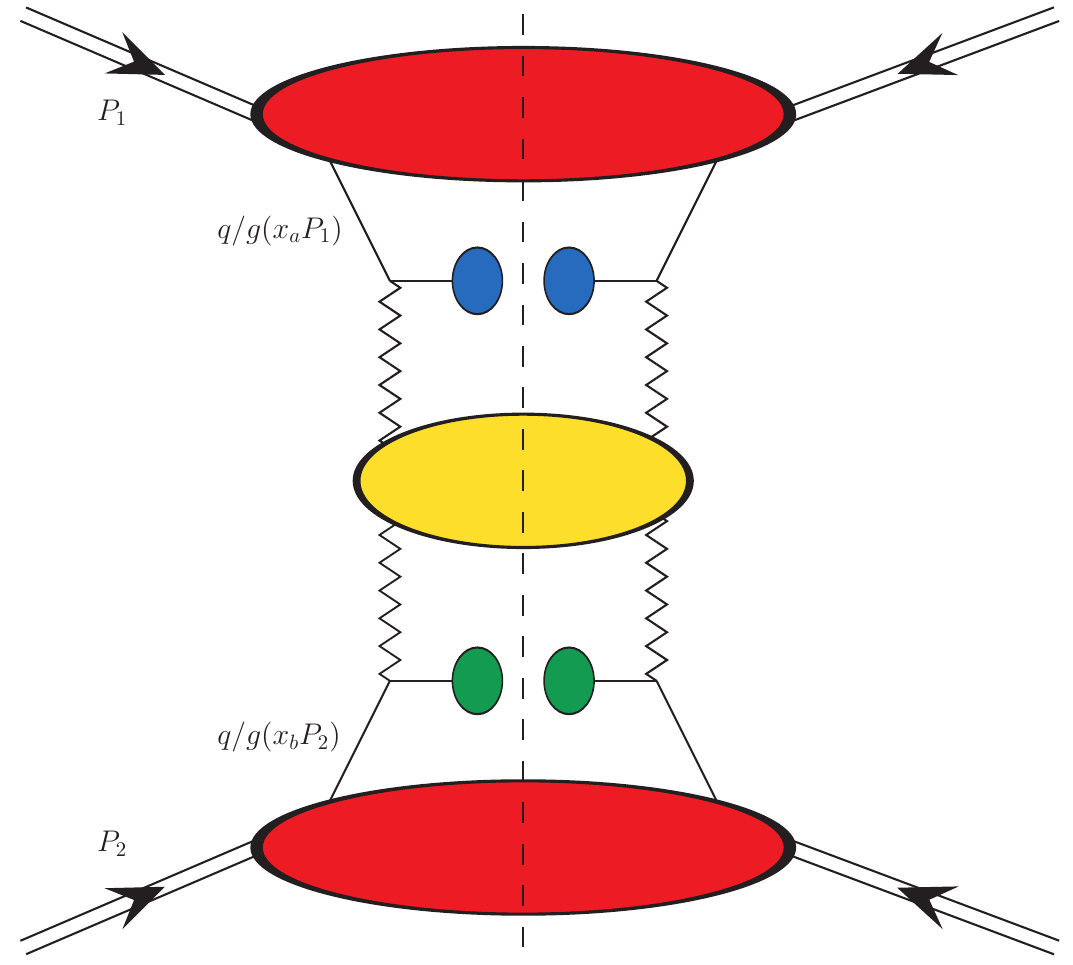}
\hfil
\includegraphics[width=0.43\textwidth]{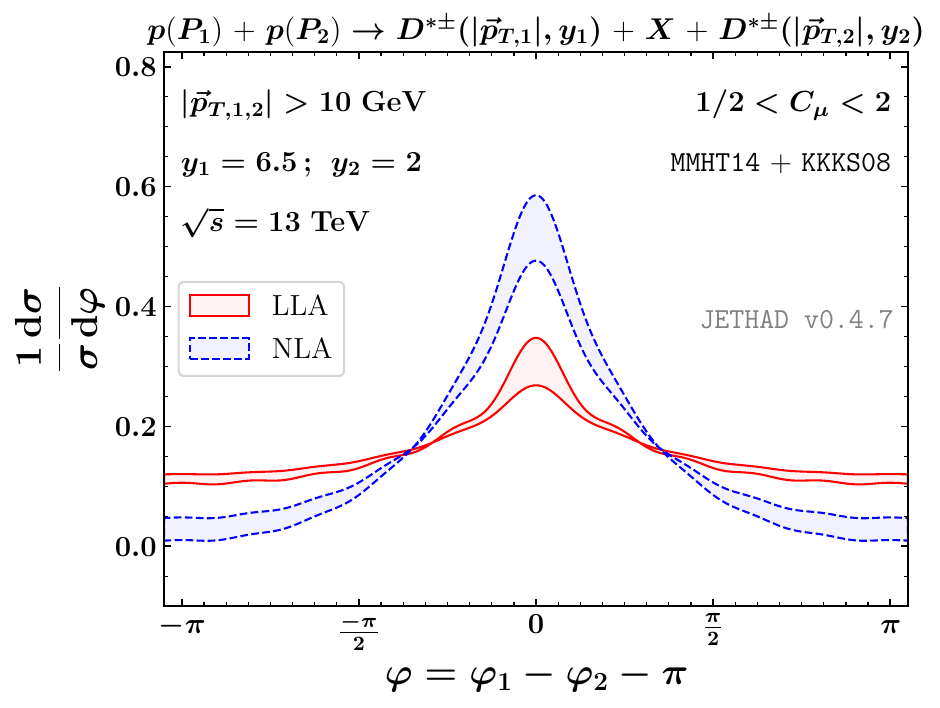}  
\caption{\label{figforward}
Left: A cross section for a forward-central hadroproduction process leading to two hadrons with large rapidity separation in the hybrid formalism. Right: Predictions of the hybrid formalism for the normalized distribution of the azimuthal angle $\varphi=\varphi_1-\varphi_2-\pi$ in the forward-central production process $pp \rightarrow$ $D^*$($y_1$) + $D^{*}$($y_2$) + $X$ with $y_1 = 6.5 $, $y_2 = 2$, and $p_{T,1},\ p_{T,2} > 10$ GeV, as testable by the FPF + ATLAS detectors in a tight timing coincidence setup. The constraints on time coincidence might be made less severe by increasing the $p_T$ cuts. The predictions are based on the resummation of BFKL logarithms in the leading logarithmic approximation (LLA) and NLA. Uncertainty bands refer to the $\mu_R$ = $\mu_F$ = $\mu_0$ = $\sqrt{p_{T,1}^2 + m_1^2} + \sqrt{p_{T,2}^2 + m_2^2}$ scale variation in the interval [1/2, 2]$\mu_0$. } 
\end{figure}
    
Probed production channels of the second class feature two identified final-state objects emitted in hard scattering and separated by a rapidity interval $\Delta y$ larger than about 2. These processes are predicted within the hybrid formalism using two partonic impact factors, convoluted with the BFKL gluon Green's function that embodies the resummation of energy logarithms in the $t$-channel. The result is then convoluted with initial-state collinear PDFs and, in the case of hadron production and identification, final-state collinear FFs (see Fig.~\ref{figforward}, left panel). 

Here, we expect to see a stabilization of the BFKL series with respect to scale variations upon inclusion of subleading logarithms, as was already observed when comparing NLL calculations with the LL ones in CMS configurations for Higgs-plus-jet~\cite{Celiberto:2020tmb}, heavy-light dijet~\cite{Bolognino:2021mrc}, and $\Lambda_c$ baryon~\cite{Celiberto:2021dzy} emissions. The ATLAS-FPF coincidence observations would allow one to explore large rapidity intervals, e.g., $\Delta y \sim 5$, with the most forward of the two emitted objects detected by the FPF, and the most central one by ATLAS. We call these channels {\it forward-central} production channels. We expect the BFKL effects to be enhanced with increasing $\Delta y$. In this setup, we can even study production channels where the most centrally detected object is a jet. Jets in fact are not visible at the FPF, but can easily be reconstructed in ATLAS. An example is shown in the right panel of Fig.~\ref{figforward}, where a $pp$ collision produces two $D^*$ mesons at $p_T$ values large enough to ensure the dominance of perturbative effects in the production process, with a $\Delta y$ value that can be explored in the FPF + ATLAS coincidence setup. The panel shows a comparison between predictions in the LLA and NLA. The fact that the NLA predicts a stronger correlation in the azimuthal plane between the two emitted identified objects than the LL approximation (the peak at $\varphi = 0$ corresponds to back-to-back emissions in the azimuthal plan) is expected and gives a strong hint of the relevance of BFKL dynamics in the considered kinematic domain. The differences between LLA and NLA at the peak would be enhanced in case of asymmetric $p_T$ cuts. Studies in this directions are underway. The practical feasibility of this kind of study will depend on the timing resolution that can be achieved. Employing harder $p_T$ cuts might be helpful to reduce pileup, which, in turn, would loosen the timing resolution requirements. The considered example confirms the importance of complementing the information of the FPF with that of ATLAS for QCD-related studies. In general, many other studies of the associated production of multiple objects will benefit from a FPF + ATLAS coincidence setup. 

\subsection{Forward Physics in Event Generators}
\label{sec:MC}

Event generators are standard tools for emulating complete events in as much detail as possible. The main components include hard and soft processes, parton showers, multiparton interactions (MPIs), hadronization, and decays. Generators can be used to predict, e.g., the inclusive neutrino flux as a function of energy and angle (see \secref{neutrinos}). However, their main strength is that they can be used to explore nontrivial correlations, say between forward and central activity.

The general-purpose LHC $pp$ physics event generators, \textsc{Herwig}~\cite{Bellm:2019zci}, \textsc{Pythia}~\cite{Sjostrand:2014zea}, and \textsc{Sherpa}~\cite{Sherpa:2019gpd}, are primarily intended and tested for the central region, say, in the rapidity region $|y| \le 5$. QCD-centered generators more common in cosmic-ray studies, such as \textsc{Sibyll}~\cite{Riehn:2019jet}, QGSJET~\cite{Ostapchenko:2019few}, DMPJET~\cite{Roesler:2000he}, and EPOS~\cite{Pierog:2019opp}, generally are better tuned in the forward region. Nevertheless there is not one that provides a good overall description of all existing data, e.g., of the LHCf neutron and $\pi^0$ spectra~\cite{LHCf:2015nel,LHCf:2015rcj}. At present, these spectra may be considered as representatives of baryon and meson forward spectra more generally, for lack of alternatives. The FPF, on the other hand, will provide information on charged meson and baryon spectra, complementing the one on neutral hadrons provided by LHCf. 

A common feature of all generators is that MPIs may occur, mainly in the central region. This means that several partons are taken out of the colliding hadrons, leaving behind a beam remnant. This remnant usually is color-connected to the activity in the rest of the event (if not, we have a diffractive topology). Each such color line can be associated with a color-confinement string that eventually will break up to produce the primary hadrons. To the first approximation, there will be one string stretched to the remnant for each (anti)quark taken out, and two for each gluon. If the remnant momentum is evenly shared between all string ends, then the hadron spectra will drop too steeply at large momenta. All generators therefore use some procedure to increase the likelihood that a single hadron can take an appreciable fraction of the remnant's momentum. 

One example is \textsc{Herwig}, where an enforced parton-shower-like backwards evolution is used to reconstruct, e.g., a quark entering an MPI as coming from a gluon that in its turn is emitted from the color line of the hardest MPI. This MPI is evolved back to come from a single valence quark, meaning that the proton beam remnant always is a single diquark, but, of course, the energy of that diquark can vary widely by the steps taken. Another example is \textsc{Sibyll}, where all non-hardest MPIs are assumed to give simple closed gluon pairs disconnected from the remnants. The latter therefore consist of at most two objects each, say a quark and a diquark. These share the energy unevenly, with the diquark taking most. Additionally the first string break of the diquark uses a special extra hard FF to ensure a hard leading baryon.

In \textsc{Pythia} more complicated beam remnants are allowed, such that the momentum can be split more ways, which would give too-soft leading-baryon spectra. Several steps are introduced to improve the situation. A color line from a remnant to one MPI may be identified to an anticolor line to another MPI, which means that this linked line can be removed from the remnant. This procedure takes care of all gluons and all same-flavor quark--antiquark pairs. In the case of multi-parton remnants, it is also possible to split off color-singlet hadrons, e.g., a four-quark remnant could be split into a baryon plus a single quark. Another complication arises when two valence quarks are kicked out into separate MPIs, which means that the baryon number (represented by a junction) drifts into the central region of the event. In most events one still is left with a diquark in a proton remnant, possibly in association with a quark or meson. Even though the diquark is made here to take the major part of the remnant's momentum, the resulting baryon spectra still are softer than data. 

\begin{figure}[tbp]
\includegraphics[width=0.325\linewidth]{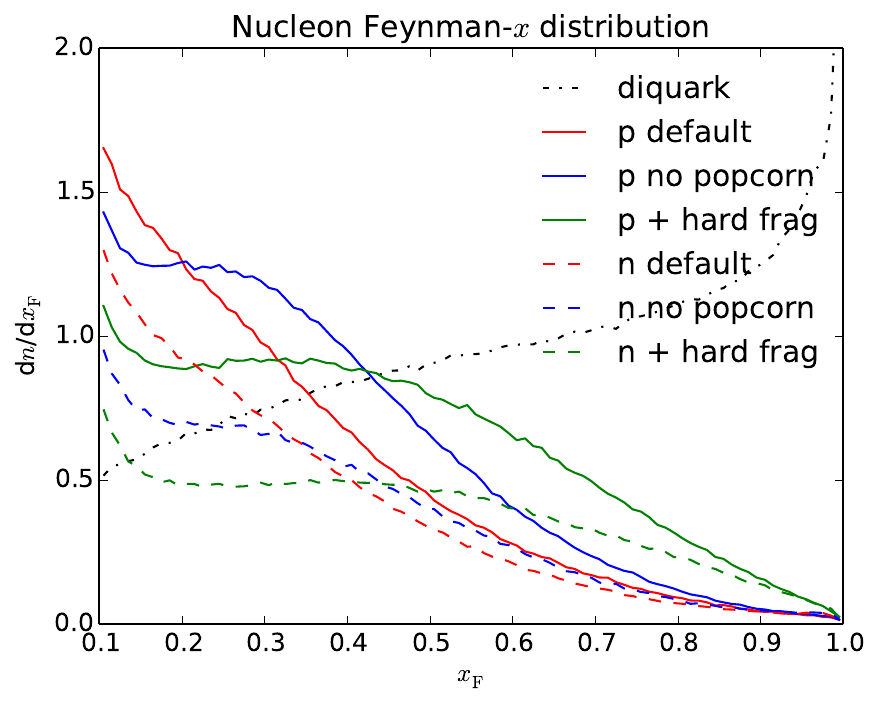}
\includegraphics[width=0.325\linewidth]{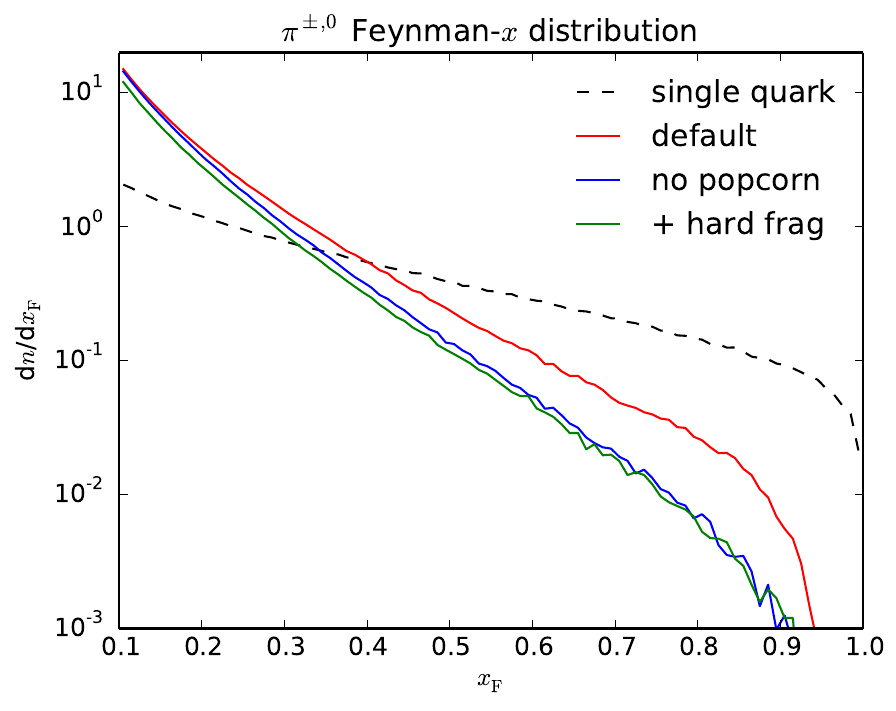}
\includegraphics[width=0.325\linewidth]{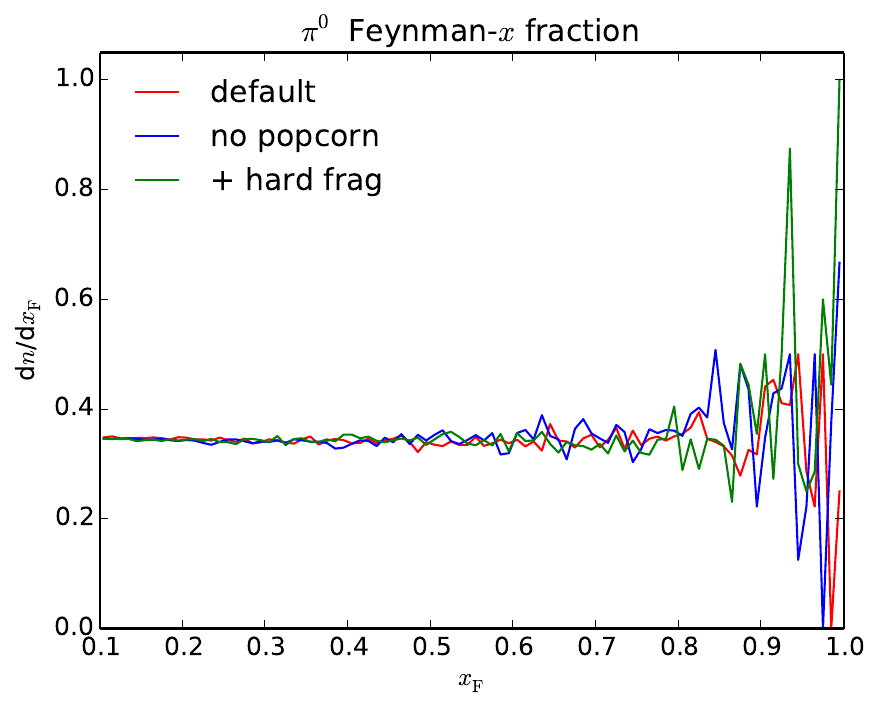}
\caption{\label{fig:xFForward} Nucleon ($p$, $n$) (left) and pion ($\pi^{\pm,0}$) 
$x_{\mathrm{F}} = 2|p_z|/E_{\mathrm{CM}}$ (center) distributions in $\sqrt{s} =$ 14~TeV $pp$ collisions, for options as presented in the text. Also shown are the spectra of remnant diquark and quark string ends. The ratio of $\pi^0$ to ($\pi^0 + \pi^+ + \pi^-$) $x_{\mathrm{F}}$ distributions is also shown (right) and leads to an approximately constant value $\sim 0.35$.}       
\end{figure}

Recently a first attempt has been made to remedy the situation~\cite{Sjostrand:2021dal}, with two new options. Firstly, the popcorn mechanism~\cite{Andersson:1984af} can be switched off for the remnant diquark. Thereby one reverts to the simpler diquark picture, where a diquark is handled as a stable unit, and the leading particle (in a flavour sense, but usually also in energy) becomes a baryon. This is in contrast to the default popcorn scenario, where  a quark--antiquark pair may be created in between the two quarks of a diquark, breaking it up. When that happens, the leading particle becomes a meson, and the baryon produced next has a lower average energy than otherwise. And secondly, a separate harder FF may be tuned for the remnant diquark, similarly to \textsc{Sibyll}. These two have about equal importance for the harder nucleons result shown in Fig.~\ref{fig:xFForward}, which are better in line with what is needed to describe the LHCf neutron data. The remnant diquark $x_{\mathrm{F}}$ spectrum shows that an even harder FF would have been possible. The pions take less energy when the neutrons take more, which also improves agreement with the LHCf $\pi^0$ spectrum, though with remaining problems. Here fragmentation of both quark and diquark string ends plays a role.

It should be noted that also the simulation of transverse momentum effects is relevant. A smaller $p_{\perp}$ spread leads to a larger number of hadrons in the forward direction. This affects the rate within the LHCf acceptance and also the FPF event rates. There are three contributing main sources: primordial $k_{\perp}$ of the MPI systems that are compensated by the remnants, relative $p_{\perp}$ kicks between the remnant constituents, and regular fragmentation $p_{\perp}$ when the strings break. 

The forward neutrino flux comes from a combination of hadron production and their subsequent decays. For lighter particles, such as $\pi^{\pm}$, $K^{\pm}$, $\Lambda/\overline{\Lambda}$, $\mu^{\pm}$, and $\tau^{\pm}$, the handling of weak decays with proper $V-A$ matrix elements offers only minor problems. Also the bulk of charm and bottom weak decays are understood, where the \textsc{EvtGen} package~\cite{Lange:2001uf} is often used as a plugin to better handle bottom decays, but some complex final states are only crudely modeled.

Charm and bottom quarks can be produced in hard processes or in initial- or final-state parton showers, both in the primary hard interaction and in subsequent MPIs. The overall production rate is quite sensitive to a number of choices, such as that of quark masses or parton distributions. Some element of tuning will therefore always be necessary to describe the data. The $c/b$ quark is always at one end of a string, and for the hadronization step it is relevant where the other end of that string is. If that end is in a beam remnant, then the $c/b$ quark is pulled forward by the string tension, and the resulting hadron typically will have a larger longitudinal momentum than the mother quark. In extreme cases, a string may even be so short that it collapses into a single hadron, like a $\Lambda_{\mathrm{c}}^+$ if a $c$ combines with a $ud$ remnant diquark, and a $\overline{\mathrm{D}}^0$ if a $\bar{c}$ combines with a $u$ remnant. Note that quarks and antiquarks may be pulled in different directions and collapse into different hadrons, giving rise to small particle--antiparticle asymmetries. Such asymmetries have been observed and are well modeled at low energies~\cite{Norrbin:2000zc}, and they are now also observed by LHCb, as already noted in \secref{ForwardCharmHybrid}.  This is an example where a full event generator can be more predictive than the semi-analytical FF approach. The latter typically parametrizes $c/b$ FFs based on LEP/SLC data, where the $c/b$ quark is always pulled backwards by the string, and so is not well equipped to address the more complicated color topologies of hadronic collisions. String effects also need to be taken into account before intrinsic charm is introduced as a way to increase forward charm production.

Event generators address a multitude of physics aspects, many of a non-perturbative nature where theory currently has little to say. This means that parameters have to be introduced and tuned to data. Typically this is a two-step process, where final-state showers and hadronization are tuned to LEP data, and then initial-state radiation, MPIs, beam remnants, and more are tuned to hadron collider data. An example is the \textsc{Pythia} Monash tune~\cite{Skands:2014pea}, where more than 50 parameters are explored. Subsequent tunes by the LHC collaborations work with a smaller subset, but they have the advantage that the collaborations have better access to and understanding of their own data. 
We expect that data extracted from the FPF will have an impact on future tunes, complementing those of the LHC experiments. This might be relevant not only for generators for LHC physics, but even for those for high-energy cosmic-ray physics.

In summary, progress is being made in the modeling of the forward region and the tuning to the limited forward data, but more remains to be done. One should add that generators can also be used to simulate subsequent neutrino interactions with matter. This is a capability that already exists, but where work is ongoing to improve the modeling, and nuclear effects could be nontrivial. Standard detector simulation should also not be forgotten, where bits and pieces of \textsc{Pythia} are already used inside various generators commonly adopted at this purpose.

\subsection{QCD Summary}
\label{sec:QCDSummary}

In this \secref{qcd}, we have explored the promise of the FPF for probing the strong interactions and the structure of the proton and nuclei.  As discussed above, the FPF will be sensitive to the very forward production of light hadrons and charmed mesons, providing access to both the very low-$x$ and the very high-$x$ regions of the colliding protons. The low-$x$ regime will shed light on many important topics in QCD, such as BFKL effects and non-linear dynamics, as well as the gluon PDF down to the ultra-low values of $x\sim 10^{-7}$, extending the coverage of other experiments. The very high-$x$ regime provides access to other open questions, including, for example, the possibility of intrinsic charm. In addition, the FPF acts as a neutrino-induced DIS experiment with TeV-scale neutrino beams. The resulting measurements of neutrino DIS structure functions will provide valuable constraints on the partonic structure of nucleons and nuclei. Finally, the FPF will also provide additional data to refine hadron production in event generators in the forward region, where they are currently relatively poorly constrained.  

In summary, QCD provides an important physics motivation for the FPF, supplementing the motivations from new physics searches and neutrino physics discussed above in \cref{sec:bsm,sec:neutrinos}, respectively.  In addition to the intrinsic interest in the strong interactions and proton and nuclear structure, the QCD studies described here will also have important implications, for example, for astroparticle physics, to which we now turn in \cref{sec:astro}.

\section{Astroparticle Physics}
\label{sec:astro}




Historically, cosmic rays and cosmic neutrinos have contributed greatly to high-energy physics, from the landmark identification of new elementary particles in the early days, to the confirmation of long-suspected neutrino oscillations, to measuring cross sections and accessing particle interactions far above current collider energies. Two recent examples that illustrate the astroparticle $\leftrightharpoons$ high-energy physics connection are (i)~the measurement of the $pp$ cross section at a center-of-mass energy of $\sqrt{s} \sim 75~{\rm TeV}$~\cite{Collaboration:2012wt, Abbasi:2015fdr, Abbasi:2020chd}, which provides evidence that the proton behaves as a black disk at asymptotically high energies~\cite{Block:2012nj, Block:2015mjw}, and (ii)~the measurements of both the CC neutrino-nucleon cross section~\cite{Aartsen:2017kpd, IceCube:2020rnc, Bustamante:2017xuy} and the NC to CC cross section ratio~\cite{Anchordoqui:2019ufu} at $\sqrt{s} \sim 1~{\rm TeV}$, which provide restrictive constraints on fundamental physics at sub-fermi distances. In this section, we will explore the synergistic links between astroparticle physics and the FPF. 

\subsection{Cosmic Ray Physics and the Muon Puzzle}

\begin{figure}[tb]
\centering
\includegraphics[width=0.485\textwidth]{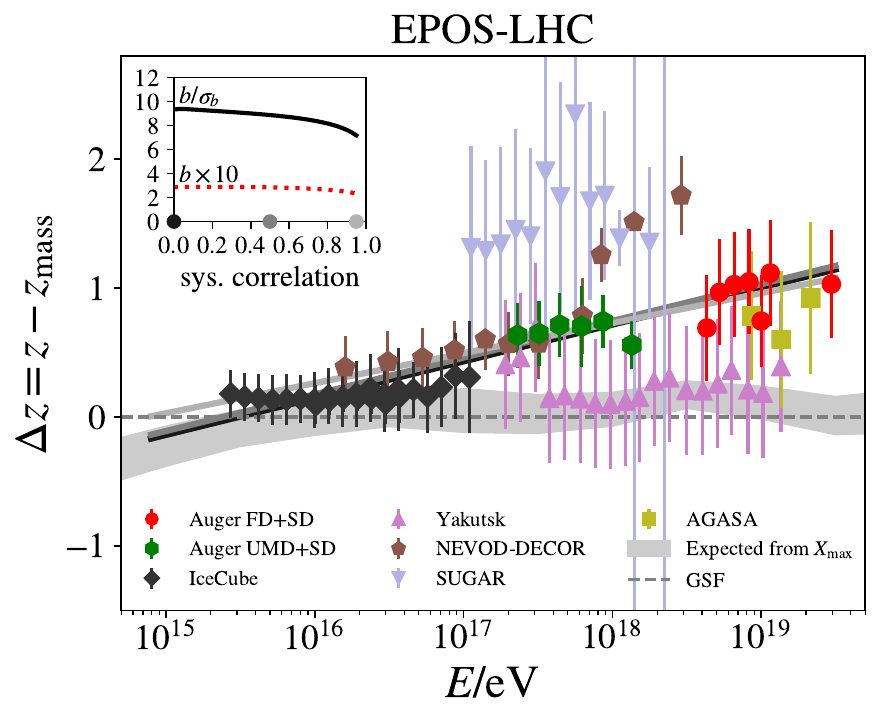}
\includegraphics[width=0.485\textwidth]{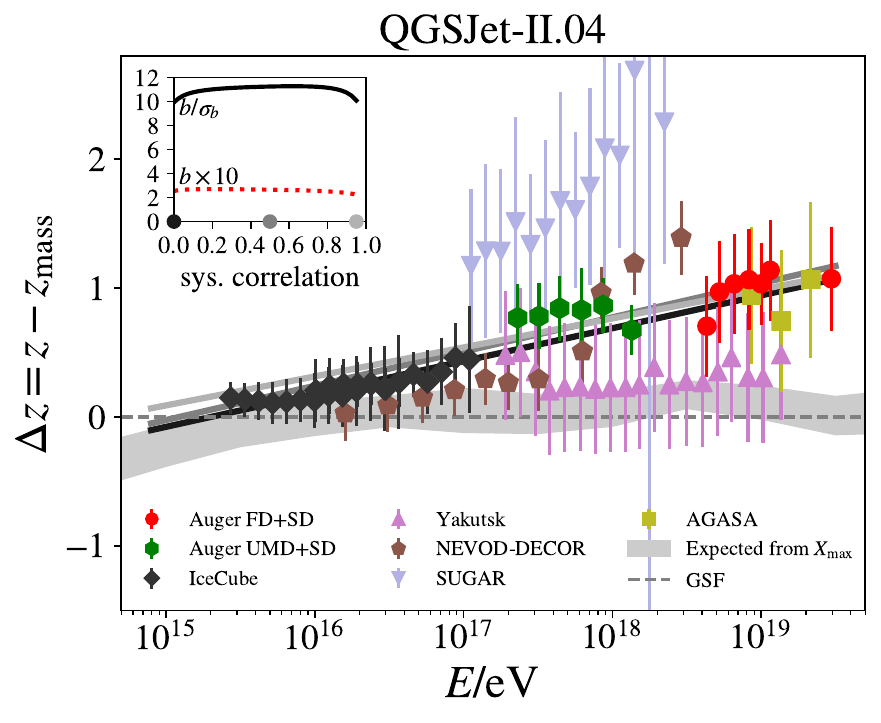}
\vspace{-0.5cm}
\caption{Linear fits to the $\Delta z=z-z_\mathrm{mass}$ distributions as a function of air shower energy from Ref.~\cite{Soldin:2021WHISP}, where $z_\mathrm{mass}$ is the number of muons predicted by a hadronic interaction model, here \textsc{EPOS-LHC} (left) and  \textsc{QGSJet-II.04} (right), assuming a mass composition of the primaries based on experimental parametrization from Ref.~\cite{Dembinski:2019uta} (global spline fit). The quantity $\Delta z$ measures the difference between the experimental data and the inferred number of muons for a given hadronic model. A positive value indicates an excess of muons in data with respect to simulations, and zero indicates a perfect match. Shown in the inset are the slope $b$ and its deviation from zero in standard deviations for an assumed correlation of the point-wise uncertainties within each experiment. Examples of the fits are shown for correlations of $0.0$, $0.5$, and $0.95$ in varying shades of gray.}
\label{fig:muon_puzzle}
\end{figure}

Cosmic rays have been measured in the Earth's atmosphere with energies exceeding $10^{11}~{\rm GeV}$, but their sources remain unclear, their acceleration mechanisms and nuclear composition are uncertain, and several features observed in the energy spectrum are not well understood~\cite{Blumer:2009jrd, Anchordoqui:2018qom, AlvesBatista:2019tlv}. Observations of cosmic rays with energies exceeding about $10^6~{\rm GeV}$ rely on indirect measurements of air showers.  To infer the energy and mass of cosmic rays from observable air shower features, for example, one has to quantitatively model the shower based on known particle physics~\cite{Kampert:2012mx}. Simulations reasonably reproduce many air shower features, but there is a longstanding deficit in the number of muons produced in air showers, which was first observed by the HiRes-MIA experiment more than 20 years ago~\cite{AbuZayyad:1999xa}. Since then, both simulations and experiments have made enormous progress, but the Muon Puzzle persists~\cite{Albrecht:2021yla}. The most unambiguous experimental evidence of the deficit was revealed in the analysis of Auger data~\cite{PierreAuger:2014ucz, PierreAuger:2016nfk}. A meta-analysis~\cite{Dembinski:2019uta, Cazon:2020zhx, Soldin:2021WHISP} of recent muon measurements from several experiments is shown in \cref{fig:muon_puzzle}. The $z$-scale is used to make different muon measurements in air showers comparable. It is approximately independent of the experimental details, but depends on the hadronic interaction model used in air shower simulations~\cite{Dembinski:2019uta}. After subtracting the expected variation due to the cosmic ray mass composition, $z_\text{mass}$, an upward trend remains, which starts at moderate center-of-mass energies of about 10\,TeV, accessible by the LHC, followed by a linear increase with the logarithm of the shower energy.

The muons seen by air shower experiments are of low energy (a few to tens of GeV). They are produced at the end of a cascade of hadronic interactions with up to 10 steps, where the dominant process is soft forward hadron production, which cannot be calculated from first principles in perturbative QCD. Effective theories are used to describe these interactions, in particular Gribov-Regge field theory. Detailed simulations~\cite{Ulrich:2010rg, Baur:2019cpv} have shown that the hadron multiplicity and, in particular, the hadron species at forward pseudorapidities of $\eta \gg 2$ have the largest impact on muon production in air showers. The sensitivity to the produced hadrons is high, and even small deviations of 5\% in the multiplicity and/or identity of the secondary hadrons have a sizeable impact on the muon production.

Proposed models that account for such deviations are based on the restoration of chiral symmetry~\cite{Farrar:2013sfa}, the production of fireballs~\cite{Anchordoqui:2016oxy}, a core-corona effect~\cite{Baur:2019cpv}, and a quark-gluon plasma~\cite{Pierog:2020ghc, Anchordoqui:2019laz}. These models have in common that the neutral particle production is suppressed with respect to the effective theories encapsulated in the current post-LHC hadronic interaction models (for example, \textsc{EPOS-LHC}~\cite{Pierog:2013ria}, \textsc{QGSJet-II.04}~\cite{Ostapchenko:2013pia}, \textsc{SIBYLL-2.3c/d}~\cite{Riehn:2017mfm, Riehn:2019jet}, and \textsc{DPMJet-III.2017}~\cite{Roesler:2000he, Cerutti:2015lcn}). This indirectly enhances the muon content at ground without altering the remainder of the shower development. Regardless of the details of the model, generally two extremes can be distinguished: a rather strong suppression occurring in the first few interactions of the air shower---reflecting some kind of threshold effect of exotic physics---or a small suppression over a large range of energies where the effect on the muon content accumulates throughout the shower development. The fit shown in \cref{fig:muon_puzzle} seems to favor the latter, as $\Delta z$ is continuously increasing with shower energy. A measurement of shower-to-shower fluctuations of the muon content~\cite{PierreAuger:2021qsd} further motivates the accumulation scenario, which, in turn, requires an effect to be visible at FPF energies.

\begin{figure}[tb]
\includegraphics[width=0.9\textwidth]{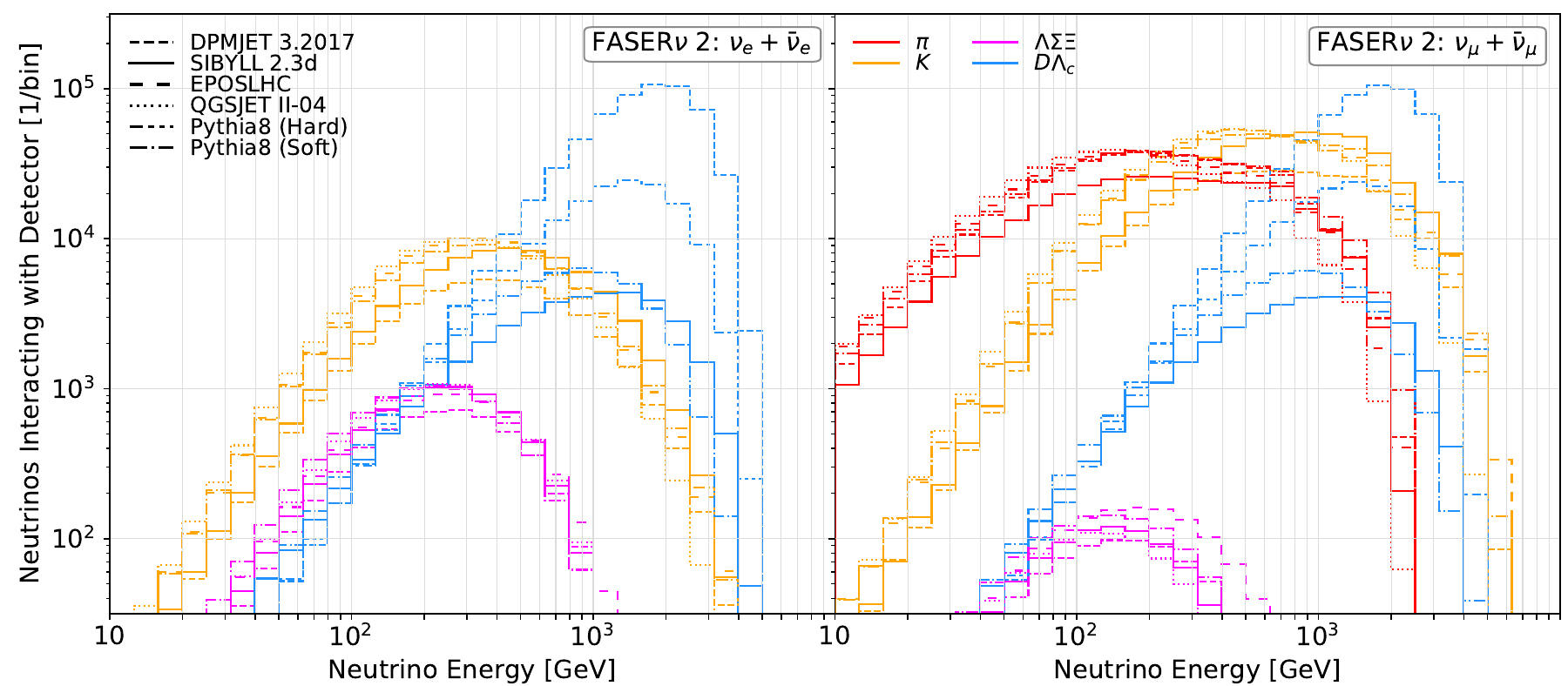}
\caption{Neutrino energy spectra for electron neutrinos (left) and muon neutrinos (right) passing through FASER$\nu$2. The vertical axis shows the number of neutrinos per energy bin that go through the detector's cross-sectional area for an integrated luminosity of $3~\iab$. The different production modes are indicated by different colors: pion decays (red), kaon decays (orange), hyperon decays (magenta), and charm decays (blue). The different line styles correspond to predictions obtained from \textsc{SIBYLL-2.3d} (solid), \textsc{DPMJET-III.2017} (short dashed), \textsc{EPOS-LHC} (long dashed), \textsc{QGSJet-II.04} (dotted), and Pythia 8.2 using soft-QCD processes (dot-dashed) and with hard-QCD processes for charm production (double-dot-dashed). Note that the predictions differ by up to a factor 2 for neutrinos from pion and kaon decays, which is much bigger than the anticipated statistical uncertainties at the FPF~\cite{Kling:2021gos}.}
\label{nu-rate}
\end{figure}

The amount of forward strangeness production seems of particular relevance~\cite{ALICE:2016fzo}. It is traced by the ratio of charged kaons to pions, for which the ratio of electron and muon neutrino fluxes is a proxy that will be measured by the FPF~\cite{Kling:2021gos}. While pions primarily decay into muon neutrinos, kaon decays contribute to both the electron and muon neutrino fluxes. In addition, as shown in \cref{nu-rate}, $\nu_e$ and $\nu_\mu$ from different sources populate different energy regions, which can be used to disentangle them. Furthermore, neutrinos from pion decay are more concentrated around the LOS than those of kaon origin, given that $m_\pi < m_K$, and thus neutrinos from pions obtain less additional transverse momentum than those from kaon decays. Thereby, the closeness of the neutrinos to the LOS, or equivalently their rapidity distribution, can be used to disentangle different neutrino origins and estimate the pion to kaon ratio. If technically feasible, a correlation of the FPF measurements with activity in ATLAS could also be interesting, as it would allow one to study the dependence of the charge ratio on the charged particle multiplicity as a function of rapidity in an extended rapidity range~\cite{CMS:2019kap}.  

In addition to neutrinos, it might also be possible to use the large number of forward-going muons that pass through the FPF experiments to constrain forward particle production. Based on {\em in situ} measurements, the muon flux at FASER is estimated to be approximately $1~\hz/\cm^2$~\cite{Ariga:2018pin}. This implies that about $2 \times 10^9$ muons will be detected by FASER in Run~3 (2022-24). The number at FASER2 in the FPF at HL-LHC (2027-36) is about 1000 times larger. However, despite the large fluxes, the potential of muon measurements is dimmed for two reasons: (i) The origin of this muon flux is currently not well understood. Besides in-flight decay of pions and kaons, muons are also produced in secondary interactions occurring in the downstream infrastructure, e.g., in hadronic showers caused by very high energy neutrons that hit the TAN. (ii) Along their trajectories to the FPF, the muons will likely pass through various LHC magnets and undergo multiple Coulomb scattering in the rock, which will both lead to a change of the muon direction. Therefore, the direction of the muon at the FPF cannot be correlated anymore with the direction of the muon at its point of production. Dedicated studies are needed to understand the origin, the trajectories, and the physics potential of muons passing through the FPF. 

As we have seen, the FPF experiments will provide complementary data on far-forward hadron production. While the LHCf experiment has previously measured the neutral pion and neutron production cross sections~\cite{LHCf:2015rcj, LHCf:2015nel, LHCf:2018gbv, Adriani:2020ujx}, the FPF experiments can make complementary measurements of the charged pion production cross section by using neutrinos and possibly also muons as proxies. A combination of data from the FPF and LHCf will constrain the hadron composition in the far-forward region. As the Muon Puzzle is assumed to be of soft-QCD origin, there is a strong connection to the QCD program of the FPF and the measurements will help to better understand multi-particle production in air showers.

\subsection{Prompt Atmospheric Neutrino Fluxes}
\label{sec:atmosphericneutrinos}

In 2013, IceCube reported the first observation of a diffuse astrophysical neutrino flux at energies above $\sim 30\,\mathrm{TeV}$~\cite{IceCube:2013cdw, IceCube:2013low}, and subsequent measurements up to energies of a few PeV~\cite{IceCube:2014stg, IceCube:2015qii, IceCube:2020acn, IceCube:2020wum, IceCube:2021rpz} have further improved the understanding of the astrophysical flux since then. However, prompt atmospheric neutrinos produced in air showers yield an important background to these measurements and introduce large uncertainties in the determination of spectral index and flux normalization. The dominant contribution to the prompt neutrino flux comes from charm hadron production and decay in the atmosphere. Calculations of the prompt neutrino fluxes~(see, e.g., Refs.~\cite{Gondolo:1995fq, Enberg:2008te, Fedynitch:2015zma, Halzen:2016thi} and Ref.~\cite{Goncalves:2021yvw}) produced in high-energy cosmic ray collisions in the atmosphere depend sensitively on the $D$ meson production cross section in $pA$ collisions, which in turn depends on both the proton and nuclear parton distributions at $Q\simeq m_c$. 

\begin{figure}[tp]
\centering
\includegraphics[width=0.49\textwidth]{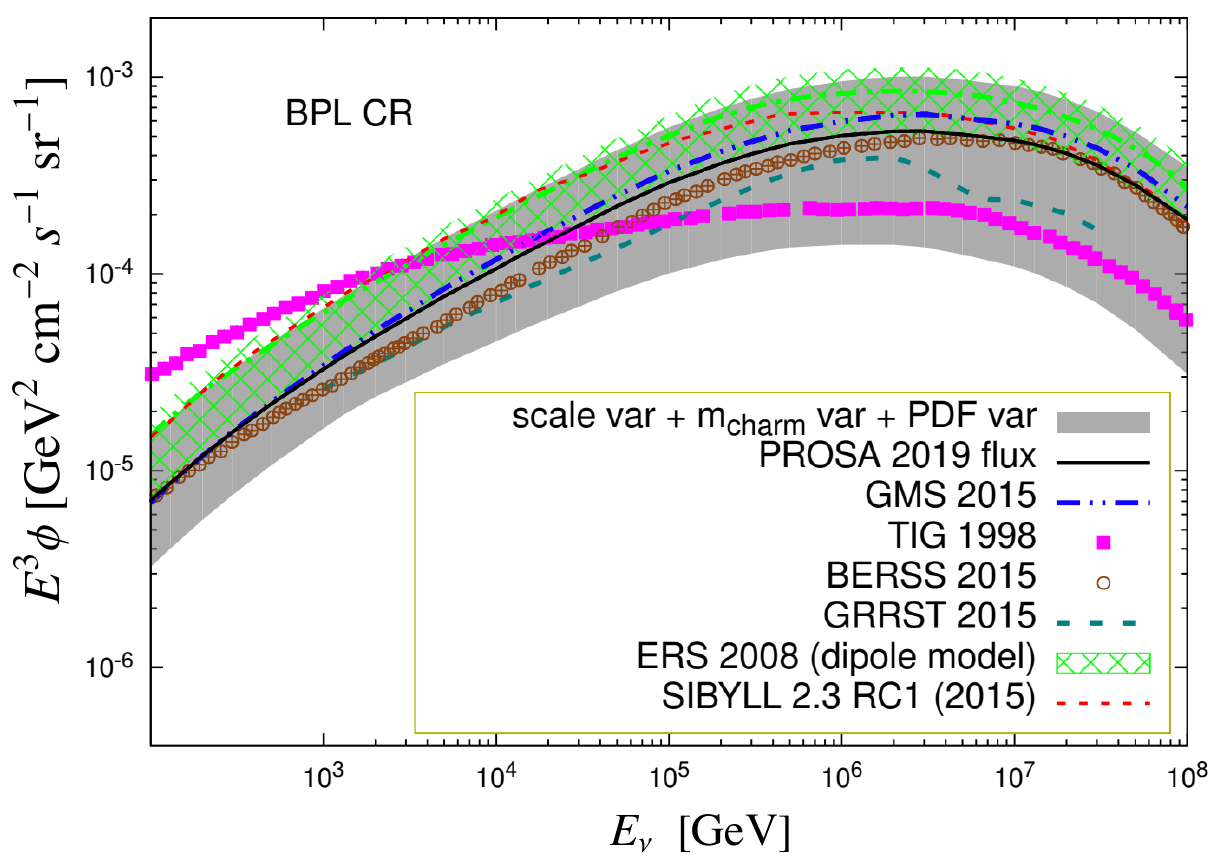}
\includegraphics[width=0.49\textwidth]{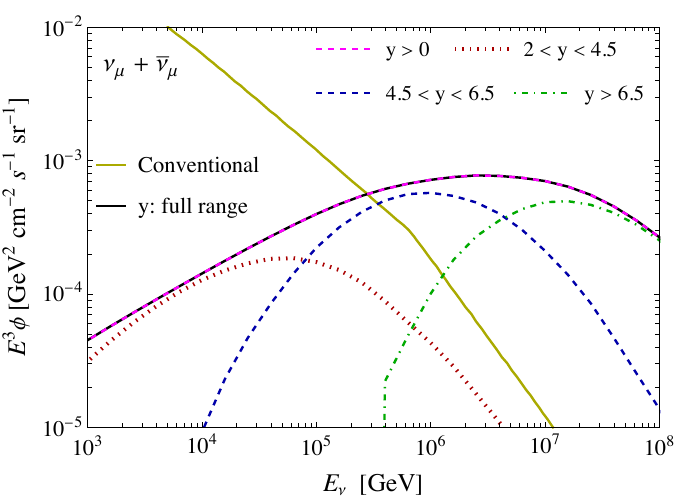} 
\caption{Left:~Comparison of the prompt atmospheric muon neutrino flux, $\phi$, as a function of the neutrino energy, $E_\nu$, assuming a broken power law (BPL) for the incident cosmic ray flux, from recent calculations~\cite{Zenaiev:2019ktw,Garzelli:2015psa,Gondolo:1995fq,Bhattacharya:2015jpa,Gauld:2015kvh,Enberg:2008te,Fedynitch:2015zma} following Ref.~\cite{Zenaiev:2019ktw}. Right:~Prompt atmospheric neutrino fluxes from different collider rapidity ranges~\cite{Jeong:2021vqp} for charm production and the conventional atmospheric neutrino flux from Ref.~\cite{Honda:2006qj}. In the prompt flux evaluation, the $pp$ charmed hadron energy distributions are scaled to account for the air target average atomic number $A=14.5$. The calculation of prompt atmospheric fluxes involves $pA$ collisions in a wide range of center-of-mass energies, including, but not limited to, LHC energies. The peak of $E^3\phi$ is particularly sensitive to collisions at LHC energies. }
    \label{fig:prompt_nuflux_comp}
\end{figure}

Predictions of the prompt flux have been made with charm production models consistent with LHC data in the forward region~\cite{Gauld:2015kvh, Gauld:2015yia, Gauld:2016kpd, Garzelli:2015psa, Bhattacharya:2015jpa, Bhattacharya:2016jce, Zenaiev:2019ktw}. The left panel of \cref{fig:prompt_nuflux_comp} shows several evaluations of the prompt atmospheric $\nu_\mu+\bar{\nu}_\mu$ flux using a broken power law incident cosmic ray flux to illustrate the range of uncertainties associated with the charm mass, PDF variation, scale variations, and model. The atmospheric flux of high-energy neutrinos ($E_{\nu} >$ 10$^5$ GeV) comes predominantly from the production of charmed mesons with large Feynman-$x$ ($x_F \sim 0.1$) and small transverse momentum. The main contribution to the charm production cross section originates from partons with very small values of parton momentum fraction, $x$, in the target air nucleons and very large $x$ values in the cosmic ray nucleons. The dominant regions of the $c\bar{c}$ phase space relevant to the prompt atmospheric neutrino flux are discussed in Ref.~\cite{Goncalves:2017lvq}. The rapidity dependence of the prompt neutrino flux as a function of neutrino energy is shown in the right panel of \cref{fig:prompt_nuflux_comp}.  The solid prompt neutrino flux curve and the $y>0$ dashed curve overlap. For neutrino energies $E_\nu \gtrsim 10^5 - 10^6~{\rm GeV}$, where the prompt flux dominates the conventional neutrino flux, prompt atmospheric neutrinos are mostly from the charm produced at rapidities $y \gtrsim 4.5$. 

Predictions of the prompt atmospheric neutrino flux can be improved by tuning the charm production models to LHC data, in particular in the forward region~\cite{Gauld:2015kvh, Gauld:2015yia, Gauld:2016kpd,Zenaiev:2019ktw}. The information to be provided by the FPF will make possible pinning down the small- and large-$x$ proton and nuclear PDFs from charm production in $pp$ and $pA$ collisions, and hence improve theoretical predictions for the neutrino flux from charm. These constraints on the proton and nuclear structure will be fully complementary to those arising from other experiments operating at the same time, in particular from the EIC~\cite{Garcia:2020jwr,Khalek:2021ulf}. Existing calculations of the prompt neutrino flux at the FPF are also very uncertain~\cite{Bai:2020ukz, Maciula:2020dxv, Kling:2021gos}. Another approach to improve the calculations of the prompt neutrino flux will be the exploitation of the recently computed NNLO QCD corrections to heavy quark production~\cite{Catani:2020kkl, Czakon:2021ohs}, which will reduce the currently limiting theoretical uncertainty arising from missing higher-order uncertainties in both current FPF neutrino flux and prompt atmospheric neutrino flux predictions.

The future FPF measurement of forward neutrinos could thereby provide critical information on perturbative charm and associated charm production at Feynman $x_F$ close to 1. These processes almost certainly yield the dominant atmospheric background for measurements of cosmic neutrinos above 100 TeV and are among the largest uncertainties in determining the spectral index and flux normalization of the diffuse astrophysical neutrino flux~\cite{IceCube:2020wum}. Measurements at the FPF will therefore also provide crucial information to reduce the uncertainties in future measurements of the astrophysical neutrino flux in large-scale neutrino telescopes.

\section{Conclusions and Outlook}
\label{sec:conclusions}

In 2022, the LHC will begin Run~3 with an expected integrated luminosity of $150~\ifb$ collected over 3 years. At the conclusion of Run~3, after a 2- to  3-year Long Shutdown, the HL-LHC will continue running for another decade, with a target integrated luminosity of $3~\iab$.  The importance of making good use of the remaining lifetime of the LHC cannot be overstated.

At present the far-forward region is largely unexplored at the LHC. For Run~3 and the HL-LHC, the total $pp$ inelastic cross section is of the order of $100~\text{mb}$, but the overwhelming majority of these inelastic collisions produce particles that disappear down the beam pipe and are undetected.  It is now clear that many physics opportunities are being missed in the far-forward region.  As an indication of promise, a small 11 kg detector, constructed from parts recycled from other experiments and placed in the far-forward region for 6 weeks, has recently detected several neutrino candidates, the first such events ever recorded at a collider. In Run~3, the FASER, FASER$\nu$, and SND@LHC detectors will begin operating on or just slightly off the beam collision axis in the far-forward region of the ATLAS IP.  Shielded from the ATLAS IP by $\sim100$ m of concrete and rock, these small detectors will detect $\sim 10,000$ neutrinos with energies in the currently almost completely unexplored window between 400 GeV and 6 TeV, and they will sensitively probe new regions of parameter space in many BSM models that predict new light, weakly-interacting particles.  

The proposed FPF will extend this nascent far-forward research program into the HL-LHC era by providing the dedicated space and infrastructure to house a suite of far-forward experiments.  As discussed in \secref{facility}, two options for the FPF site are currently under study. In the first, UJ12, an existing cavern along the LHC tunnel, would be enlarged with alcoves to house up to 3 experiments along the beam collision axis.  UJ12 is roughly 500 m from the ATLAS IP on the Swiss side, near the current location of the FASER and FASER$\nu$ detectors.  The second option is to construct a purpose-built facility, consisting of a new cavern and shaft on the French side of ATLAS.  This option would provide a space roughly 617 to 682 m from the ATLAS IP along the beam collision axis for a large number of experiments.  Both options are shielded from the ATLAS IP by at least 100 m of rock and concrete, providing excellent locations for both SM and BSM studies.  The UJ12 Alcoves site is the less expensive option, while the purpose-built facility would provide far more flexibility during both the construction and operation phases, as discussed in \secref{facility}.

In \secref{experiments}, we have presented several possible experiments that could be housed in the FPF.  The FPF will provide an ideal location for upgrades of existing experiments, which are currently located in tunnels where the available space and infrastructure are severe limitations, as well as for new technologies.  The experiments discussed include FASER2, a magnetic spectrometer and tracking detector targeting searches for new long-lived particles; FASER$\nu$2, an emulsion detector designed to detect a million flavor-tagged neutrinos during the HL-LHC era; AdvSND, consisting of two electronic detectors, one located in the FPF and targeting charm physics, tau neutrinos, and tests of lepton universality, and a near detector, placed closer to the IP and at smaller pseudorapidities to overlap with LHCb's coverage and reduce systematic uncertainties; FLArE, a $\sim 10$-tonne liquid argon TPC, designed for neutrino studies and dark matter searches; and FORMOSA, an experiment dedicated to searches for mCPs and related signatures.

Many models of BSM physics predict the existence of new light and weakly interacting particles. In \secref{bsm}, we discussed the potential for experiments at the FPF to discover these new states. We have shown that these experiments will be sensitive to a variety of signatures associated with these new particles, including (i) the displaced decays of new long-lived particles, e.g., dark photons, dark scalars, HNLs, and ALPs, inside the FASER2 detector; (ii) the scattering of new stable particles, e.g., dark matter, with the dense neutrino detectors FASER$\nu$2, AdvSND, and FLArE; and (iii) the anomalous energy depositions induced by mCPs at FORMOSA. For each signature, we have presented the associated reach for selected benchmark models and discussed how the FPF can help to address outstanding problems in particle physics, such as the particle nature of DM, the origin of neutrino masses, the strong CP problem, the hierarchy problem, the matter-antimatter asymmetry of the universe, and inflation, as well as currently unresolved experimental anomalies.

In \cref{sec:neutrinos}, we discussed the FPF's potential for neutrino physics. The FPF will utilize the LHC's beam of high-energy neutrinos and observe neutrino interactions with unprecedented statistics in the several hundreds of GeV to TeV energy range. For example, the FASER$\nu$2 detector with a target mass of 20 tonnes will detect as many as $\mathcal{O}(10^5)$ electron neutrinos, $\mathcal{O}(10^6)$ muon neutrinos, and $\mathcal{O}(10^3)$ tau neutrinos in the HL-LHC era.  This will allow one to measure the neutrino cross section for deep-inelastic, resonant, and quasi-elastic neutrino-nucleus scatterings, as well as neutrino-electron scattering at TeV energies; to test lepton universality in neutrino scattering; to probe quark-hadron duality in the weak sector through shallow inelastic scatterings; and to study final-state hadronic effects in neutrino interactions. Finally, FPF neutrino measurements will also have the potential to discover or constrain BSM neutrino physics, such as non-standard interactions, neutrino dipole moments, and sterile neutrino oscillations. 

In~\cref{sec:qcd}, we discussed QCD challenges and opportunities, ranging from the concrete exploration of different factorization frameworks not easily accessible at the major LHC central detectors, to the possibility of bounding (nuclear) PDFs in regions not yet constrained by any other experiment, and better investigating the interplay between perturbative and non-perturbative QCD elements. All of these aspects are of great relevance for our understanding of QCD and future collider experiments.  In addition, they will provide more reliable and less uncertain descriptions of the interactions of cosmic rays with the atmosphere, which, in turn, will sharpen and possibly help resolve a number of longstanding astroparticle physics puzzles, as discussed in~\cref{sec:astro}. We envisage the opportunity of joined analyses of FPF data together with complementary data from the HL-LHC phase and forthcoming colliders, such as the EIC and the LHeC, which together will advance our conceptual understanding of QCD theory.

Studying the compatibility of FPF data with those from high-energy astroparticle physics experiments, such as very large volume neutrino telescopes and extended air shower observatories, will provide additional insights on the virtues and limitations of the SM. Together with the possibility of distinguishing neutrinos from antineutrinos and recognizing different flavors, a crucial element for the success of this program is the possibility of extending the neutrino rapidity coverage of the FPF towards values at least as low as $\sim 6.5$. This will provide information that is complementary to the LHC central detectors, which are sensitive to hadrons and charged leptons at smaller rapidities, but are not capable of measuring individual neutrinos. The possibility of combining data from the FPF and ATLAS through specific triggering schemes and timing techniques currently under development will further enhance QCD opportunities, opening the road for the detection of new channels involving simultaneously a very forward neutrino and a more central object, not accessible at the FPF by itself. The possibility of measuring/reconstructing Feynman $x$ distributions will also be very welcome, especially in view of astrophysical applications. 

Organized interest in the FPF began just over a year ago with the Snowmass Letter of Interest~\cite{SnowmassFPF} and continued in two dedicated FPF meetings in November 2020~\cite{FPFKickoffMeeting} and May 2021~\cite{FPF2ndMeeting}.  In this work, we have summarized the current status.  Further work will address key aspects, including specifying the FPF location and cost and designing the experiments that will be housed in the FPF.  We anticipate additional meetings to discuss progress on the FPF in the coming months and a more comprehensive paper that will be submitted to the Snowmass process in Spring 2022, detailing progress in quantifying the FPF's diverse physics capabilities and exploring its unique physics potential~\cite{Feng:2022inv}.

\section*{Acknowledgements}

We thank the participants of the FPF meetings and the Snowmass working groups for discussions that have contributed both directly and indirectly to this study.  We gratefully acknowledge the invaluable support of the CERN Physics Beyond Colliders study group and the work of CERN technical teams related to civil engineering studies (SCE-DOD), safety discussions (HSE-OHS, HSE-RP, EP-DI-SO), integration (EN-ACE), and discussions on services (EN-CV, EN-EL, EN-AA) and simulations (SY-STI). 
L.~A.~Anchordoqui is supported by U.S.~National Science Foundation (NSF) Grant PHY-2112527.
A.~Ariga is supported by JSPS KAKENHI Grant JP20K23373 and the European Research Council (ERC) under the European Union's Horizon 2020 research and innovation programme (Grant 101002690).
T.~Ariga acknowledges support from JSPS KAKENHI Grant JP19H01909. 
B.~Batell is supported by U.S.~Department of Energy (DOE) Grant DE–SC0007914 and by PITT PACC.
The work of J.~Bramante is supported by NSERC.
F.~G.~Celiberto acknowledges support from the INFN/NINPHA project and thanks the Universit\`a degli Studi di Pavia for the warm hospitality. 
G.~Chachamis is supported by the Funda\c{c}\~ao para a Ci\^encia e a Tecnologia (Portugal) under project CERN/FIS-PAR/0024/2019 and contract ``Investigador auxiliar FCT - individual Call CEECIND/03216/2017 and acknowledges funding from the European Union’s Horizon 2020 research and innovation programme (Grant 824093).
P.~B.~Denton is supported by DOE Grant DE-SC0012704.
Y.~Du was supported by National Science Foundation of China (NSFC) Grants 12022514 and 11875003.
The work of J.~L.~Feng~is supported in part by NSF Grants PHY-1915005 and PHY-2111427, Simons Investigator Award \#376204, Simons Foundation Grant 623683, and Heising-Simons Foundation Grants 2019-1179 and 2020-1840.
The work of M.~Fieg, T.~B.~Smith, and Y.-D.~Tsai is supported in part by NSF Grant PHY-1915005.
M.~Fucilla, M.~M.~A.~Mohammed, and A.~Papa acknowledge support from the INFN/QFT@COLLIDERS project.
The work of M.~V.~Garzelli was supported in part by the Bundesministerium f\"ur Bildung und Forschung (BMBF). 
V.~P.~Goncalves was partially financed by the Brazilian funding agencies CNPq, FAPERGS, and INCT-FNA (process number 464898/2014-5).
M.~Guzzi is supported in part by NSF Grants PHY-1820818 and PHY-2112025.
J.~C.~Helo acknowledges support from ANID FONDECYT-Chile Grant 1201673 and from ANID Millennium Science Initiative Program ICN2019 044.
K.~Jodłowski is supported by the National Science Centre, Poland, Grant 2015/18/A/ST2/00748.
Ahmed~Ismail and R.~Mammen Abraham are supported by DOE Grant DE-SC0016013.
Ameen~Ismail is supported by NSERC (reference number 557763) and by NSF Grant PHY-2014071.
Y.~S.~Jeong is supported by the National Research Foundation of Korea (NRF) grant funded by the Korea government (MSIT: Ministry of Science and ICT) (2021R1A2C1009296).
The work of F.~Kling is supported by DOE Grant DE-AC02-76SF00515 and by the Deutsche Forschungsgemeinschaft under Germany's Excellence Strategy -- EXC 2121 Quantum Universe -- 390833306.
P.~Nadolsky is supported by DOE Grant DE-SC0010129.
H.~Otono acknowledges support from JSPS KAKENHI Grant JP20H01919.
V.~Pandey is supported by DOE Grant DE-SC-0009824.
M.~H.~Reno is supported in part by DOE Grant DE-SC-0010113.
The work of A.~Ritz is supported by NSERC, Canada.
The work of D.~Sengupta is supported by NSF Grant PHY-1915147. 
T.~Sj\"ostrand is supported by Swedish Research Council Grant 2016-05996. 
D.~Soldin acknowledges support from the NSF Grant PHY-1913607.
S.~Trojanowski is supported by the grant ``AstroCeNT: Particle Astrophysics Science and Technology Centre'' carried out within the International Research Agendas programme of the Foundation for Polish Science financed by the European Union under the European Regional Development Fund, and by the Polish Ministry of Science and Higher Education through its scholarship for young and outstanding scientists (decision 1190/E-78/STYP/14/2019).
K.~Xie is supported by DOE Grant DE-FG02-95ER40896, NSF Grant PHY-1820760, and in part by PITT PACC.
Parts of this work have been supported by Fermi Research Alliance, LLC under Contract DE-AC02-07CH11359 with the DOE.


\bibliography{references}

\end{document}